\documentclass[11pt,a4paper]{article}
\pdfoutput=1

\usepackage{dcolumn}
\usepackage{bm}
\usepackage{amsthm}
\usepackage{setspace}
\usepackage[normalem]{ulem}
\usepackage{relsize}
\usepackage{cancel}
\usepackage{verbatim}
\usepackage{enumitem}
\usepackage{mathtools}
\usepackage{empheq}
\usepackage{dsfont}
\usepackage{amsmath}
\usepackage{slashed}

\usepackage{graphicx}
\usepackage{caption}
\usepackage{subcaption}

\usepackage{jheppub}

\allowdisplaybreaks

\newcommand{\numthis}{\stepcounter{equation}\tag{\theequation}}

\usepackage{amsthm}

\newcommand{\Lie}[1]{\mathcal{L}_{#1}\,}
\newcommand{\Lieb}[1]{\bar{\mathcal{L}}_{#1}\,}

\newcommand{\wn}{\bar\nabla}

\newcommand{\ve}[2]{e^{#1}{}_{#2}} 

\newcommand{\dels}{\delta_\sigma^W}
\newcommand{\epslif}{\varepsilon_\text{lif}}
\newcommand{\epstil}{\tilde\varepsilon}

\newcommand{\tr}{\operatorname{Tr}}

\newcolumntype{I}[1]{>{\centering\arraybackslash$}m{#1}<{$}}

\title{Lifshitz Anomalies, Ward Identities and Split Dimensional Regularization}

\author{Igal Arav,}
\author{Yaron Oz,}
\author{Avia Raviv-Moshe}
\affiliation{Raymond and Beverly Sackler School of Physics and Astronomy, Tel-Aviv University, 55 Haim Levanon street, Tel-Aviv, 69978, Israel}
\emailAdd{aravigal@post.tau.ac.il}
\emailAdd{yaronoz@post.tau.ac.il}
\emailAdd{aviaravi@mail.tau.ac.il}

\abstract{
We analyze the structure of the stress-energy tensor correlation functions in Lifshitz field theories and construct the corresponding anomalous Ward identities.
We develop a framework for calculating the anomaly coefficients
that employs a split dimensional regularization and the pole residues.
We demonstrate the procedure by calculating the free scalar Lifshitz scale anomalies in $2+1$ spacetime
dimensions. We find that the analysis of the regularization dependent trivial terms requires a curved spacetime description without a foliation structure.
We discuss potential ambiguities in Lifshitz scale anomaly definitions.
}

\keywords{Anomalies in Field and String Theories, Space-Time Symmetries}

\arxivnumber{1612.03500}

\begin{document}
\maketitle
\flushbottom

\section{Introduction}

Lifshitz scale symmetry is a symmetry under which space and time scale differently, 
\begin{equation}
t \to \lambda^z t, \qquad x^i \to \lambda x^i, \qquad i=1,\ldots,d  \ ,
\label{scale}
\end{equation}
where $z$ is the dynamical critical exponents which represents the anisotropy between space and time, and $d$ is the number of spatial dimensions. 
Non-relativistic field theories that exhibit a Lifshitz scale symmetry have attracted much attention in recent years.
In nature Lifshitz scaling is a property of certain low energy systems of condensed matter, which exhibit quantum criticality (see for example \cite{Sachdev:2011cup}). The study of holographic Lifshitz systems has been initiated e.g.\ in 
\cite{Son:2008ye,Kachru:2008yh}

Similary to the relativistic trace anomalies (for brief reviews see e.g.\  \cite{Duff:1993wm,Deser:1996na}), non-relativistic quantum field theories may exhibit Lifshitz scale anomalies, where the classical Lifshitz symmetries are broken by the quantum corrections.
A cohomological analysis provides a general framework to determine the possible structures
of Lifshitz scale anomalies \cite{Arav:2014goa, Arav:2016xjc,Pal:2016rpz,Auzzi:2015fgg} (see also \cite{Griffin:2011xs,Baggio:2011ha} for 
a $z=2$ example). Naturally, it is of importance to calculate the anomaly coefficients, which carry information
about the quantum field theory.
A heat kernel calculation of anomaly coefficients of a Lifshitz scalar in $2+1$ spacetime dimensions
has been performed in \cite{Baggio:2011ha}, and in $3+1$ dimensions using a different regularization scheme
in \cite{Adam:2009gq}.
The aim of this work is to develop a general scheme for such field theory calculations.
In order to do that we will  first analyze the general structure of correlation functions of the stress-energy tensor in Lifshitz field theories and
analyze the corresponding anomalous Ward identities.
We will encounter a subtle ambiguity in the definition of the anomaly coefficients and will clarify its meaning.
Next, we will develop a general framework for calculating the anomaly coefficients.
It consists of two elements: a split dimensional regularization  \cite{Leibbrandt:1996np,Leibbrandt:1997kh}
where space (momentum) and time (frequency) integrals are 
regulated separately, and pole residue calculations  \cite{Anselmi:2007ri} that allow to extract the anomaly
coefficients without  a full calculation of the correlation functions.
 In order to demonstrate the power of the latter
we will calculate using the pole residues the trace anomaly coefficients of a relativistic scalar in two and four
spacetime dimensions and obtain the known results.
We will then apply the complete framework  to calculate the anomaly coefficients of a Lifshitz scalar in $2+1$ spacetime dimensions.
The results agree with the heat kernel calculation  \cite{Baggio:2011ha}. Following the discussion in \cite{Arav:2016xjc}, we will show that the analysis of the regularization dependent trivial terms arising from this calculation requires a curved spacetime description that violates the Frobenius condition (and therefore has no foliation structure).

This paper is organized as follows. In section \ref{WardIdentities} we present the Ward identities corresponding to the symmetries of the field theories we consider, in terms of the expectation value of the conserved currents in curved spacetime, and in terms of the flat space correlation functions of the stress-energy tensor. We also discuss a possible ambiguity in the Lifshitz anomaly coefficients.
In section \ref{RegAndRenorm} we explain the method by which we use split dimensional regularization to extract the Lifshitz anomaly coefficients from the flat space field theory correlation functions. In both sections we start by reviewing the case of a conformal field theory and then discuss the Lifshitz case.  In section \ref{sec:ConformalAnom} we apply 
the aforementioned method of extracting scale anomaly coefficients from the pole residues of correlation functions in dimensional regularization to calculate the trace anomaly coefficients for a free relativistic scalar, both in two and four spacetime dimensions. In section \ref{Sec:z2LishitzFreeScalarResults} we 
use this method in its split dimensional regularization version to calculate the anomaly coefficients for a $z=2$ free Lifshitz scalar in $2+1$ dimensions.
We conclude in section \ref{SummaryandOutlook}.

\section{Ward Identities and Correlation Functions}
\label{WardIdentities}

In this section we present the form of the Ward identities corresponding to the symmetries of the field theories we consider, both in terms of the expectation values of the conserved currents over curved spacetime, and of the flat space correlation functions of these currents. These include the anomalous scaling symmetry as well as the other symmetries that are assumed to be non-anomalous. The anomalous Ward identities will later be used to extract the anomaly coefficients from the flat space correlation functions. 
We start by reviewing the conformal case, and then discuss the Lifshitz (non-relativistic) case.
 
\subsection{Review of the Conformal Case}
\label{ConformalWardIdentities}

\subsubsection{Symmetries and Ward Identities In Curved Spacetime}

Consider a conformal field theory in $d$ dimensions. The theory can be coupled to a curved spacetime manifold equipped with a background metric $g_{\mu\nu}$ (or alternatively vielbein structure $\ve{a}{\mu}$). Suppose this theory is described by the classical action $S(g_{\mu\nu},\{\phi\})$, where $\{\phi\}$ stands for the dynamic fields of the theory. This action reduces to the flat space action of the theory when $g_{\mu\nu} \rightarrow \eta_{\mu\nu}$, and is invariant under the following symmetries:
\begin{enumerate}
\item Diffeomorphisms, given by:
\begin{equation}
\delta^D_\xi g_{\mu\nu} = \nabla_\mu \xi_\nu + \nabla_\nu \xi_\mu,
\end{equation}
or in vielbein notation (here we also include local Lorentz transformations):
\begin{equation}
\delta^D_\xi \ve{a}{\mu}  = \xi^\nu \nabla_\nu \ve{a}{\mu} + \nabla_\mu \xi^\nu \ve{a}{\nu},
\qquad
\delta^L_\alpha \ve{a}{\mu}  =  - \alpha^a{}_b \ve{b}{\mu}.
\end{equation}

\item Weyl transformations, given by:
\begin{equation}
\dels g_{\mu\nu} = 2\sigma g_{\mu\nu},
\qquad
\dels \ve{a}{\mu} = \sigma \ve{a}{\mu}.
\end{equation}
\end{enumerate}

The stress-energy tensor is defined as the variation of the action with respect to the metric:
\begin{equation}
\label{eq:StressTensorDef}
{T^{\mu \nu }} = \frac{2}{{\sqrt{|g|} }}\frac{{\delta S}}{{\delta {g_{\mu \nu }}}},
\end{equation}
where $ g \equiv \det g_{\mu\nu} $.
The stress-energy tensor satisfies classical Ward identities corresponding to the above symmetries.
From diffeomorphism invariance (and local Lorentz invariance) we have the conservation and symmetry of the stress-energy tensor:
\begin{equation}\label{CurvedConfWard:DiffWardIdent}
\nabla_\mu T^{\mu\nu} = 0, \qquad T^{\mu\nu}=T^{\nu\mu},
\end{equation}
and from Weyl invariance we get the traceless property:
\begin{equation}\label{CurvedConfWard:WeylWardIdent}
T^\mu_\mu = 0.
\end{equation}
When conformal anomalies are present, the expectation value of the stress-energy tensor no longer satisfies identity \eqref{CurvedConfWard:WeylWardIdent}. Instead it is modified:
\begin{equation}\label{CurvedConfWard:AnomWeylWardIdent}
\langle T^\mu_\mu \rangle = \mathcal{A},
\end{equation}
where $\mathcal{A}(g_{\mu\nu})$ is a local scalar function of the metric. The possible form of $\mathcal{A}$ is restricted by the Wess-Zumino consistency condition to a linear combination of possible expressions (see e.g.\ \cite{Deser:1993yx,Deser:1996na,Bonora:1985cq,Bonora:1983ff,Boulanger:2007st} for details): the Euler density of the background manifold $E_d$ (A-type anomaly), the various Weyl invariant densities of the manifold (B-type anomalies) and other terms that can be cancelled by adding local counterterms to the curved space effective action of the theory (trivial terms). 

In $d=2$ dimensions, the only such expression is the Euler density $E_2 = R$ (where $R$ is the Ricci scalar of the manifold), so that:
\begin{equation}
\mathcal{A} = \beta R .
\end{equation}
In $d=4$ dimensions, these expressions consist of the Euler density $E_4$ (A-type anomaly), the Weyl tensor squared $W^2$ (B-type anomaly) and $\Box R$ (trivial term):
\begin{equation}
\label{eq:GeneralExpreForAnomalyRel4d}
\mathcal{A} = \beta_a W^2 + \beta_b E_4 + \beta_c \Box R .
\end{equation}
While these expressions are universal, the value of their coefficients depends on the content of the theory. Note that the coefficients of the trivial terms may depend on the regularization scheme, but those of the anomalies are regularization independent. One method to obtain the value of these coefficients is to extract them from the flat space $n$-point correlation functions of the stress-energy tensor.

For later reference, for a free scalar in $d=2$ dimensions, the anomaly coefficient has been determined to be (see e.g.
\cite{Polchinski}):
\begin{equation}
\label{eq:CoeefsRel2d}
\beta=-\frac{1}{48\pi}.
\end{equation}
For a free scalar in $d=4$ dimensions, the coefficients have been determined to be\footnote{The coefficient $\beta_c$ here corresponds to a dimensional regularization scheme. The values cited here of the coefficients $\beta, \beta_a, \beta_b, \beta_c$ correspond to our conventions for the definitions of the stress-energy tensor and the Riemann tensor.} 
(see \cite{Capper:1974ic,Duff:1977ay,Duff:1993wm,Birrell:1982ix,Osborn:1993cr,Coriano:2012wp} and references therein): 
\begin{equation}
\label{eq:CoeefsRel4d}
{\beta _a} = \frac{3}{2}\frac{1}{{2880{\pi ^2}}}, \quad {\beta _b} =  - \frac{1}{2}\frac{1}{{2880{\pi ^2}}}, \quad {\beta _c} =   \frac{1}{{2880{\pi ^2}}}.
\end{equation}

\subsubsection{Ward Identities For Correlation Functions}

We denote by $W(g_{\mu\nu})$ the effective action of the conformal field theory on curved spacetime, defined by:
\begin{equation}
\label{eq:GeneratingFunctional}
e^{i\,W(g_{\mu\nu})} \equiv \int D\phi\, e^{i\,S(g_{\mu\nu})} ,
\end{equation}
where $\phi$ stands for the dynamic fields in the theory. 
Two types of correlation functions can be defined for the stress-energy tensor: First, the connected Feynman $n$-point correlation functions, given by the expectation value\footnote{In the following sections we may omit the F subscript when referring to Feynman correlation functions.}:
\begin{equation}\label{CorrConfWard:FeynmanCorrFuncDef}
\begin{split}
&\left\langle \mathcal{O}_1(x_1) \ldots  \mathcal{O}_n (x_n) \right\rangle_F \equiv \\
&\qquad \frac{1}{Z_0} \int D\phi\,\mathcal{O}_1(x_1) \ldots  \mathcal{O}_n (x_n)\, e^{i\,S} - \,\text{Non-connected terms},
\end{split}
\end{equation}
where $\mathcal{O}_1,\ldots,\mathcal{O}_n$ are local functions of the dynamic and the background fields, and $Z_0$ is the flat space partition function. 
Alternatively, one may define the correlation functions as variations of the effective action $W$ with respect to the metric:
\begin{equation}
\label{eq:nPointFunctionDef}
\left\langle T^{\mu _1 \nu _1}\left(x_1\right)\ldots T^{\mu _n \nu _n}\left( {{x_n}} \right) \right\rangle _W \equiv  \frac{(-i)^{n-1} 2^n}{\sqrt{|g(x_1)|}\ldots\sqrt{|g(x_n)|}} {\frac{{{\delta ^n}W}}{{\delta {g_{{\mu _1}{\nu _1}}}\left( {{x_1}} \right)...\delta {g_{{\mu _n}{\nu _n}}}\left( {{x_n}} \right)}}} .
\end{equation}
We will refer to these as the variational correlation functions.
Since the stress-energy tensor itself depends on the metric, these two types of correlation function do not coincide. One obtains the following relations between the two types of two point functions and three point functions in flat spacetime\footnote{In deriving these formulas, one point functions have been dropped as they correspond to massless ``tadpole'' diagrams and therefore vanish in the flat space limit in a dimensional regularization scheme.} (see \cite{Coriano:2012wp}):
\begin{align}\label{eq:2pointDefs2}
&\left\langle T^{\mu\nu}(x) T^{\rho\sigma}(y) \right\rangle_W = \left\langle T^{\mu\nu}(x) T^{\rho\sigma}(y) \right\rangle_F , \\
\begin{split}\label{eq:3pointDefs}
&  \langle T^{\mu\nu}(x) T^{\rho\sigma}(y)T^{\alpha\beta}(z) \rangle_W = 
\langle T^{\mu\nu}(x) T^{\rho\sigma}(y)T^{\alpha\beta}(z) \rangle_F \\
&\quad
- 8i \left[ \left\langle \frac{\delta^2 S}{\delta g_{\mu\nu}(x)\delta g_{\rho\sigma}(y)}\frac{\delta S}{\delta g_{\alpha\beta}(z)} \right\rangle_F 
+\left\langle \frac{\delta^2 S}{\delta g_{\mu\nu}(x)\delta g_{\alpha\beta}(z)}\frac{\delta S}{\delta g_{\rho\sigma}(y)} \right\rangle_F \right.\\
&\qquad\qquad\qquad\qquad\qquad\qquad\qquad\qquad
\left.+\left\langle \frac{\delta^2 S}{\delta g_{\rho\sigma}(y)\delta g_{\alpha\beta}(z)}\frac{\delta S}{\delta g_{\mu\nu}(x)} \right\rangle_F \right] .
\end{split}
\end{align}

The Ward identities satisfied by the variational correlation functions can be obtained by taking derivatives of the curved spacetime Ward identities \eqref{CurvedConfWard:DiffWardIdent}--\eqref{CurvedConfWard:WeylWardIdent} (or \eqref{CurvedConfWard:AnomWeylWardIdent} in the presence of conformal anomalies) with respect to the background metric. These identities have been derived in \cite{Coriano:2012wp, Osborn:1993cr,Erdmenger:1996yc} for Euclidean signature. We repeat them here modified to our use of Lorentzian signature. 
For diffeomorphism invariance, one obtains the following identities for two and three point correlation functions:
\begin{align}
\label{eq:ConRelativistic2Point}
&\partial^x_\nu\left\langle  T^{\mu\nu}(x) T^{\rho\sigma}(y)\right\rangle_W = 0 , \\
\begin{split}
\label{eq:ConRelativistic3Point}
&\partial^x_\nu \langle  T^{\mu\nu}(x) T^{\rho\sigma}(y) T^{\alpha\beta}(z) \rangle_W \\
&\qquad + i \left[ \langle T^{\rho\sigma}(x)T^{\alpha\beta}(z)\rangle_W\, \partial^\mu\delta(x-y) + \langle T^{\alpha\beta}(x) T^{\rho\sigma}(y) \rangle_W\, \partial^\mu \delta(x-z) \right]\\
&\qquad - i \left[ \eta^{\mu\rho} \langle T^{\nu\sigma}(x) T^{\alpha\beta}(z) \rangle_W + \eta^{\mu\sigma} \langle T^{\nu\rho}(x) T^{\alpha\beta}(z) \rangle_W \right] \partial_\nu \delta(x-y)\\
&\qquad - i \left[ \eta^{\mu\alpha} \langle T^{\nu\beta}(x) T^{\rho\sigma}(y) \rangle_W + \eta^{\mu\beta} \langle T^{\nu\alpha}(x) T^{\rho\sigma}(y) \rangle_W \right] \partial_\nu \delta(x-z) = 0.
\end{split}
\end{align}
For the (anomalous) Weyl invariance, one obtains the following identities:
\begin{align}\label{CorrConfWard:WardIdentForVarCorrWeyl2Point}
&\eta_{\mu\nu} \left\langle T^{\mu\nu}(x) T^{\rho\sigma}(y) \right\rangle_W = -2i \left. \frac{\delta\mathcal{A}(x)}{\delta g_{\rho\sigma}(y)} \right|_\text{flat} , \\
\begin{split}\label{CorrConfWard:WardIdentForVarCorrWeyl3Point}
&\eta_{\mu\nu} \langle T^{\mu\nu}(x) T^{\rho\sigma}(y) T^{\alpha\beta}(z) \rangle_W 
-2i \langle T^{\rho\sigma}(y) T^{\alpha\beta}(z)\rangle_W \left[\delta(x-y) + \delta(x-z)\right]   \\
&\qquad + i \eta^{\rho\sigma}\eta_{\mu\nu} \langle T^{\mu\nu}(x)T^{\alpha\beta}(z) \rangle_W\, \delta(x-y)
+ i \eta^{\alpha\beta}\eta_{\mu\nu} \langle T^{\mu\nu}(x)T^{\rho\sigma}(y) \rangle_W\, \delta(x-z)\\
&\qquad = -4 \left.\frac{\delta^2 \mathcal{A}(x)}{\delta g_{\rho\sigma}(y) \delta g_{\alpha\beta}(z)}\right|_\text{flat} .
\end{split}
\end{align}
Using identities \eqref{CorrConfWard:WardIdentForVarCorrWeyl2Point}--\eqref{CorrConfWard:WardIdentForVarCorrWeyl3Point}, one can extract the conformal anomaly coefficients from the flat space correlation functions.

\subsection{The Lifshitz Case}
\label{LifshitzWardIdentities}

\subsubsection{Symmetries and Ward Identities In Curved Spacetime}
\label{subsubsec:LifshitzSymWardCuvedSpacetime}

Consider a non-relativistic field theory in $d+1$ spacetime dimensions with a Lifshitz scale symmetry of the form 
(\ref{scale}). Assume this theory can be coupled to a curved spacetime manifold equipped with a metric $ g_{\mu\nu} $ (or alternatively vielbein structure $\ve{a}{\mu}$) and a 1-form $t_\alpha$ corresponding to the time direction at each point (or the normalized $n_\alpha$: $n_\alpha n^\alpha=-1$), as described in \cite{Arav:2014goa,Arav:2016xjc} \footnote{As in \cite{Arav:2016xjc} we do not assume that $t_\mu$ satisfies the Frobenius condition. This description is therefore more general than one using an ADM-like decomposition.}. Suppose this theory is described by the classical action $S(g_{\mu\nu},t_\alpha,\{\phi\})$, where $\{\phi\}$ are the dynamic fields of the theory (or alternatively $S(\ve{a}{\mu},t_\alpha,\{\phi\})$ in vielbein formalism). Further assume that this action can be defined such that it reduces to the flat space action of the theory when\footnote{See appendix \ref{app:conventions} for our notations and conventions.} $ g_{\mu\nu} \rightarrow \delta_{\mu\nu} = \operatorname{diag}(-1,1,1,\ldots), \, t_\alpha\rightarrow (1,0,0\ldots)$, and is invariant under the following symmetries:

\begin{enumerate}
\item Time-direction preserving diffeomorphisms (TPD). These are the curved spacetime generalization of space rotations. As explained in \cite{Arav:2014goa,Arav:2016xjc}, in our covariant notation these take the form of standard diffeomorphisms, given by:
\begin{equation}
\delta^D_\xi g_{\mu\nu} = \nabla_\mu \xi_\nu + \nabla_\nu \xi_\mu,
\quad
\delta^D_\xi t_\alpha = \Lie{\xi} t_\alpha = \xi^\beta \nabla_\beta t_\alpha + \nabla_\alpha \xi^\beta t_\beta  ,
\end{equation}
or in vielbein notation (here we also include local Lorentz transformations):
\begin{equation}\label{CurvedLifshitzWard:TPDTransVielbein}
\begin{split}
\delta^D_\xi \ve{a}{\mu} & = \xi^\nu \nabla_\nu \ve{a}{\mu} + \nabla_\mu \xi^\nu \ve{a}{\nu},
\qquad
\delta^D_\xi t^a = \xi^\nu \nabla_\nu t^a,
\\
\delta^L_\alpha \ve{a}{\mu} & =  - \alpha^a{}_b \ve{b}{\mu},
\qquad
\delta^L_\alpha t^a = -\alpha^a{}_b t^b .
\end{split}
\end{equation}

\item Anisotropic Weyl transformations. These are the local generalization of Lifshitz scaling, given by:
\begin{equation}
\begin{split}
& \dels t_\alpha    = 0 ,
\qquad
 \dels (g^{\alpha \beta} t_\alpha t_\beta)  = -2\sigma z (g^{\alpha \beta} t_\alpha t_\beta),
\\
& \dels P_{\alpha\beta}  = 2 \sigma P_{\alpha\beta},
\qquad
 \dels n_\alpha  = z  \sigma n_\alpha, \qquad \delta^W_\sigma n^\alpha  = - z  \sigma n^\alpha,
\end{split}
\end{equation}
where $P_{\mu\nu} = g_{\mu\nu} + n_\mu n_\nu$ is the spatial projector, or alternatively using the vielbeins:
\begin{equation}\label{CurvedLifshitzWard:WeylTransVielbein}
\begin{split}
& \dels (n_a \ve{a}{\mu}) = z \sigma n_a \ve{a}{\mu},
\qquad \dels (P_b^a \ve{b}{\mu}) = \sigma P_b^a \ve{b}{\mu},
\\
& \dels t^b = -z\sigma t^b,
\qquad  \dels n^b=0 .
\end{split}
\end{equation}
\end{enumerate}

We define various field theory currents as the variation of the action with respect to the background fields. The stress-energy tensor can be defined in two possible ways, either using the metric or vielbein descriptions:
\begin{equation}\label{CurvedLifshitzWard:StressTensorDef}
T^{\mu\nu}_{(g)} \equiv \left. \frac{2}{\sqrt{|g|}} \frac{\delta S}{\delta g_{\mu\nu}} \right|_{t_\alpha} ,
\qquad
T_{(e)} {}^\mu {}_a \equiv \left. \frac{1}{e} \frac{\delta S}{\delta \ve{a}{\mu}} \right|_{t^a}.
\end{equation}
We also define the variation of the action with respect to the time direction 1-form:
\begin{equation}
J^\alpha \equiv \left. \frac{1}{\sqrt{|g|}}\frac{\delta S}{\delta t_\alpha} \right|_{g_{\mu\nu}} =
\frac{1}{e} e^{b\alpha} \left. \frac{\delta S}{\delta t^b} \right|_{\ve{a}{\mu}} ,
\end{equation}
as well as its normalized version:
\begin{equation}
\hat{J}^\alpha \equiv \sqrt{|g^{\mu\nu}t_\mu t_\nu|}J^\alpha \, ,
\end{equation}
Note that unlike the relativistic case, these two definitions of the stress-energy tensor do not coincide. Instead they are related by the following identity:
\begin{equation}
T_{(e)}^{\mu\nu} = T_{(g)}^{\mu\nu} + J^\mu t^\nu .
\end{equation}

From the symmetries and these definitions, one can obtain the classical Ward identities these currents satisfy over curved spacetime (see \cite{Arav:2014goa,Arav:2016xjc}). From TPD invariance follow the identities:
\begin{equation}
\nabla_\mu T_{(g)}^\mu{}_\nu =
\hat{J}^\mu \nabla_\nu n_\mu - \nabla_\mu(\hat{J}^\mu n_\nu),
\end{equation}
or equivalently in terms of $T_{(e)}^{\mu\nu}$:
\begin{equation}\label{CurvedLifshitzWard:TPDWardIdentTe}
\begin{split}
T_{(e)[\mu \nu]} &= \hat J_{[\mu}n_{\nu]} ,\\
\nabla_\mu T_{(e)}{}^\mu{}_\nu &= 
\hat{J}^\mu \nabla_\nu n_\mu .
\end{split}
\end{equation}
These identities imply that in the flat spacetime limit, $T_{(g)}^{\mu\nu}$ is symmetric but not conserved, while $T_{(e)}^{\mu\nu}$ is conserved but not symmetric. The latter then corresponds to the conserved energy and momentum currents of the flat space theory, and we therefore choose to use it and the vielbein description in the following sections\footnote{In the following sections we omit the $(e)$ subscript for brevity, so that $T_{(e)}^{\mu\nu}$ will be denoted simply as $T^{\mu\nu}$.}.

From anisotropic Weyl symmetry, one obtains the following Ward identity:
\begin{equation}\label{CurvedLifshitzWard:WeylWardIdent}
D \equiv D^{\mu\nu} T^{(g)}_{\mu\nu} = D^{\mu\nu} T^{(e)}_{\mu\nu} = 0 ,
\end{equation} 
where $ D^{\mu\nu} \equiv P^{\mu\nu}-z\,n^\mu n^\nu$. 
Note that similarly to the relativistic case, assuming that a Weyl-invariant coupling of the theory to curved spacetime exists is equivalent to assuming that the flat space stress-energy tensor can be improved to satisfy equation \eqref{CurvedLifshitzWard:WeylWardIdent} (and still remain conserved). 

In a field theory for which TPD invariance is not anomalous, the expectation value of the stress-energy tensor satisfies the Ward identities \eqref{CurvedLifshitzWard:TPDWardIdentTe}. However, when the Lifshitz scaling symmetry is anomalous, its expectation value no longer satisfies identity \eqref{CurvedLifshitzWard:WeylWardIdent}. Instead it is modified:
\begin{equation}\label{CurvedLifshitzWard:AnomWeylWardIdent}
\langle D \rangle \equiv D^{\mu\nu} \langle T^{(e)}_{\mu\nu} \rangle = \mathcal{A},
\end{equation}
where $\mathcal{A} (\ve{a}{\mu}, t^a) $ is a TPD invariant and local function of the background fields. The possible form of $\mathcal{A}$ is restricted by the Wess-Zumino consistency condition to a linear combination of possible expressions (analogous to $E_4$, $W^2$ and $ \Box R$ of the $(3+1)$-dimensional conformal case), and was obtained for several cases in \cite{Baggio:2011ha,Griffin:2011xs,Arav:2014goa,Arav:2016xjc,Pal:2016rpz}. While these expressions are universal, the value of their coefficients depends on the content of the theory.  Like the conformal case, the coefficients of the trivial terms may depend on the regularization scheme, but those of the anomalies are regularization independent. Our goal in this work is to extract these coefficients from the flat space $n$-point correlation functions of the stress-energy tensor.

For the case of a Lifshitz field theory in $2+1$ dimensions with $z=2$, which is the case we study in section \ref{Sec:z2LishitzFreeScalarResults}, it was shown in \cite{Arav:2016xjc} that generally (when the Frobenius condition is not assumed) the possible expressions in $\mathcal{A}$ consist of an infinite set of linearly independent anomalies and trivial terms, with increasing number of derivatives. However, the only anomalies that possibly contribute to the three point correlation functions of the stress-energy tensor (and therefore relevant to the calculations we perform here) are given by the following expressions:
\begin{equation}
\mathcal A^{(2,0,0)} = \tr(K_S^2) - \frac{1}{2} K_S^2 ,
\end{equation}
in the two derivatives sector, and:
\begin{equation}\label{eq:4DerAnomaliesFullExpr}
\begin{split}
\mathcal A_2^{(0,4,0)} = & \, K_A \Lie{n}^2 K_A + K_A K_S \Lie{n} K_A ,\\
\mathcal A_3^{(0,4,0)} = & \, \tilde K_S^{\alpha\beta} (a_\alpha \wn_\beta  K_A + \wn_\alpha \wn_\beta K_A) ,\\
\mathcal A_4^{(0,4,0)} = & \,
\left( \widehat{R} + \wn_\alpha a^\alpha \right)^2,
\end{split}
\end{equation}
in the four derivatives sector. The various geometrical structures in these expressions are defined as follows:
\begin{itemize}
\item $(K_S)_{\mu\nu} \equiv \frac{1}{2}\Lie{n} P_{\mu\nu}$ is a generalization of the extrinsic curvature of the foliation induced by $n_\mu$ to the non-Frobenius case. We also define $K_S \equiv (K_S)^\mu_\mu$, $\tr(K_S^2) \equiv (K_S)_{\mu\nu}(K_S)^{\mu\nu}$ and $(\tilde K_S)_{\alpha\beta} \equiv \bar \epsilon_{(\alpha}{}^\gamma\, K^S_{\beta)\gamma} $, where $\bar\epsilon^{\mu\nu} \equiv n_\alpha \epsilon^{\alpha\mu\nu} $. 
\item $(K_A)_{\mu\nu} \equiv P_\mu^{\mu'} P_\nu^{\nu'} \nabla_{[\mu'} n_{\nu']}$ is antisymmetric and vanishes when $n_\mu$ satisfies the Frobenius condition. We also define $ K_A \equiv \frac{1}{2} K_{\mu\nu}^A \bar \epsilon^{\mu\nu}$.
\item $a_\mu \equiv \Lie{n} n_\mu$ is the acceleration vector associated with $n_\mu$.
\item $\wn_\mu$ is a space projected covariant derivative.\
\item $\widehat{R}$ is a generalization of the Ricci scalar of the foliation induced by $n_\mu$ to the non-Frobenius case.
\end{itemize}
For further discussion and more detailed definitions see \cite{Arav:2016xjc} and appendix \ref{app:LifCurvedSpacetimeMultTimDims}. The possible trivial terms relevant to our calculations are listed in appendix \ref{SecVarPhis}.

\subsubsection{Ward Identities For Correlation Functions}
\label{sec:WardIdentitiesCorrLifshitz}
Let us denote by $ W(\ve{a}{\mu}, t^a) $ the effective action of the Lifshitz field theory on curved spacetime, defined by:
\begin{equation}\label{CorrLifshitzWard:EffActionDef}
e^{i\,W(\ve{a}{\mu}, t^a)} = \int D\phi \, e^{i\,S(\ve{a}{\mu},t_\alpha,\{\phi\})} ,
\end{equation}
where $\phi$ again stands for the dynamic fields in the theory.
We define two types of correlation functions for the stress-energy tensor. First, the connected Feynman $n$-point correlation functions, 
defined similarly to the relativistic case \eqref{CorrConfWard:FeynmanCorrFuncDef},
where $\mathcal{O}_1,\ldots,\mathcal{O}_n$ are local functions of the dynamic and the background fields, and $Z_0$ is the flat space partition function. Alternatively, we define the correlation functions given by variations of the effective action $W$ with respect to the background fields:
\begin{equation}
\left\langle T^{\mu_1}{}_{a_1}(x_1) \ldots T^{\mu_n}{}_{a_n}(x_n) \right\rangle_W \equiv
(-i)^{n-1}\, \frac{1}{e(x_1)} \ldots \frac{1}{e(x_n)}\, \frac{\delta^n W}{\delta\ve{a_1}{\mu_1}(x_1) \ldots \delta\ve{a_n}{\mu_n}(x_n)}.
\end{equation}
We will refer to these as the variational correlation functions.
Like the relativistic case, these two definitions for the correlation functions do not coincide, since the curved space field theory currents depend on the background fields. Instead these two types of correlation functions are related via relations, that are obtained by differentiating equation \eqref{CorrLifshitzWard:EffActionDef}. For the two and three point functions, one obtains:
\begin{align}
&\left\langle T^\mu{}_a (x) T^\rho{}_b(y) \right\rangle_W = 
-\frac{i}{e(x)e(y)} \left\langle  \frac{\delta^2 S}{\delta\ve{a}{\mu}(x)\delta\ve{b}{\rho}(y)} \right\rangle_F
 + \left\langle T^\mu{}_a (x) T^\rho{}_b(y) \right\rangle_F , \\
 \begin{split}
&\left\langle T^\mu{}_a (x) T^\rho{}_b(y) T^\alpha{}_c(z)  \right\rangle_W =
-\frac{1}{e(x)e(y)e(z)} \left\langle  \frac{\delta^3 S}{\delta\ve{a}{\mu}(x)\delta\ve{b}{\rho}(y)\delta\ve{c}{\alpha}(z)} \right\rangle_F \\
&\quad - \frac{i}{e(x)e(y)} \left\langle \frac{\delta^2 S}{\delta\ve{a}{\mu}(x)\delta\ve{b}{\rho}(y)} T^\alpha{}_c(z) \right\rangle_F 
 - \frac{i}{e(x)e(z)} \left\langle \frac{\delta^2 S}{\delta\ve{a}{\mu}(x)\delta\ve{c}{\alpha}(z)} T^\rho{}_b(y) \right\rangle_F \\
&\quad - \frac{i}{e(y)e(z)} \left\langle \frac{\delta^2 S}{\delta\ve{b}{\rho}(y)\delta\ve{c}{\alpha}(z)} T^\mu{}_a(x) \right\rangle_F
+ \left\langle T^\mu{}_a (x) T^\rho{}_b(y) T^\alpha{}_c(z)  \right\rangle_F .
 \end{split}
\end{align}

In the flat space limit, these relations reduce to the following:\footnote{Note that the one point functions correspond to ``tadpole'' diagrams with no dimensionful parameter, and therefore vanish in the flat space limit.}
\begin{align}\label{CorrLifshitzWard:VarToFeynmanCorr2Point}
&\left\langle T^\mu{}_a (x) T^\rho{}_b(y) \right\rangle_W = \left\langle T^\mu{}_a (x) T^\rho{}_b(y) \right\rangle_F , \\
 \begin{split}\label{CorrLifshitzWard:VarToFeynmanCorr3Point}
&\left\langle T^\mu{}_a (x) T^\rho{}_b(y) T^\alpha{}_c(z)  \right\rangle_W =
\left\langle T^\mu{}_a (x) T^\rho{}_b(y) T^\alpha{}_c(z)  \right\rangle_F
 - i \left\langle \frac{\delta^2 S}{\delta\ve{a}{\mu}(x)\delta\ve{b}{\rho}(y)} T^\alpha{}_c(z) \right\rangle_F \\ 
&\qquad\qquad\qquad\qquad\enspace
 - i \left\langle \frac{\delta^2 S}{\delta\ve{a}{\mu}(x)\delta\ve{c}{\alpha}(z)} T^\rho{}_b(y) \right\rangle_F 
 - i \left\langle \frac{\delta^2 S}{\delta\ve{b}{\rho}(y)\delta\ve{c}{\alpha}(z)} T^\mu{}_a(x) \right\rangle_F.
 \end{split}
\end{align}
These relations make it clear that, while the flat space Feynman correlation functions depend only on the field theory currents (given as the first derivative of the action with respect to the background fields), the variational correlation functions depend on higher derivatives of the action as well. The implication is that, if there is more than one way to couple the field theory to curved spacetime while preserving all symmetries discussed in subsection \ref{subsubsec:LifshitzSymWardCuvedSpacetime} and leaving the flat space currents the same, then any such coupling would produce the same Feynman correlation functions but different variational correlation functions. As we will show, this leads to a possible ambiguity in the relation between the flat space Feynman correlation function and the consistent anomalies obtained from the Wess-Zumino analysis, which does not occur in the relativistic case.

One method to derive Ward identities directly for the flat space Feynman correlation functions is via a change of variables in the path integral. Suppose that the operator $\delta$ implements an infinitesimal transformation that corresponds to a symmetry of the theory, and transforms both the dynamic and the background fields so that $\delta S = 0$. Let $\delta_{\text{dyn}}$ denote an operation that transforms only the dynamic fields, leaving the background fields unchanged, and similarly let $\delta_{\text{bg}}$ transform only the background fields, so that the following is satisfied:
\begin{equation}
\begin{alignedat}{3}
\delta_{\text{dyn}} \phi &= \delta\phi, &\qquad \delta_{\text{dyn}} \ve{a}{\mu} &= 0, &\qquad \delta_{\text{dyn}} t^a &= 0, \\
\delta_{\text{bg}} \phi &= 0, &\qquad \delta_{\text{bg}} \ve{a}{\mu} &= \delta \ve{a}{\mu}, &\qquad \delta_{\text{bg}} t^a &= \delta t^a,
\end{alignedat}
\end{equation}
and $\delta = \delta_{\text{dyn}} + \delta_{\text{bg}}$. These operators are explicitly given by:
\begin{equation}
\label{CorrLifshitzWard:ExplicitExpressionForDeltaOperators}
\delta_{\text{dyn}} = \int \delta\phi \left.\frac{\delta}{\delta\phi}\right|_{\ve{a}{\mu}, t^a},
\qquad
\delta_{\text{bg}} = \int \delta\ve{a}{\mu} \left. \frac{\delta}{\delta\ve{a}{\mu}} \right|_{\phi,t^a} +
\delta t^a \left. \frac{\delta}{\delta t^a} \right|_{\phi,\ve{a}{\mu}} .
\end{equation}
By performing a change of variables $\phi \rightarrow \tilde\phi = \phi + \delta\phi$  in the path integral \eqref{CorrConfWard:FeynmanCorrFuncDef}, and assuming for now that that the symmetry corresponding to $\delta$ is non-anomalous (so that the change of integration measure does not contribute in the context of the dimensional regularization scheme we are using here), we obtain:
\begin{equation}
\int D\phi\, \delta_\text{dyn} \left[ \mathcal{O}_1(x_1) \ldots  \mathcal{O}_n (x_n)\, e^{i\,S} \right] = 0.
\end{equation}
Noting that $\delta_\text{dyn} S = - \delta_\text{bg} S$, the following identity can then be derived:
\begin{multline}\label{CorrLifshitzWard:WardIdentForFeynmanCorrGeneral}
\left\langle (\delta_\text{bg} S) \mathcal{O}_1(x_1) \ldots \mathcal{O}_n(x_n) \right\rangle_F + i \left\langle \delta_\text{dyn}\mathcal{O}_1(x_1) \ldots \mathcal{O}_n(x_n) \right\rangle + \ldots \\
+ i \left\langle \mathcal{O}_1(x_1) \ldots \delta_\text{dyn} \mathcal{O}_n(x_n) \right\rangle = 0 .
 \end{multline} 
Using TPD (as given in \eqref{CurvedLifshitzWard:TPDTransVielbein}) in identity \eqref{CorrLifshitzWard:WardIdentForFeynmanCorrGeneral} and choosing $ \xi^\sigma(w) =  \delta^\sigma_a\, \delta(w-x) $, we get the following Ward identities for the flat space two and three point Feynman correlation functions of the stress-energy tensor:\footnote{See appendix \ref{LifshitzNotation} for details on our notations for the Ward identities.}
\begin{align}
\label{eq:ConWard2Point}
(\mathcal{I}^{(2)}_D)^\rho_{ab}(x,y) &\equiv  \left\langle (\partial_\mu T^\mu{}_a(x)) T^\rho{}_b(y) \right\rangle_F = 0 , \\
\begin{split}
\label{eq:ConWard3Point}
(\mathcal{I}^{(3)}_{D})^{\rho\alpha}_{abc}(x,y,z)  &\equiv \left\langle (\partial_\mu T^\mu{}_a(x)) T^\rho{}_b(y) T^\alpha{}_c(z) \right\rangle_F 
- i \left\langle (\delta^D_{\text{dyn}}T^\rho{}_b)(x,y) \, T^\alpha{}_c(z) \right\rangle_F \\
&\qquad\qquad\qquad\qquad\qquad\qquad\quad
 -i \left\langle T^\rho{}_b(y) (\delta^D_{\text{dyn}} T^\alpha{}_c)(x,z)  \right\rangle_F = 0,
 \end{split}
\end{align}
where for a scalar field $\phi$, for example, we have $ (\delta^D_\text{dyn} \phi) (x,w) = \delta(w-x) \partial_a \phi(w) $.
Similarly, when applying identity \eqref{CorrLifshitzWard:WardIdentForFeynmanCorrGeneral} to the anisotropic Weyl transformation \eqref{CurvedLifshitzWard:WeylTransVielbein} and setting $\sigma(w)=\delta(w-x)$, we obtain the following identities for the flat space correlation functions:
\begin{align}\label{CorrLifshitzWard:WardIdentForFeynmanCorrWeyl2Point}
(\mathcal{I}^{(2)}_W)^\rho_b(x,y) &\equiv \left\langle D(x) T^\rho{}_b(y) \right\rangle_F = 0 , \\
\begin{split}\label{CorrLifshitzWard:WardIdentForFeynmanCorrWeyl3Point}
(\mathcal{I}^{(3)}_W)^{\rho\alpha}_{bc}(x,y,z) &\equiv
\left\langle D(x)T^\rho{}_b(y)T^\alpha{}_c(z) \right\rangle_F
+ i \left\langle (\delta^W_\text{dyn} T^\rho{}_b)(x,y) T^\alpha{}_c (z) \right\rangle_F\\
&\qquad\qquad\qquad\qquad\qquad\qquad\quad
+ i \left\langle T^\rho{}_b(y) (\delta^W_\text{dyn} T^\alpha{}_c) (x,z) \right\rangle_F = 0,
\end{split}
\end{align} 
where for a Lifshitz scalar\footnote{Here and in the following sections, ``Lifshitz scalar'' will refer to a scalar $\phi$ with a second order time derivative kinetic term in the action as used in section \ref{Sec:z2LishitzFreeScalarResults}, so that it has a Lifshitz dimension of $[\phi]=\frac{d-z}{2}$.
} 
$\phi$ we have $(\delta^W_\text{dyn} \phi)(x,w) = \left( \frac{z-d}{2} \right) \delta(w-x) \phi(w) $.
In the presence of Lifshitz scaling anomalies, expressions \eqref{CorrLifshitzWard:WardIdentForFeynmanCorrWeyl2Point}--\eqref{CorrLifshitzWard:WardIdentForFeynmanCorrWeyl3Point} will not vanish, and in general will be equal to some linear combinations of contact terms instead. 

An alternative method to derive the Ward identities for the variational correlation functions is by taking derivatives of the curved spacetime Ward identities \eqref{CurvedLifshitzWard:TPDWardIdentTe} and \eqref{CurvedLifshitzWard:WeylWardIdent} (or \eqref{CurvedLifshitzWard:AnomWeylWardIdent} in the presence of Lifshitz anomalies) with respect to the background fields. This method allows one to directly relate the flat space correlation functions to derivatives of the consistent anomalies as derived in \cite{Arav:2016xjc}, and extract the anomaly coefficients. Taking the first and second derivatives of identity \eqref{CurvedLifshitzWard:AnomWeylWardIdent} in the flat space limit, we obtain the following Ward identities for the two and three point variational correlation functions:
\begin{align}\label{CorrLifshitzWard:WardIdentForVarCorrWeyl2Point}
(\mathcal{W}^{(2)}_W)^\rho_b(x,y) &\equiv
D^a_\mu \left\langle T^\mu{}_a(x) T^\rho{}_b(y) \right\rangle_W = 
-i \left. \frac{\delta\mathcal{A}(x)}{\delta\ve{b}{\rho}(y)}\right|_{\text{flat}} , \\
\begin{split}\label{CorrLifshitzWard:WardIdentForVarCorrWeyl3Point}
(\mathcal{W}^{(3)}_W)^{\rho\alpha}_{bc}(x,y,z) &\equiv
D^a_\mu \left\langle T^\mu{}_a(x) T^\rho{}_b(y) T^\alpha{}_c(z) \right\rangle_W \\ 
&\quad
-i (\delta^\rho_\mu D^a_b - \delta^\rho_b D^a_\mu) \delta(x-y) \left\langle T^\mu{}_a(x) T^\alpha{}_c(z) \right\rangle_W \\
&\quad
-i (\delta^\alpha_\mu D^a_c - \delta^\alpha_c D^a_\mu) \delta(x-z) \left\langle T^\mu{}_a(x) T^\rho{}_b(y) \right\rangle_W \\
& = - \left. \frac{\delta^2\mathcal{A}(x)}{\delta\ve{b}{\rho}(y)\delta\ve{c}{\alpha}(z)} \right|_\text{flat} . 
\end{split}
\end{align}
Using the relations \eqref{CorrLifshitzWard:VarToFeynmanCorr2Point}--\eqref{CorrLifshitzWard:VarToFeynmanCorr3Point} in these expressions,  
$\mathcal{W}^{(2)}_W$,$\mathcal{W}^{(3)}_W$ can be written as linear combinations of the expressions in the Ward identities \eqref{CorrLifshitzWard:WardIdentForFeynmanCorrWeyl2Point}--\eqref{CorrLifshitzWard:WardIdentForFeynmanCorrWeyl3Point} and other expressions that vanish in the absence of Lifshitz anomalies:
\begin{align}
(\mathcal{W}^{(2)}_W)^\rho_b(x,y) &= (\mathcal{I}^{(2)}_W)^\rho_b(x,y) , \\
\begin{split}
(\mathcal{W}^{(3)}_W)^{\rho\alpha}_{bc}(x,y,z) &= 
(\mathcal{I}^{(3)}_W)^{\rho\alpha}_{bc}(x,y,z) 
- i \mathcal{J}^{\rho\alpha}_{bc}(x,y,z) - i \mathcal{J}^{\alpha\rho}_{cb}(x,z,y)\\
&\quad + i \delta^\rho_b \delta(x-y) (\mathcal{I}^{(2)}_W)^\alpha_c(x,z)
+ i \delta^\alpha_c \delta(x-z) (\mathcal{I}^{(2)}_W)^\rho_b(x,y)\\
&\quad
-i \mathcal{K}^{\rho\alpha}_{bc}(x,y,z),
\end{split}
\end{align}
where $\mathcal{J}^{\rho\alpha}_{bc}(x,y,z)$ and $ \mathcal{K}^{\rho\alpha}_{bc}(x,y,z) $ are given by:
\begin{align}
\begin{split}\label{CorrLifshitzWard:JTermDef}
\mathcal{J}^{\rho\alpha}_{bc}(x,y,z) &\equiv 
D^a_\mu \left\langle \frac{\delta^2 S}{\delta\ve{a}{\mu}(x)\delta\ve{b}{\rho}(y)} T^\alpha{}_c (z)\right\rangle_F 
+ D^a_b \delta(x-y) \left\langle T^\rho{}_a(x) T^\alpha{}_c(z) \right\rangle_F \\
&\qquad\qquad\qquad\qquad\qquad\qquad\qquad\quad 
+ \left\langle (\delta^W_\text{dyn} T^\rho{}_b)(x,y) T^\alpha{}_c (z) \right\rangle_F ,
\end{split}\\
\label{CorrLifshitzWard:KTermDef}
\mathcal{K}^{\rho\alpha}_{bc}(x,y,z) &\equiv
\left\langle D(x) \frac{\delta^2 S}{\delta\ve{b}{\rho}(y)\delta\ve{c}{\alpha}(z)} \right\rangle_F .
\end{align}

When no anomalies are present, the expressions $\mathcal{I}^{(2)}_W$, $\mathcal{I}^{(3)}_W$ and $\mathcal{K}$ are expected to vanish due to identity \eqref{CorrLifshitzWard:WardIdentForFeynmanCorrGeneral}. The expression $\mathcal{J}$ can also be shown to vanish in this case, by noting that:\footnote{Like the Ward identities \eqref{CorrLifshitzWard:WardIdentForFeynmanCorrWeyl2Point}--\eqref{CorrLifshitzWard:WardIdentForVarCorrWeyl3Point}, this identity may acquire contact terms on its RHS after a renormalization procedure.}
\begin{equation}
\left\langle (\delta^W_\text{bg} + \delta^W_\text{dyn} - \delta^W) T^\rho{}_b (x,y) \, T^\alpha{}_c(z) \right\rangle_F = 0 ,
\end{equation}
and using the following expression for $ (\delta^W_\text{bg} T^\rho{}_b) (x,y) $ (where again $\sigma(w)=\delta(w-x)$):
\begin{equation}
(\delta^W_\text{bg} T^\rho{}_b) (x,y) = D^a_\mu \left.\frac{\delta T^\rho{}_b(y)}{\delta\ve{a}{\mu}(x)}\right|_\text{flat} 
= D^a_\mu \left.\frac{\delta^2 S}{\delta\ve{a}{\mu}(x)\delta\ve{b}{\rho}(y)}\right|_\text{flat} - (d+z) \delta(x-y) T^\rho{}_b(y),
\end{equation}
and the anisotropic Weyl scaling properties of $T^\rho{}_b$:
\begin{equation}
(\delta^W T^\rho{}_b) (x,y) = -(d+z)\delta(x-y) T^\rho{}_b(y) 
- D^a_b \delta(x-y) T^\rho{}_a(y).
\end{equation}
This scaling property can be derived by using the definition of the stress-energy tensor \eqref{CurvedLifshitzWard:StressTensorDef}, expressing the operator $\delta^W$ in terms of the anisotropic Weyl transformations of the background and dynamic fields as in \eqref{CorrLifshitzWard:ExplicitExpressionForDeltaOperators}, exchanging the order of derivatives and using the anisotropic Weyl invariance of the classical action $S$.\footnote{This derivation assumes that the anisotropic Weyl transformation of the dynamic fields $\delta^W \phi$ does not explicitly depend on the vielbeins $\ve{a}{\mu}$. This is indeed the case whenever the dynamic fields transform covariantly under anisotropic Weyl transformations, i.e.\ $\delta^W \phi = s \sigma \phi$ for some $s$. }

\subsubsection{Ambiguity of the Anomaly Coefficients?}

As mentioned earlier, unlike the relativistic case, the relation between the flat space Feynman correlation functions and the Lifshitz anomaly coefficients (that is, the coefficients on the RHS of equation \eqref{CurvedLifshitzWard:AnomWeylWardIdent} of the various anomalous terms as obtained from the Wess-Zumino consistency conditions) may be ambiguous in some cases. This is due to the Weyl invariant coupling of the theory to curved spacetime being non-unique. 

As an example, consider any Lifshitz invariant theory for which $d=z$, that contains a real Lifshitz scalar $\phi$, so that the field $\phi$ is dimensionless. In the absence of anomalies, one expects the two point function $\left\langle D(x) \phi^2(y) \right\rangle_F$ to vanish. When Lifshitz anomalies are present, it will instead be equal to some contact term, and from dimensional analysis we have:
\begin{equation}
\left\langle D(x) \phi^2(y) \right\rangle_F = i c \,\delta(x-y) ,
\end{equation}
where $c$ is some constant.

Suppose that $\mathcal{A}_0(\ve{a}{\mu},t^a)$ is any local functional of the background fields with the following 2 properties: 
\begin{enumerate}
\item $\mathcal{A}_0$ is second order in the background fields, that is $ \left. \mathcal{A}_0 \right|_\text{flat} = \left.\frac{\delta\mathcal{A}_0}{\delta\ve{a}{\mu}}\right|_\text{flat} = \left.\frac{\delta\mathcal{A}_0}{\delta t^a}\right|_\text{flat} = 0 $, but $ \left.\frac{\delta^2 \mathcal{A}_0}{\delta\ve{a}{\mu}\delta\ve{b}{\rho}} \right|_\text{flat} \neq 0$.
\item $\mathcal{A}_0$ is a Weyl invariant density of dimension $d+z$, that is:
\begin{equation}
\dels \mathcal{A}_0 = - (d+z) \sigma \mathcal{A}_0 .
\end{equation}
\end{enumerate}
Next, consider adding to the curved spacetime classical action $S(\ve{a}{\mu},t^a)$ a term of the form:
\begin{equation}
S_0 \equiv \int d^{d+1}w\, e \mathcal{A}_0(w) \phi^2(w).
\label{ambiguity}
\end{equation}
The new action $ \tilde{S} = S + S_0 $ is still invariant under TPD and anisotropic Weyl transformations, and it coincides with $S$ in flat spacetime. Moreover, the flat spacetime conserved currents derived from $\tilde{S}$ (including the stress-energy tensor) are the same as those derived from $S$. As a result, the anomalous Ward identity expressions $ \mathcal{I}^{(2)}_W $, $ \mathcal{I}^{(3)}_W $ (as defined in \eqref{CorrLifshitzWard:WardIdentForFeynmanCorrWeyl2Point}--\eqref{CorrLifshitzWard:WardIdentForFeynmanCorrWeyl3Point}) remain unchanged in flat spacetime. However, expression $ \mathcal{W}^{(3)}_W $ does change. 
Due to the assumed properties of $\mathcal{A}_0$, the term $S_0$ satisfies:
\begin{equation}
D^a_\mu \left. \frac{\delta^2 S_0}{\delta\ve{a}{\mu}(x)\delta\ve{b}{\rho}(y)} \right|_\text{flat} = 0,
\end{equation}
and therefore does not contribute to $\mathcal{J}$ (as defined in \eqref{CorrLifshitzWard:JTermDef}) either. The only contribution of $S_0$ to the anomalous identity $\mathcal{W}^{(3)}_W$ is from the expression $\mathcal{K}$ (as defined in \eqref{CorrLifshitzWard:KTermDef}). It follows that the change in $\mathcal{W}^{(3)}_W$ due to the $S_0$ term is given by:
\begin{equation}
\begin{split}
(\Delta\mathcal{W}^{(3)}_W)^{\rho\alpha}_{bc}(x,y,z) 
&= -i \left\langle D(x) \frac{\delta^2 S_0}{\delta\ve{b}{\rho}(y)\delta\ve{c}{\alpha}(z)} \right\rangle_F\\
&= -i \int d^{d+1}w \frac{\delta^2 \mathcal{A}_0(w)}{\delta\ve{b}{\rho}(y)\delta\ve{c}{\alpha}(z)} \left\langle D(x) \phi^2(w) \right\rangle_F \\
&= -i \int d^{d+1}w \frac{\delta^2 \mathcal{A}_0(w)}{\delta\ve{b}{\rho}(y)\delta\ve{c}{\alpha}(z)} ic \delta(x-w) \\
&= c \, \frac{\delta^2 \mathcal{A}_0(x)}{\delta\ve{b}{\rho}(y)\delta\ve{c}{\alpha}(z)} .
\end{split}
\end{equation}
If we choose $ \mathcal{A}_0 = a \tilde{\mathcal{A}} $, where $\tilde{\mathcal{A}}$ is a possible B-type anomaly of the theory and $a$ is a constant, we conclude that $S_0$ contributes an additional $ - ac$ to the coefficient of the $\tilde{\mathcal{A}}$ anomaly, without changing the action or the current operators of the flat space theory.\footnote{In fact, the same argument can be made using the more general term $ S_0 \equiv \int d^{d+1}w\, e \mathcal{A}_0(w) \phi^n(w) $.}

The freedom to add such a term to the action and thereby change the corresponding anomaly coefficient was previously pointed out in \cite{Griffin:2012qx} for a purely spatial anomaly of a free Lifshitz scalar in $d=z=2$, but it is in fact more general. It could be applied, for example, to any of the consistent anomalies in the $d=z=2$ case (which are all B-type, see \cite{Griffin:2011xs,Baggio:2011ha,Arav:2014goa,Arav:2016xjc}), for any theory that contains a Lifshitz scalar. This freedom may suggest that, unlike the relativistic case, knowing the flat space action, the flat space currents and their Feynman correlation functions is not enough in these cases to determine the consistent anomaly coefficients -- one needs to specify the full curved spacetime action, and by choosing different couplings of the theory to curved spacetime one may obtain any value for any of the B-type anomaly coefficients. 

Stated differently, knowing all the Feynman correlation functions
of the flat space field theory is not sufficient in order to determine the curved spacetime action. 
In order to construct the full curved spacetime action we have to know
all the variational correlations functions and the ambiguity is in the relation between these two types
of correlation functions, the variational and Feynman.
This ambiguity can be avoided if we add another ingredient to the discussion.
The $\phi$ field in (\ref{ambiguity}) is a Log-correlated field and is therefore ill defined when we take the large volume
limit. Consistency of the quantum field theory at infinite volume forbids such an operator in the correlation
functions.\footnote{We thank Z. Komargodski for this comment.} The ambiguity may still have consequences
for field
theories on a  finite volume spacetime, or with other modifications of the IR physics that take care of the Log divergence.
We leave this for future studies. 

In the following sections, we use split dimensional regularization to calculate the Lifshitz anomaly coefficients for a free Lifshitz scalar in $2+1$ dimensions and $z=2$. The above discussion implies that a specific coupling of the theory to curved spacetime needs to be specified. However, since we are performing the calculation for a free theory in an infinite volume and no physical IR regulator, we will use the minimal coupling (which is Weyl invariant in this case) and will not allow for Weyl invariant couplings of the form \eqref{ambiguity}.

\section{The Regularization and Renormalization Method}
\label{RegAndRenorm}

In this section we present the split dimensional regularization scheme we employ and explain the method by which we use it to extract the Lifshitz anomaly coefficients from the field theory correlation functions. We start by reviewing the relativistic conformal case for reference, and then explain the non-relativistic Lifshitz case.

\subsection{Review of Dimensional Regularization in the Conformal Case}
\label{ReviewOfDimensionalRegConformalCase}

\subsubsection{Conformal Anomaly From Dimensional Regularization}

In the standard relativistic dimensional regularization scheme, one defines the theory and calculates various quantities in a general dimension $d$, and then analytically continues the obtained expressions to dimension $d=d^\text{phys} - \varepsilon$, where $d^\text{phys}$ is the physical dimension. Suppose that $ I_{(n)} (d,p_i,m)$ is some $n$-point correlation function written in momentum space and calculated to one-loop order in perturbation theory using the corresponding 1PI Feynman diagrams, where $p^\mu_i \, (i=1,\ldots, n)$ are external momenta and $m$ is an IR mass regulator. Generally after analytic continuation of the dimension, $I_{(n)}$  will take the form:\footnote{Beyond one-loop order, one has to first cancel the possible subdivergences using the appropriate counterterms. The pole in $\varepsilon$ can then be of higher order. In the following sections we focus on a free theory, and therefore on the one-loop case.}
\begin{equation}
I_{(n)}(\varepsilon,p_i,m) = \frac{1}{\varepsilon} f(\varepsilon,p_i,m) ,
\end{equation}
where $f(\varepsilon,p_i,m)$ is an expression which is regular around $\varepsilon=0$, and is a linear combination of terms of the form: 
\begin{equation}\label{RegAndRenConf:GeneralFormOfTermInCorrFunction}
g(\varepsilon,p_i,m)\, p_{i_1}^{\mu_1} p_{i_2}^{\mu_2}\ldots\eta^{\nu_1\nu_2}\eta^{\nu_3\nu_4}\ldots \,, 
\end{equation}
where $ g(\varepsilon,p_i,m) $ is a scalar expression.
Expanding around $\varepsilon=0$ we have:
\begin{equation}
\begin{split}
I_{(n)}(\varepsilon,p_i,m) &= \frac{1}{\varepsilon} f(0,p_i,m) + \frac{\partial}{\partial\varepsilon} f(0,p_i,m) + O(\varepsilon)\\
&\equiv \frac{1}{\varepsilon} I_{(n)}^{(\text{res})}(p_i,m) + I_{(n)}^{(\text{ren})}(p_i,m) + O(\varepsilon),
\end{split}
\end{equation}
where $  I_{(n)}^{(\text{res})} $ is the residue of the $\varepsilon$ pole, and $ I_{(n)}^{(\text{ren})} \equiv \lim_{\varepsilon\to0} \left[ I_{(n)} - \frac{1}{\varepsilon} I_{(n)}^{(\text{res})} \right] $ is the renormalized correlation function.\footnote{We use a minimal subtraction renormalization scheme.} It is a well known property of relativistic field theories that the residue $I_{(n)}^{(\text{res})}$ is always a polynomial in the external momenta and the mass regulator (see for example \cite{Weinberg:1959nj,Itzykson:1980rh,Collins:1984xc}). This can be shown by taking derivatives of $I_{(n)}$ with respect to the external momenta (and mass regulator) enough times so that the corresponding Feynman diagram no longer diverges. As long as there are no IR divergences in the physical dimension, one can safely take the limit $m\to 0$ in $I_{(n)}^{(\text{ren})}$ to obtain the physical renormalized correlation function\footnote{If no IR divergences occur, $I_{(n)}$ is regular when $m\to0$, and $I^{(\text{res})}_{(n)}$ is polynomial in $m$ and therefore also regular. Therefore  $I_{(n)}^{(\text{ren})}$ is regular in this limit too, and the order of taking the limit $m\to 0$ and renormalizing does not matter.} (the correlation functions of the stress-energy tensor in the cases studied here are indeed free of IR divergences, even in $d=2$ dimensions, as will be explained in subsection \ref{SeriesExp}).

A useful property of scale anomalies is that, in some cases, one can calculate them from the $\varepsilon$ pole residue alone: Suppose the theory has a symmetry (such as a scaling symmetry) that is not explicitly broken by the dimensional regularization scheme itself, with a corresponding Ward identity of the form:
\begin{equation}\label{RegAndRenConf:GeneralFormOfWardIdentity}
T(\varepsilon) \left[I_k\right] = 0,
\end{equation}
where $\{I_k\}$ is a set of correlation functions and $ T(\varepsilon) $ is a linear operator that takes expressions of the form \eqref{RegAndRenConf:GeneralFormOfTermInCorrFunction} to expressions of the same form, and may or may not depend on the dimension. Since the symmetry is not broken by dimensional regularization, the unrenormalized correlation functions satisfy identity \eqref{RegAndRenConf:GeneralFormOfWardIdentity}. Therefore we can deduce:
\begin{equation}
\frac{1}{\varepsilon} T(\varepsilon)\left[I_k^{(\text{res})}\right] = - T(\varepsilon)\left[I_k^{(\text{ren})}\right] + O(\varepsilon).
\end{equation} 
The anomalous Ward identity is then given by:
\begin{equation}\label{RegAndRenConf:AnomalyFromRes}
T(0)\left[I_k^{(\text{ren})}\right] = - \lim_{\varepsilon\to0}\left( \frac{1}{\varepsilon} T(\varepsilon)\left[I_k^{(\text{res})}\right]  \right).
\end{equation}
Since the LHS of equation \eqref{RegAndRenConf:AnomalyFromRes} is finite, we can immediately draw two conclusions from it: 
\begin{enumerate}
\item $T(\varepsilon)\left[I_k^{(res)}\right] \sim O(\varepsilon)$.
\item If $T$ does not depend on $\varepsilon$ then $T\left[I_k^{(res)}\right] = 0$ and there is no anomaly. For example, the operator $T$ corresponding to diffeomorphism invariance does not explicitly introduce new factors that depend on $d$, and therefore as long as the dimensional regularization itself does not break this symmetry, it will not be anomalous.
\end{enumerate}

\subsubsection{Expansion in the External Momenta}
Equation \eqref{RegAndRenConf:AnomalyFromRes} allows one to calculate the anomalous Ward identity from the $\varepsilon$ pole residue. This is useful, since there is no need to calculate the full correlation functions in order to extract their divergent part -- it can be obtained simply by expanding the Feynman diagram integrand in powers of the external momenta. Suppose the integrand is $ h(p_i,q,m) $, where $q$ is the internal loop momentum and $h$ has a mass dimension $l$ and therefore satisfies:
\begin{equation}
h(\lambda p_i, \lambda q, \lambda m) = \lambda^l h(p_i,q,m).
\end{equation}
Rescaling $q$ and $m$ by a factor of $1/\lambda$ where $\lambda \to 0$, we get:
\begin{equation}
h\left(p_i,\frac{q}{\lambda},\frac{m}{\lambda}\right) = \lambda^{-l} h(\lambda p_i, q, m).
\end{equation}
We next expand in powers of $\lambda$ around $\lambda=0$ to obtain:\footnote{Formally, this expansion is done after performing a Wick rotation to Euclidean signature, however one obtains the same results by performing the expansion first and Wick rotating only in the last step when evaluating the integrals \eqref{RegAndRenConf:KnownIntGamma}.}
\begin{equation}
\begin{split}
h\left(p_i,\frac{q}{\lambda},\frac{m}{\lambda}\right)& = 
\lambda^{-l} \left[ \sum_{k=0}^{k_0} \frac{\lambda^k}{k!} p_{i_1}^{\mu_1}  p_{i_2}^{\mu_2} \ldots  p_{i_k}^{\mu_k} \frac{\partial^k h}{\partial p_{i_1}^{\mu_1} \partial p_{i_2}^{\mu_2}\ldots \partial p_{i_k}^{\mu_k}} (0,q,m) 
+ O(\lambda^{k_0+1}) \right] \\
&= \sum_{k=0}^{k_0} \frac{1}{k!} p_{i_1}^{\mu_1}  p_{i_2}^{\mu_2} \ldots  p_{i_k}^{\mu_k} \frac{\partial^k h}{\partial p_{i_1}^{\mu_1} \partial p_{i_2}^{\mu_2}\ldots \partial p_{i_k}^{\mu_k}} \left(0,\frac{q}{\lambda},\frac{m}{\lambda}\right) 
+ O(\lambda^{k_0-l+1}) .
\end{split}
\end{equation}
Defining:
\begin{equation}
\tilde{h}(p_i,q,m) \equiv \sum_{k=0}^{k_0} \frac{1}{k!} p_{i_1}^{\mu_1}  p_{i_2}^{\mu_2} \ldots  p_{i_k}^{\mu_k} (h^{(k)})^{i_1\ldots i_k}_{\mu_1\ldots \mu_k}(q,m), 
\end{equation}
where
\begin{equation}
(h^{(k)})^{i_1\ldots i_k}_{\mu_1\ldots \mu_k}(q,m) \equiv \frac{\partial^k h}{\partial p_{i_1}^{\mu_1} \partial p_{i_2}^{\mu_2}\ldots \partial p_{i_k}^{\mu_k}} \left(0,q,m\right),
\end{equation}
we conclude that the integral over $h(p_i,q,m) - \tilde{h}(p_i,q,m)$ has a divergence degree\footnote{Note that $m$ always appears alongside $q$ in these 1PI diagrams, and therefore scaling $q$ and $m$ together here still gives the correct divergence degree in $q$.} of $ d+l-k_0-1 $. If we choose $k_0=d^\text{phys}+l$, the integral over  $h(p_i,q,m) - \tilde{h}(p_i,q,m)$ converges in $d=d^\text{phys}$ dimensions,  and therefore the integrals over $ h(p_i,q,m) $ and $ \tilde{h}(p_i,q,m) $ have the same $\varepsilon$ pole. 

Thus in order to calculate the pole residue, the only integrals left to evaluate are the ones over the expressions $ (h^{(k)})^{i_1\ldots i_k}_{\mu_1\ldots \mu_k}(q,m) $. Each of these expressions is a linear combination of terms of the form:
\begin{equation}
\eta_{\nu_1\nu_2}\eta_{\nu_3\nu_4}\ldots \frac{(q^2)^a q_{\mu_1}q_{\mu_2}\ldots q_{\mu_b}}{[q^2-m^2 + i\epsilon ]^s}.
\end{equation}
When performing the integration over $q$ we may use the Lorentz symmetry of the integral to make the standard replacement:
\begin{equation}\label{RegAndRenConf:SphericalSymmReplacement}
q_{\mu_1}q_{\mu_2}\ldots q_{\mu_b} \Rightarrow
\begin{cases}
0, & b=2n-1 \\
\prod\limits_{j=0}^{n} \frac{1}{d+2j} \, G^{(n)}_{\mu_1\mu_2\ldots\mu_{2n}}, & b=2n
\end{cases},
\end{equation}
where $ G^{(n)}_{\mu_1\mu_2\ldots\mu_{2n}} $ is a sum of all possible unique products of $n$ metric factors, i.e.:
\begin{equation}\label{RegAndRenConf:SymmMetricProdDef}
 G^{(n)}_{\mu_1\mu_2\ldots\mu_{2n}} = \eta_{\mu_1\mu_2}\eta_{\mu_3\mu_4}\ldots\eta_{\mu_{2n-1}\mu_{2n}} + \text{All possible permutations}.
\end{equation}
Using \eqref{RegAndRenConf:SphericalSymmReplacement}, the integral over $ \tilde{h}(p_i,q,m) $ can be written in terms of the following known integral (see e.g.\ \cite{Narison:1980ti,'tHooft:1972fi,Collins:1984xc}):
\begin{equation}
\label{RegAndRenConf:KnownIntGamma}
\begin{split}
X(d,r,s,m) &\equiv
\int {\frac{{{d^d}q}}{{{{\left( {2\pi } \right)}^d}}}} \frac{{{{\left( {{q^2}} \right)}^r}}}{{{{\left[ {{q^2} - {m^2} + i\epsilon} \right]}^s}}} \\
&= \frac{i(-1)^{r-s}}{{{{\left( {4\pi } \right)}^{d/2}}}}\frac{{\Gamma \left( {r + d/2} \right)\Gamma \left( {s - r - d/2} \right)}}{{\Gamma \left( {d/2} \right)\Gamma \left( s \right)}}{\left( {{m^2}} \right)^{r - s + d/2}},
\end{split}
\end{equation}
where $k=l-2r+2s\leq d^\text{phys}+l$ (so that $ d^\text{phys}+2r-2s \geq 0 $). Note that the integral \eqref{RegAndRenConf:KnownIntGamma} contributes to the pole residue a term proportional to $ m^{2r-2s+d^\text{phys}} $. Since terms with $ 2r-2s+d^\text{phys} > 0 $ vanish in the $m\to 0$ limit, one only needs to compute terms in the expansion with $ 2r-2s+d^\text{phys} = 0 $ (those of order $ k = d^\text{phys}+l $) in order to obtain the $\varepsilon$ pole residue. For these terms, the integral \eqref{RegAndRenConf:KnownIntGamma} has the following pole:
\begin{equation}\label{RegAndRenConf:KnownIntPole}
X(d,r,s,m) = \frac{i(-1)^{r-s}}{\varepsilon} \frac{2}{(4\pi)^{d^\text{phys}/2}}\frac{\Gamma(r+d^\text{phys}/2)}{\Gamma(d^\text{phys}/2)\Gamma(s)} + O(1).
\end{equation}
Using this procedure of expanding the Feynman diagram integrand in external momenta, computing the pole residue via equations \eqref{RegAndRenConf:SphericalSymmReplacement}--\eqref{RegAndRenConf:KnownIntPole} and applying formula \eqref{RegAndRenConf:AnomalyFromRes}, one can calculate the anomalous Ward identities corresponding to scale symmetry without calculating the full correlation functions. This will be especially useful in the Lifshitz case, where the Feynman diagrams are more difficult to fully evaluate.

\subsection{Split Dimensional Regularization and the Lifshitz Case}
\label{SplitDimensionalRegAndTheLifshitzCase}

\subsubsection{Lifshitz Anomalies From Split Dimensional Regularization}

In the non-relativistic case, in analogy to the relativistic case, we use a split dimensional regularization scheme (first suggested in \cite{Leibbrandt:1996np,Leibbrandt:1997kh} for regularizing gauge theories in the Coulomb gauge). We follow the general scheme defined in \cite{Anselmi:2007ri}, in the context of non-relativistic field theories. We start by defining the theory in a general number of time dimensions $d_t$ and space dimensions $d_s$ (while keeping the critical dynamical exponent $z$ constant). 

In flat spacetime, the theory is defined on a manifold $ \mathcal{M} = \mathcal{M}_t \times \mathcal{M}_s $, where $ \mathcal{M}_t $ is a $d_t$-dimensional time manifold and $ \mathcal{M}_s  $ is a $d_s$-dimensional space manifold, such that is invariant both under rotations in the time manifold and in the space manifold separately. 
Spacetime coordinates will be denoted by $ x^\mu = (x^{\hat\mu},x^{\bar\mu}) $ where $ \hat\mu = 1,\ldots,d_t $ are time indexes, $ \bar\mu = 1,\ldots,d_s $ are space indexes and $\mu=1,\ldots,d_t+d_s$ are spacetime indexes. 
We define a flat metric $ \hat \delta_{\hat\mu\hat\nu} = \operatorname{diag}(-1,\ldots,-1)$ on $ \mathcal{M}_t$, and $ \bar \delta_{\bar\mu\bar\nu} = \operatorname{diag}(1,\ldots,1)$ on $\mathcal{M}_s$. We also define the time projector on $\mathcal{M}$ as $ \hat\delta_{\mu\nu} = \operatorname{diag}(\hat\delta_{\hat\mu\hat\nu},0) $ and similarly the space projector as $ \bar\delta_{\mu\nu} = \operatorname{diag}(0,\bar\delta_{\bar\mu\bar\nu})$, so that $ \delta^\mu_\nu = \hat\delta^\mu_\nu + \bar\delta^\mu_\nu $. Given a vector $v^\mu$ on $\mathcal{M}$, we denote its time projection by $\hat v^\mu \equiv \hat\delta^\mu_\nu v^\nu$, and its space projection by $\bar v^\mu \equiv \bar\delta^\mu_\nu v^\nu$.\footnote{See Appendix \ref{app:conventions} for our conventions and notations.}

Similarly to the relativistic case, we calculate various expressions for general $d_t$ and $d_s$ values, and then analytically continue them to non-integer dimensions $ d_t = 1 - \varepsilon_t $ and $ d_s = d_s^\text{phys} - \varepsilon_s $, where $ d_s^\text{phys} $ is the number of physical space dimensions. Suppose that $ I_{(n)}(d_t,d_s,p_i,m) $ is some $n$-point correlation function calculated to one-loop order in perturbation theory using the corresponding 1PI Feynman diagrams, where $p_i^\mu (i=1,\ldots\,n)$ are external momenta and $m$ is an IR mass regulator. Generally after analytic continuation of the dimensions, $I_{(n)}$ will take the form (see \cite{Anselmi:2007ri}):
\begin{equation}
I_{(n)}(\varepsilon_t,\varepsilon_s,p_i,m) = \frac{1}{\epslif} f(\varepsilon_t,\varepsilon_s,p_i,m),
\end{equation}
where $\epslif \equiv z \varepsilon_t + \varepsilon_s $ and $ f(\varepsilon_t,\varepsilon_s,p_i,m) $ is an expression which is regular around $ \varepsilon_t = \varepsilon_s = 0$ and is a linear combination of terms of the form:
\begin{equation}\label{RegAndRenLif:GeneralFormOfTermInCorrFunction}
g(\varepsilon_t,\varepsilon_s,p_i,m) \hat p_{i_1}^{\mu_1} \hat p_{i_2}^{\mu_2}\ldots \bar p_{j_1}^{\nu_1} \bar p_{j_2}^{\nu_2}\ldots \hat\delta^{\rho_1\rho_2}\hat\delta^{\rho_3\rho_4}\ldots \bar\delta^{\sigma_1\sigma_2}  \bar\delta^{\sigma_3\sigma_4}\ldots,
\end{equation}
where $g(\varepsilon_t,\varepsilon_s,p_i,m)$ is a scalar expression (with respect to time and space rotations). 
Since there are two different dimensional regularization parameters in this case, unlike the relativistic case, in order to renormalize the expression one must choose a particular way in which one takes the limit $(\varepsilon_t,\varepsilon_s)\to(0,0)$, with each choice leading to a different renormalized expression. Define $ \tilde\varepsilon(\varepsilon_t,\varepsilon_s) $ to be some coordinate on the two dimensional regularization parameter space such that $ \tilde\varepsilon(0,0)=0$ and such that the transformation $ (\varepsilon_t,\varepsilon_s) \rightarrow (\epslif,\tilde\varepsilon) $ is regular and invertible around $ \varepsilon_t=\varepsilon_s=0 $. Expanding $ f(\varepsilon_t,\varepsilon_s,p_i,m) $ in $\epslif$ while keeping $\tilde\varepsilon$ constant, we have:
\begin{equation}\label{RegAndRenLif:RenormDef}
\begin{split}
I_{(n)}(\epslif,\epstil,p_i,m) &= \frac{1}{\epslif} f(0,\epstil,p_i,m) + \left.\frac{\partial f}{\partial\epslif}\right|_{\epstil} (0,\epstil,p_i,m) + O(\epslif) \\
&= \frac{1}{\epslif} f(0,\epstil,p_i,m) + \left.\frac{\partial f}{\partial\epslif}\right|_{\epstil} (0,0,p_i,m) + O(\epslif) + O(\epstil) \\
&\equiv \frac{1}{\epslif} I_{(n)}^{(\text{res})}(\epstil,p_i,m) + I_{(n)}^{(\text{ren})}(p_i,m) + O(\epslif) + O(\epstil),
\end{split}
\end{equation}
where $I_{(n)}^{(\text{res})}$ is the residue of the $\epslif$ pole, and $ I_{(n)}^{(\text{ren})}\equiv \lim_{(\epslif,\epstil)\to 0}\left[ I_{(n)}-\frac{1}{\epslif}I_{(n)}^{(\text{res})} \right] $ is the renormalized correlation function. Like the relativistic case, the residue $ I_{(n)}^{(\text{res})} (\epstil) $ is a polynomial in the external momenta and the mass regulator for any value of $\epstil$ (see \cite{Anselmi:2007ri}), and therefore represents contact terms in coordinate space. This can be shown by taking derivatives of $ I_{(n)} $ with respect to the external momenta (and mass regulator) enough times so that the corresponding Feynman diagram integral no longer diverges. As long as there are no IR divergences in the physical dimension, one can safely take the limit $m\to 0$ in $I_{(n)}^{(\text{ren})}$ to obtain the physical renormalized correlation function (the correlation functions of the stress-energy tensor in the case studied here are indeed free of IR divergences, as will be explained in section \ref{Sec:z2LishitzFreeScalarResults}).

As mentioned earlier, the renormalized function $ I_{(n)}^{(\text{ren})} $ depends on the choice of the parameter $\epstil$ that is kept constant as we take the limit $(\varepsilon_t,\varepsilon_s)\to(0,0)$. Suppose we instead choose a different parameter $ \epstil'(\varepsilon_t,\varepsilon_s)$, that corresponds to the renormalized correlation function $ \left(I_{(n)}^{(\text{ren})}\right)' $. Then we have the following relation between $ I_{(n)}^{(\text{ren})} $ and $ \left(I_{(n)}^{(\text{ren})}\right)' $:
\begin{equation}\label{RegAndRenLif:RegChangeEffectOnCorr}
\begin{split}
I_{(n)}^{(\text{ren})} &= \left.\frac{\partial f}{\partial\epslif}\right|_{\epstil=\text{const}}(0,0,p_i,m)\\
& = \left.\frac{\partial f}{\partial\epslif}\right|_{\epstil'=\text{const}}(0,0,p_i,m) +
\left. \frac{\partial f}{\partial \epstil'} \right|_{\epslif=\text{const}}(0,0,p_i,m) \left.\frac{\partial\epstil'}{\partial\epslif} \right|_{\epstil=\text{const}}(0,0)\\
&= \left(I_{(n)}^{(\text{ren})}\right)' - \alpha \left. \frac{\partial f}{\partial \epstil'} \right|_{\epslif=\text{const}}(0,0,p_i,m),
\end{split}
\end{equation}
where $\alpha \equiv - \left.\frac{\partial\epstil'}{\partial\epslif} \right|_{\epstil}(0,0) $. Note that, since $f(0,\epstil',p_i,m)$ is a polynomial in $p_i$ and $m$ for any value of $ \epstil' $, so is $ \left. \frac{\partial f}{\partial \epstil'} \right|_{\epslif}(0,0,p_i,m) $. Therefore $ I_{(n)}^{(\text{ren})} $ and $ \left(I_{(n)}^{(\text{ren})}\right)' $ differ from each other by contact terms, as expected from a change in the renormalization scheme. Also note that the possible change in the renormalized expressions as a result of the choice of $\epstil$ is completely described by the single parameter $\alpha$. In order to account for all possible choices, we leave $\alpha$ as a free parameter in our calculations, and use the following choice of $\epstil$:
\begin{equation}\label{RegAndRenLif:EpsTilChoiceDef}
\epstil = \varepsilon_t + \frac{\alpha}{1+z\alpha} \varepsilon_s,
\end{equation}
which is chosen such that moving from $\epstil$ to $\epstil' = \varepsilon_t$ we get the factor $ \left.\frac{\partial\epstil'}{\partial\epslif} \right|_{\epstil} = \left.\frac{\partial\varepsilon_t}{\partial\epslif} \right|_{\epstil} = -\alpha $. The inverse transformation from $(\epslif,\epstil) $ to $(\varepsilon_t,\varepsilon_s)$ is given by:
\begin{equation}
\begin{split}
\varepsilon_t &= -\alpha \epslif + (1+z\alpha) \epstil  , \\
\varepsilon_s &= (1+z\alpha)(\epslif - z \epstil).
\end{split}
\end{equation}

As in the relativistic case, we can calculate the Lifshitz scale anomaly coefficients from the $\epslif$ pole residue. Suppose the theory has a symmetry (such as a Lifshitz scale symmetry) that is not explicitly broken by the split dimensional regularization scheme itself, with a corresponding Ward identity of the form:
\begin{equation}\label{RegAndRenLif:GeneralFormOfWardIdentity}
T(\epslif,\epstil)[I_k] = 0,
\end{equation} 
where $\{I_k\}$ is a set of correlation functions and $T(\epslif,\epstil)$ is a linear operator that takes expressions of the form \eqref{RegAndRenLif:GeneralFormOfTermInCorrFunction} to expressions of the same form, and may or may not depend explicitly on the time and space dimensions. Since the symmetry is not broken by the regularization, the unrenormalized correlation functions satisfy identity \eqref{RegAndRenLif:GeneralFormOfWardIdentity}. We therefore have from \eqref{RegAndRenLif:RenormDef}:
\begin{equation}
\frac{1}{\epslif} T(\epslif,\epstil)\left[ I_k^{(\text{res})} \right] = -  T(\epslif,\epstil)\left[ I_k^{(\text{ren})} \right] + O(\epslif) + O(\epstil).
\end{equation}
The anomalous Ward identity is then given by:
\begin{equation}\label{RegAndRenLif:AnomalyFromRes}
T(0,0)\left[ I_k^{(\text{ren})} \right] = - \lim_{(\epslif,\epstil)\to 0} \left( \frac{1}{\epslif} T(\epslif,\epstil)\left[ I_k^{(\text{res})}(\epstil) \right] \right).
\end{equation}
Since the LHS of equation \eqref{RegAndRenLif:AnomalyFromRes} is finite, we can again draw the following conclusions:
\begin{enumerate}
\item $ T(\epslif,\epstil)\left[ I_k^{(\text{res})}(\epstil) \right] \sim O(\epslif) $.
\item If $T$ does not depend on $ \epslif $ and $\epstil$ then $ T\left[ I_k^{(\text{res})}(\epstil) \right] =0$, and there is no anomaly. Therefore, as long as TPD invariance is not explicitly broken by the split dimensional regularization scheme (as is the case with the free scalar we consider in the following sections), we don't expect it to be anomalous.\footnote{Like in the relativistic case, the operator $T$ that corresponds to conservation of the stress-energy tensor does not introduce any factors that depend on $d_t$ or $d_s$.} 
\end{enumerate}

It is important to consider the consequences of changing the choice of the parameter $\epstil$ (and thereby the renormalization) on the anomalous Ward identity \eqref{RegAndRenLif:AnomalyFromRes}. By changing our choice from $\epstil'$ to $\epstil$, we know from \eqref{RegAndRenLif:RegChangeEffectOnCorr} that the change in the anomalous Ward identity is given by:
\begin{equation}
T(0,0)\left[ I_k^{(\text{ren})} \right] = T(0,0)\left[ \left(I_{(n)}^{(\text{ren})}\right)' \right] - \alpha\, T(0,0) \left[ \frac{\partial f}{\partial\epstil'}(0,0) \right].
\end{equation}
Since $ \frac{\partial f}{\partial\epstil'}(0,0) $ is a local expression (a polynomial in the external momenta), the term $ T(0,0) \left[ \frac{\partial f}{\partial\epstil'}(0,0) \right] $ represents a trivial solution of the WZ consistency condition (one that can be cancelled by a local counterterm).  We therefore expect only the coefficients of trivial terms to depend on $\alpha$. This is consistent with the general expectation that only coefficients of trivial terms can be regularization dependent.

\subsubsection{Expansion in the External Momenta}

Equation \eqref{RegAndRenLif:AnomalyFromRes} allows us to calculate the anomalous Ward identity from the $\epslif$ pole residue. This is especially useful in the Lifshitz case, since the denominators of the propagators are generally polynomials of degree $2z$, and the Feynman diagram integrals are therefore more difficult to fully evaluate than they are in the relativistic case. Their divergent parts, however, can again be obtained simply by expanding the Feynman diagram integrands in powers of the external momenta. Suppose the integrand is $ h(\hat p_i,\bar p_i, \hat q, \bar q,m) $, where $q$ is the internal loop momentum and $h$ has a Lifshitz dimension\footnote{In cases where the integrand does not have a uniform Lifshitz dimension, one can always write it as a sum of terms with uniform Lifshitz dimensions and calculate the pole residue of each of them separately.} $l$ and therefore satisfies:
\begin{equation}
h(\lambda^z \hat p_i,\lambda \bar p_i, \lambda^z \hat q, \lambda \bar q, \lambda^z m) = \lambda^l h(\hat p_i,\bar p_i, \hat q, \bar q,m).
\end{equation}
Rescaling $\bar q$ by a factor of $\frac{1}{\lambda}$, and $\hat q$ and $m$ by a factor of $\frac{1}{\lambda^z}$ (where $\lambda\to 0$) we get:
\begin{equation}
h\left(\hat p_i,\bar p_i,\frac{\hat q}{\lambda^z},\frac{\bar q}{\lambda},\frac{m}{\lambda^z}\right) = \lambda^{-l} h(\lambda^z \hat p_i,\lambda \bar p_i,\hat q,\bar q,m).
\end{equation}
We next expand $ h(\lambda^z \hat p_i,\lambda \bar p_i,\hat q,\bar q,m) $ in powers of $\lambda$ to obtain:
\begin{equation}
\begin{split}
h\left(\hat p_i,\bar p_i,\frac{\hat q}{\lambda^z},\frac{\bar q}{\lambda},\frac{m}{\lambda^z}\right) &= \lambda^{-l} \left[ \sum_{k=0}^{k_0} \lambda^k\, h^{(k)}(\hat p_i, \bar p_i, \hat q, \bar q, m)
 + O(\lambda^{k_0+1}) \right]\\
&= \sum_{k=0}^{k_0} h^{(k)}\left(\hat p_i, \bar p_i, \frac{\hat q}{\lambda^z}, \frac{\bar q}{\lambda}, \frac{m}{\lambda^z} \right) + O(\lambda^{k_0-l+1}),
\end{split}
\end{equation}
where $ h^{(k)}(\hat p_i, \bar p_i, \hat q, \bar q, m) $ is a polynomial in the external momenta, given by:
\begin{equation}
\begin{split}
&h^{(k)}(\hat p_i, \bar p_i, \hat q, \bar q, m)\\
&\qquad \equiv
 \sum_{rz+s=k} \frac{1}{r!\,s!}\, \hat p_{i_1}^{\mu_1}\ldots\hat p_{i_r}^{\mu_r} \bar p_{j_1}^{\nu_1}\ldots \bar p_{j_s}^{\nu_s} \frac{\partial^{r+s} h}{ \partial\hat p_{i_1}^{\mu_1}\ldots\partial\hat p_{i_r}^{\mu_r} \partial\bar p_{j_1}^{\nu_1}\ldots \partial\bar p_{j_s}^{\nu_s}}(0,0,\hat q, \bar q, m).
\end{split}
\end{equation}
Defining:
\begin{equation}
\tilde{h}(\hat p_i,\bar p_i,\hat q,\bar q,m) \equiv \sum_{k=0}^{k_0} h^{(k)}(\hat p_i,\bar p_i,\hat q,\bar q,m),
\end{equation}
we conclude that the integral over $ h(\hat p_i,\bar p_i,\hat q,\bar q,m) - \tilde{h}(\hat p_i,\bar p_i,\hat q,\bar q,m) $ has a divergence degree of $ zd_t + d_s +l-k_0-1 $. If we choose $ k_0 = d_s^\text{phys} + z + l$ , the integral over $ h - \tilde{h} $ converges in $d_s^\text{phys} + 1$ dimensions, and therefore the integrals over $ h(\hat p_i,\bar p_i,\hat q,\bar q,m) $ and $ \tilde{h}(\hat p_i,\bar p_i,\hat q,\bar q,m) $ have the same $\epslif$ pole.

Thus in order to calculate the pole residue, the only integrals left to evaluate are the ones over the polynomial coefficients in the expressions $ \tilde{h}(\hat p_i,\bar p_i,\hat q,\bar q,m) $. Each of these is a linear combination of terms of the form:
\begin{equation}
\hat\delta_{\rho_1\rho_2}\hat\delta_{\rho_3\rho_4}\ldots\bar\delta_{\sigma_1}\bar\delta_{\sigma_2}\ldots \frac{(\hat q^2)^a (\bar q^2)^b \hat q_{\mu_1}\ldots \hat q_{\mu_c} \bar q_{\nu_1}\ldots \bar q_{\nu_d}}{\left[\hat q^2 + \kappa\left(\bar q^2 \right)^z + m^2 + i\epsilon \right]^I}.
\end{equation} 
When performing the integrations over $\hat q$ and $ \bar q$, we may use the time rotation and space rotation symmetries to make the replacements given in equations \eqref{RegAndRenConf:SphericalSymmReplacement}--\eqref{RegAndRenConf:SymmMetricProdDef} separately for $ \hat q$ products (using the $\hat\delta_{\mu\nu} $ metric) and for $\bar q$ products (using the $\bar\delta_{\mu\nu} $ metric).
The integral over $ \tilde{h}(\hat p_i,\bar p_i,\hat q,\bar q,m) $ can then be written in terms of the following known integral (see \cite{Anselmi:2007ri}):
\begin{equation}
\label{RegAndRenLif:GeneralZIntegral}
\begin{array}{l}
X\left( {d_t,d_s,r,s,I,z} \right) \equiv \int {\frac{{{d^{{d_t}}}\hat{q} }}{{{{\left( {2\pi } \right)}^{{d_t}}}}}\int {\frac{{{d^{{d_s}}}\bar{q}}}{{{{\left( {2\pi } \right)}^{{d_s}}}}}\frac{{{{\left( {{{\hat{q}} ^2}} \right)}^r}{{\left( {{{\bar{q}}^2}} \right)}^s}}}{{{{\left[ {{{\hat{q}}^2} +{\kappa {\left( {{{\bar{q}}^2}} \right)}^z} + {m^2}+i\epsilon} \right]}^I}}}} }  = \\ \\
 = \frac{{{i^{d_t}\kappa^{ - \left( {2s + {d_s}} \right)/2z}}{{\left( {{m^2}} \right)}^{r - I + s/z + \left( {{d_t} + {d_s}/z} \right)/2}}}}{{z{{\left( {4\pi } \right)}^{\left( {{d_t} + {d_s}} \right)/2}}}}\frac{{\Gamma \left( {\frac{{2s + {d_s}}}{{2z}}} \right)\Gamma \left( {\frac{{2r + {d_t}}}{2}} \right)}}{{\Gamma \left( {\frac{{{d_t}}}{2}} \right)\Gamma \left( {\frac{{{d_s}}}{2}} \right)\Gamma \left( I \right)}}\Gamma \left( {I - r - \frac{s}{z} - \frac{{z{d_t} + {d_s}}}{{2z}}} \right),
\end{array}
\end{equation}
where $ k = l - 2zr - 2s + 2zI \leq d_s^\text{phys} + z +l $ (so that $ d_s^\text{phys}+z+2zr+2s-2zI \geq 0 $). Note that the integral \eqref{RegAndRenLif:GeneralZIntegral} contributes to the pole residue a term proportional to $ m^{2r-2I+2s/z+1+d_s^\text{phys}/z} $. Since terms with $ 2r-2I+2s/z+1+d_s^\text{phys}/z > 0 $ vanish in the $m\to 0$ limit, one only needs to compute terms in the expansion with $ 2r-2I+2s/z+1+d_s^\text{phys}/z = 0 $ (those of order $ k = d_s^\text{phys} + z + l $) in order to obtain the $\epslif$ pole residue. For these terms, the integral \eqref{RegAndRenLif:GeneralZIntegral} has the following pole:
\begin{equation}
\label{RegAndRenLif:GeneralZIntegralPole}
X(d_t,d_s,r,s,I,z) = \frac{i^{d_t}}{\epslif}\, \frac{2 \kappa^{-(2s+d_s)/2z}}{(4\pi)^{(d_t+d_s)/2}} 
\frac{\Gamma\left(\frac{2s+d_s}{2z}\right)\Gamma\left(\frac{2r+d_t}{2}\right)}{\Gamma\left(\frac{d_t}{2}\right)\Gamma\left(\frac{d_s}{2}\right)\Gamma(I)} + O(1).
\end{equation}
Note that since $r, s$ and $I$ appear in the correlation functions as non-negative integers, it is possible to relate the different poles $X(d_t,d_s,r,s,I,z)$ appearing in the correlation function of a given order for a fixed value of $z$ and with various values of $r, s, I$ using the recursive property of Gamma functions $\Gamma(n+1)=n\Gamma(n)$.

In conclusion, this procedure of expanding the Feynman diagram integrand in external momenta, computing the pole residue via equations \eqref{RegAndRenConf:SphericalSymmReplacement}--\eqref{RegAndRenConf:SymmMetricProdDef}, \eqref{RegAndRenLif:GeneralZIntegral}--\eqref{RegAndRenLif:GeneralZIntegralPole} and applying formula \eqref{RegAndRenLif:AnomalyFromRes} enables us to calculate the anomalous Ward identities corresponding to Lifshitz scale symmetry without calculating the full correlation functions. In the following sections we use this procedure to calculate the anomaly coefficients for the case of a free $z=2$ Lifshitz scalar in $2+1$ dimensions.

\section{The Conformal Scalar Field and its Weyl Anomalies}
\label{sec:ConformalAnom}
 
In this section we review the calculation of the conformal anomaly coefficients for the relativistic free and massless scalar field using dimensional regularization and the procedure described in section \ref{RegAndRenorm}. We include this example as a reference for the calculation of Lifshitz anomaly coefficients for the non-relativistic Lifshitz scalar given in section \ref{Sec:z2LishitzFreeScalarResults} using a similar procedure.
The calculation was performed both for two and four spacetime dimensions.
In both cases the Weyl anomalies agree with the known results found in literature \cite{Polchinski,Capper:1974ic,Duff:1977ay,Duff:1993wm,Birrell:1982ix,Coriano:2012wp,Osborn:1993cr}. 

In order to calculate the conformal anomalies for the free scalar, one must first define the flat space stress-energy tensor such that it satisfies the Ward identities \eqref{CurvedConfWard:DiffWardIdent}--\eqref{CurvedConfWard:WeylWardIdent} (i.e.\ it is conserved, symmetric and traceless). One way to do this is to define the theory over a curved spacetime manifold such that the action is invariant under both diffeomorphisms and Weyl transformations (see section \ref{WardIdentities}), and derive the stress-energy tensor from the curved spacetime action using the definition \eqref{eq:StressTensorDef}.

The conformal coupling of a free relativistic scalar field $\phi$ to curved spacetime is given by the following action (see e.g.\ \cite{Birrell:1982ix}):
\begin{equation}
\label{eq:RelativisticScalar}
S = \int {{d^d}x\sqrt{|g|} \left[ {\frac{1}{2}{\partial _\mu }\phi {\partial ^\mu }\phi  - \frac{{d - 2}}{{4\left( {1 - d} \right)}}R{\phi ^2}} \right]}   ,
\end{equation}
where $d$ is the number of spacetime dimensions, and $R$ is the Ricci scalar of the background manifold.  
This action in indeed diffeomorphism and Weyl-invariant.
The improved stress-energy tensor calculated from this action using the definition in equation \eqref{eq:StressTensorDef} is (see \cite{Deser:1996na}):
\begin{equation}
\label{eq:ImprovedRelStressTensor}
{{ T}_{\mu \nu }} = -{\partial _\mu }\phi {\partial _\nu }\phi  + \frac{1}{2}{\eta _{\mu \nu }}\left( {{\partial _\rho }\phi {\partial ^\rho }\phi } \right) - \frac{{d - 2}}{{4\left( {1 - d} \right)}}\left( {{\partial _\mu }{\partial _\nu } - {\eta _{\mu \nu }}{\partial ^2}} \right){\phi ^2}.
\end{equation}
Using the equations of motion, one can check that it indeed satisfies the conservation and tracelessness Ward identities: 
\begin{align}
\label{eq:ZeroRelCon}
{\partial ^\mu }{{ T}_{\mu \nu }} &= 0, \\
\label{eq:ZeroRelTraceless}
 T_\mu^\mu  &= 0,
\end{align}
for any number of spacetime dimensions $d$.\footnote{The stress-energy tensor satisfies these identities as operator equations, taking into account the equations of motion for $\phi$. Its renormalized correlation functions, however, satisfy the corresponding identities only up to local contact terms, as mentioned in sections \ref{WardIdentities} and \ref{RegAndRenorm}.} 

\subsection{The Relativistic Free Scalar in Two Dimensions}

We start with the calculation of the Weyl anomaly of a relativistic free scalar field in two spacetime dimensions from the two point correlation function of the stress-energy tensor.
We demonstrate two ways of performing the calculation: 
First, by preforming the full calculation of the one-loop diagram using dimensional regularization. 
Second, using the procedure of extracting only the divergent part of the diagram and using it to compute the anomaly coefficients, as described in section \ref{RegAndRenorm}.
The results agree with the known ones from the literature.

\subsubsection{The Full Calculation}
\label{FullRelCalc}

In the first way of calculating the anomaly, the two point correlation function of the stress-energy tensor is fully calculated using standard dimensional regularization. The Feynman rules and diagram used for the calculation are given in appendix~\ref{FeynRulesRelDiagRel}. 
 
The full evaluation of the expression that corresponds to the diagram \eqref{eq:RelativisticTwoPointGeneralExp} was performed using the massless integral formulas given in appendix~\ref{MasslessIntegrals}.
The final result for the two point correlation function is given by the following expression:
\begin{equation}
\label{eq:TwoPoint2DFullRes}
\begin{aligned}
&\left\langle {{{ T}_{\mu \nu }}\left( p \right){{ T}_{\rho \sigma }}\left( { - p} \right)} \right\rangle  = \\
&\qquad -\frac{1}{2}\left[ {{p^4}{\eta _{\mu \nu }}{\eta _{\rho \sigma }}\left( { - \frac{1}{{18}}\left( {3 + 7\varepsilon } \right)} \right){I_1} + {p^4}\left( {{\eta _{\mu \rho }}{\eta _{\nu \sigma }} + {\eta _{\mu \sigma }}{\eta _{\nu \rho }}} \right)\left( {\frac{1}{{12}} + \frac{\varepsilon }{9}} \right){I_1} + } \right.\\
 &\qquad \quad+ {p^2}\left( {{\eta _{\mu \nu }}{p_\rho }{p_\sigma } + {\eta _{\rho \sigma }}{p_\mu }{p_\nu }} \right)\left( {\frac{1}{{18}}\left( {3 + 7\varepsilon } \right)} \right){I_1} - \frac{\varepsilon }{6}{p_\mu }{p_\nu }{p_\rho }{p_\sigma }{I_1} + \\ 
&\qquad \quad\left. { + {p^2}\left( {{\eta _{\mu \rho }}{p_\nu }{p_\sigma } + {\eta _{\mu \sigma }}{p_\rho }{p_\nu } + {\eta _{\rho \nu }}{p_\mu }{p_\sigma } + {\eta _{\nu \sigma }}{p_\rho }{p_\mu }} \right)\left( { - \frac{1}{{36}}\left( {3 + 4\varepsilon } \right)} \right){I_1}} \right],
\end{aligned}
\end{equation}
where $p$ is the external momentum of the diagram, $\varepsilon$ is defined by $d=2-\varepsilon$ and the basic integral $I_1$ is given by:
\begin{equation}
\label{eq:TwoPoint2DFullResI1}
\begin{aligned}
{I_1}\left( {d,p} \right) & = 
\int {\frac{{{d^d}q}}{{{{\left( {2\pi } \right)}^d}}}} \frac{1}{{{q^2}}}\frac{1}{{{{\left( {p - q} \right)}^2}}} \\
 &= (-1)^{d/2-2}\, \frac{{i\Gamma \left( {d/2 - 1} \right)\Gamma \left( {d/2 - 1} \right)\Gamma \left( {2 - d/2} \right)}}{{{{\left( {4\pi } \right)}^{d/2}}\Gamma \left( {d - 2} \right)}}{\left( {{p^2}} \right)^{d/2 - 2}}.
\end{aligned}
\end{equation}
One can expand the integral $I_1$ around the physical dimension $d^\text{phys}=2$ to obtain:
\begin{equation}
\label{eq:I12d}
{I_1}\left( {d = 2 - \varepsilon ,p} \right) =  \frac{1}{\varepsilon} \left(\frac{i}{\pi}{p^{ - 2}}\right) + O\left( 1 \right).
\end{equation}
It is easy to verify that the result in equation \eqref{eq:TwoPoint2DFullRes} indeed satisfies the conservation Ward identity \eqref{eq:ConRelativistic2Point},
which in Fourier space takes the form:\footnote{See equations \eqref{eq:Fourier2PoingConv} and \eqref{eq:Fourier3PoingConv} for our Fourier conventions.}
\begin{equation}
\label{eq:Conser2PointWard}
{p_\mu }\left\langle {{ T^{\mu \nu }(p)}{ T^{\rho \sigma }(-p)}} \right\rangle  = 0.
\end{equation}
As expected, this identity holds separately on the finite part and on the pole part of \eqref{eq:TwoPoint2DFullRes} and is not anomalous. In addition one can verify that, when tracing over the full unrenormalized expression \eqref{eq:TwoPoint2DFullRes}, the Ward identity corresponding to Weyl invariance is satisfied:
\begin{equation}
\label{eq:RelTraceless}
\eta^{\mu\nu} \left\langle { T_{\mu\nu} \left( p \right){{ T}_{\rho \sigma }}\left( { - p} \right)} \right\rangle  = 0.
\end{equation}
However, this identity is only satisfied on the pole and finite parts of expression \eqref{eq:TwoPoint2DFullRes} together. 
Thus after performing renormalization of the correlation function we have from equation \eqref{RegAndRenConf:AnomalyFromRes}: 
\begin{equation}
\label{eq:Anomaly}
{\eta^{\mu\nu} \left\langle { T_{\mu\nu} \left( p \right){{ T}_{\rho \sigma }}\left( { - p} \right)} \right\rangle ^{\text{(ren)}}} =  - \lim_{\varepsilon\to 0} \left( \frac{1}{\varepsilon }{\eta^{\mu\nu} \left\langle { T_{\mu\nu} \left( p \right){{ T}_{\rho \sigma }}\left( { - p} \right)} \right\rangle ^{\text{(res)}}} \right).
\end{equation}
The $\varepsilon$ pole residue of expression \eqref{eq:TwoPoint2DFullRes} is given by:
\begin{equation}
\label{eq:Pole2d}
\begin{aligned}
&{\left\langle {{{ T}_{\mu \nu }}\left( p \right){{ T}_{\rho \sigma }}\left( { - p} \right)} \right\rangle ^{\text{(res)}}} = \frac{i}{{2\pi }}\left[ {\frac{3}{{18}}{p^2}{\eta _{\mu \nu }}{\eta _{\rho \sigma }} - \frac{1}{{12}}{p^2}\left( {{\eta _{\mu \rho }}{\eta _{\nu \sigma }} + {\eta _{\mu \sigma }}{\eta _{\nu \rho }}} \right) + } \right.\\ 
 &\qquad - \frac{3}{{18}}\left( {{\eta _{\mu \nu }}{p_\rho }{p_\sigma } + {\eta _{\rho \sigma }}{p_\mu }{p_\nu }} \right)\left. { + \frac{1}{{12}}\left( {{\eta _{\mu \rho }}{p_\nu }{p_\sigma } + {\eta _{\mu \sigma }}{p_\rho }{p_\nu } + {\eta _{\rho \nu }}{p_\mu }{p_\sigma } + {\eta _{\nu \sigma }}{p_\rho }{p_\mu }} \right)} \right].
\end{aligned}
\end{equation}
Tracing over \eqref{eq:Pole2d} (in $d=2-\varepsilon$ dimensions) and using equation \eqref{eq:Anomaly} we get: 
\begin{equation}
{\eta ^{\mu \nu }}{\left\langle {{{ T}_{\mu \nu }}\left( { - p} \right){{ T}_{\rho \sigma }}\left( p \right)} \right\rangle ^{\text{(ren)}}} =
\frac{i}{12\pi} \left( p^2 \eta_{\rho\sigma} - p_\rho p_\sigma \right),
\end{equation}
which is the well known anomalous Ward identity in two dimensions (see e.g.\ \cite{Polchinski,Deser:1993yx}).

\subsubsection{Poles Calculation}
\label{SeriesExp}
Using the procedure described in subsection \ref{ReviewOfDimensionalRegConformalCase}, the pole residue of the two point correlation function of the stress-energy tensor can also be computed without evaluating the full expression.
Starting from the expression for the two point function given in  \eqref{eq:RelativisticTwoPointGeneralExp}, we expand the integrand in powers of the external momentum and extract the terms proportional to $m^0$ (where $m$ is an IR mass regulator).
The expression for the two point function contains only five different types of integrals in this case. Their relevant $\varepsilon$ poles around two spacetime dimensions, obtained using expansion in the external momentum $p$, are given by the following expressions:
\begin{align}
\label{eq:RelSerEx1}
&\quad\int {\frac{{{d^d}q}}{{{{\left( {2\pi } \right)}^d}}}\frac{1}{{\left[ {{q^2} - {m^2}+i\epsilon} \right]}}} \frac{1}{{\left[ {{{\left( {q - p} \right)}^2} - {m^2}+i\epsilon} \right]}} = O(1),\\ 
\label{eq:RelSerEx2}
&\quad\int {\frac{{{d^d}q}}{{{{\left( {2\pi } \right)}^d}}}\frac{{{q_\mu }}}{{\left[ {{q^2} - {m^2}+i\epsilon} \right]}}} \frac{1}{{\left[ {{{\left( {q - p} \right)}^2} - {m^2}+i\epsilon} \right]}} = O(1),\\
\begin{split}
\label{eq:RelSerEx3}
 {I_{\mu \nu }}\left( p \right) & \equiv \int {\frac{{{d^d}q}}{{{{\left( {2\pi } \right)}^d}}}\frac{{{q_\mu }{q_\nu }}}{{\left[ {{q^2} - {m^2}+i\epsilon} \right]}}} \frac{1}{{\left[ {{{\left( {q - p} \right)}^2} - {m^2}+i\epsilon} \right]}}  \\  
&={I_{\mu \nu }}\left( {p = 0} \right) + O(1) = \frac{-i}{{8\pi }}{\eta_{\mu \nu }}\Gamma \left( {1 - \frac{d}{2}} \right){\left( {\frac{1}{{{m^2}}}} \right)^{1 - \frac{d}{2}}} + O(1),
\end{split}\\
\begin{split}
\label{eq:RelSerEx4}
{I_{\mu \nu \rho }}\left( p \right) & \equiv \int {\frac{{{d^d}q}}{{{{\left( {2\pi } \right)}^d}}}\frac{{{q_\mu }{q_\nu }{q_\rho }}}{{\left[ {{q^2} - {m^2}+i\epsilon} \right]}}} \frac{1}{{\left[ {{{\left( {q - p} \right)}^2} - {m^2}+i\epsilon} \right]}} \\
&= {p^\alpha }{\left[ {\frac{{d{I_{\mu \nu \rho }}\left( p \right)}}{{d{p^\alpha }}}} \right]_{p = 0}} + O(1) \\
&=\frac{{(-i)\left( {{p_\rho }{\eta_{\mu \nu }} + {p_\mu }{\eta_{\rho \nu }} + {p_\nu }{\eta_{\rho \mu }}} \right)}}{{16\pi }}\Gamma \left( {1 - \frac{d}{2}} \right){\left( {\frac{1}{{{m^2}}}} \right)^{1 - \frac{d}{2}}} + O(1),
\end{split}\\
\begin{split}
\label{eq:RelSerEx5}
{I_{\mu \nu \rho \sigma }}\left( p \right) & \equiv \int {\frac{{{d^d}q}}{{{{\left( {2\pi } \right)}^d}}}\frac{{{q_\mu }{q_\nu }{q_\rho }{q_\sigma }}}{{\left[ {{q^2} - {m^2}+i\epsilon} \right]}}} \frac{1}{{\left[ {{{\left( {q - p} \right)}^2} - {m^2}+i\epsilon} \right]}}\\ 
&= \frac{1}{{2!}}{p^\alpha }{p^\beta }{\left[ {\frac{{{d^2}{I_{\mu \nu \rho \sigma }}\left( p \right)}}{{d{p^\alpha }d{p^\beta }}}} \right]_{p = 0}} + O(1) \\
&= \left( {  \frac{i}{{96\pi }}{J_{\mu \nu \rho \sigma }} - \frac{i}{{24\pi }}{K_{\mu \nu \rho \sigma }}} \right)\Gamma \left( {1 - \frac{d}{2}} \right){\left( {\frac{1}{{{m^2}}}} \right)^{1 - \frac{d}{2}}} + O(1),
\end{split}
\end{align}
where:
\begin{align}
&{J_{\mu \nu \rho \sigma }} \equiv {p^2}\left( {{\eta_{\mu \nu }}{\eta_{\rho \sigma }} + {\eta_{\mu \sigma }}{\eta_{\rho \nu }} + {\eta_{\mu \rho }}{\eta_{\nu \sigma }}} \right),\\
&{K_{\mu \nu \rho \sigma }} \equiv {p_\rho }{p_\sigma }{\eta_{\mu \nu }} + {p_\rho }{p_\nu }{\eta_{\mu \sigma }} + {p_\rho }{p_\mu }{\eta_{\nu \sigma }} + {p_\sigma }{p_\mu }{\eta_{\nu \rho }} + {p_\sigma }{p_\nu }{\eta_{\mu \rho }} + {p_\mu }{p_\nu }{\eta_{\rho \sigma }}.
\end{align}

Note that although some of these expressions \eqref{eq:RelSerEx1}--\eqref{eq:RelSerEx5} are individually IR divergent at $d=2$, the total expression for the two point function of the stress-energy tensor (and indeed any correlation functions of the stress-energy tensor) is not IR divergent in this case. This follows from the form of the action \eqref{eq:RelativisticScalar} and the improved stress-energy tensor \eqref{eq:ImprovedRelStressTensor}: The action at $d=2$ contains only derivatives of the field $\phi$, and only the $O(\varepsilon)$ part of it contains $\phi$ with no derivatives. It follows that the stress-energy tensor (and any other operator constructed by taking variations of the action with respect to the metric) has the same structure -- at $d=2$ it contains only derivatives of $\phi$. These operators are therefore not Log correlated at $d=2$, and their correlation functions do not diverge in the IR. In terms of dimensionally regulated Feynman diagrams, for any propagator in the 1-loop diagram with momentum $q^\mu$, the $d=2$ part of each of its adjacent vertexes is of order $O(q)$. Thus the terms in the integrand of the diagram that contribute to the IR divergence when $q\to 0$ are of order $O(\varepsilon^2)$, and therefore vanish as $\varepsilon\to 0$ after integration is performed. This justifies the assumption made in subsection \ref{ReviewOfDimensionalRegConformalCase}, that the correlation functions of the stress-energy tensor are free of IR divergences. A similar argument will be valid for the Lifshitz case, as will be explained in the next section.

Substituting these poles in the general expression for the two point function \eqref{eq:RelativisticTwoPointGeneralExp}, we find exactly the same pole residue for the two point correlation function as in equation \eqref{eq:Pole2d}, and therefore, by the use of \eqref{eq:Anomaly}, the same anomalous Ward identity.

\subsection{The Relativistic Free Scalar in Four Dimensions}

We next turn to the example of a relativistic free scalar field in four spacetime dimensions. Again using dimensional regularization and the procedure described in subsection \ref{ReviewOfDimensionalRegConformalCase} with $d=4-\varepsilon$, the anomalous contributions to the Ward identities \eqref{CorrConfWard:WardIdentForVarCorrWeyl2Point} and \eqref{CorrConfWard:WardIdentForVarCorrWeyl3Point} can be computed from the $\varepsilon$ pole residues of the two and three point correlation functions of the stress-energy tensor.

The details needed for the calculation of the correlation functions as defined in equation \eqref{eq:nPointFunctionDef}, including the Feynman rules for the vertexes and the expressions for the diagrams, are given in appendix \ref{FeynRulesRelDiagRel}. The pole residues were extracted via a power expansion in the external momenta as described in subsection \ref{ReviewOfDimensionalRegConformalCase}. Since these calculations involve a very large number of terms, they were performed using a computer script that was written for this purpose. 

The pole residue of the two point correlation function, as obtained from the power expansion in the external momentum $p$, is given by:
\begin{equation}
\label{eq:2P4dPole}
\begin{array}{l}
{\left\langle {{{ T}_{\mu \nu }}\left( p \right){{ T}_{\rho \sigma }}\left( { - p} \right)} \right\rangle ^{\text{(res)}}} = \\ 
\qquad -\frac{i}{{16{\pi ^2} }}\frac{1}{9}\left[ { - \frac{1}{{10}}{p^4}{\eta _{\mu \nu }}{\eta _{\rho \sigma }} + \frac{3}{{20}}{p^4}\left( {{\eta _{\mu \rho }}{\eta _{\nu \sigma }} + {\eta _{\mu \sigma }}{\eta _{\nu \rho }}} \right) } \right.\\ 
\qquad + \frac{1}{{10}}{p^2}\left( {{\eta _{\mu \nu }}{p_\rho }{p_\sigma } + {\eta _{\rho \sigma }}{p_\mu }{p_\nu }} \right) + \frac{1}{5}{p_\mu }{p_\nu }{p_\rho }{p_\sigma } \\ 
\qquad \left. { - \frac{3}{{20}}{p^2}\left( {{\eta _{\mu \rho }}{p_\nu }{p_\sigma } + {\eta _{\mu \sigma }}{p_\rho }{p_\nu } + {\eta _{\rho \nu }}{p_\mu }{p_\sigma } + {\eta _{\nu \sigma }}{p_\rho }{p_\mu }} \right)} \right].
\end{array}
\end{equation}
It can be easily checked that this expression satisfies the conservation identity \eqref{eq:Conser2PointWard}.
However, tracing over the indexes $\mu,\nu$ in \eqref{eq:2P4dPole} yields an expression that does not vanish for a general dimension $d$. Using equation \eqref{RegAndRenConf:AnomalyFromRes}, one obtains:    
\begin{equation}
{\eta ^{\mu \nu }}{\left\langle {{{ T}_{\mu \nu }}\left( { - p} \right){{ T}_{\rho \sigma }}\left( p \right)} \right\rangle ^{\text{(ren)}}} = - \frac{i}{{1440{\pi ^2}}}{p^2}\left( {{p_{\rho}p_\sigma} - {\eta _{\rho \sigma }}{p^2}} \right).
\end{equation} 
This is in agreement with the anomalous Ward identity  \eqref{CorrConfWard:WardIdentForVarCorrWeyl2Point}, due to the trivial $\Box R$ term in the Weyl cohomology of the relativistic theory in four dimensions (see e.g.\ \cite{Osborn:1993cr,Coriano:2012wp}). 

The pole residue of the three point correlation function ${\left\langle {{T_{\mu \nu }}\left( k \right){T_{\rho \sigma }}\left( q \right){T_{\alpha \beta }}\left( p \right)} \right\rangle^{(\text{res})}_{W}}$ can also be calculated using expansion in the external momenta. However, the final result is too long to be shown here.  
As expected, this pole residue and the pole residue of the two point function together satisfy the following conservation identity:
\begin{equation}
\label{eq:Conser3PointWard}
\begin{array}{l}
{k_\nu }\left\langle {{T^{\mu \nu }}\left( k \right){T^{\rho \sigma }}\left( q \right){T^{\alpha \beta }}\left( p \right)} \right\rangle_W  = \\
\qquad i {p^\mu }\left\langle {{T^{\rho \sigma }}{T^{\alpha \beta }}} \right\rangle \left( q \right) +i {q^\mu }\left\langle {{T^{\rho \sigma }}{T^{\alpha \beta }}} \right\rangle \left( p \right)\\ 
 \qquad -i {p_\nu }\left[ {{\eta ^{\mu \beta }}\left\langle {{T^{\nu \alpha }}{T^{\rho \sigma }}} \right\rangle \left( q \right) + {\eta ^{\mu \alpha }}\left\langle {{T^{\nu \beta }}{T^{\rho \sigma }}} \right\rangle \left( q \right)} \right] \\
 \qquad -i {q_\nu }\left[ {{\eta ^{\mu \rho }}\left\langle {{T^{\nu \sigma }}{T^{\alpha \beta }}} \right\rangle \left( p \right) + {\eta ^{\mu \sigma }}\left\langle {{T^{\nu \rho }}{T^{\alpha \beta }}} \right\rangle \left( p \right)} \right],
\end{array}
\end{equation}
which is the Fourier transformed version of the conservation Ward identity \eqref{eq:ConRelativistic3Point}.
From these pole residues, one can use equation \eqref{RegAndRenConf:AnomalyFromRes} to calculate the LHS of the Fourier transformed version of the anomalous Ward identity of the three point correlation function \eqref{CorrConfWard:WardIdentForVarCorrWeyl3Point}, given by:
\begin{equation}
\label{eq:WardWeylRel}
\begin{aligned}
&{\eta _{\mu \nu }}{\left\langle {{T^{\mu \nu }}\left( k \right){T^{\rho \sigma }}\left( q \right){T^{\alpha \beta }}\left( p \right)} \right\rangle _W^{\text{(ren)}}} 
- 2i\left\langle {{T^{\rho \sigma }}\left( { - q} \right){T^{\alpha \beta }}\left( q \right)} \right\rangle^{\text{(ren)}}  \\
& \qquad  
- 2i\left\langle {{T^{\rho \sigma }}\left( { - p} \right){T^{\alpha \beta }}\left( p \right)} \right\rangle^{\text{(ren)}} =
{{{-4\left[ \delta^2 \left(\sqrt{|g|}\mathcal{A}\right)\left( {q,p}\right) \right]}^{\rho \sigma \alpha \beta }} 
} ,
\end{aligned}
\end{equation}
where ${{{\left[ \delta^2 \left(\sqrt{|g|} \mathcal{A}\right)\left( {q,p} \right) \right]}^{\rho \sigma \alpha \beta }}}$ represents the Fourier transformed second variation of the anomalous contribution to the Ward identity \eqref{CurvedConfWard:AnomWeylWardIdent} with respect to the metric (see appendix \ref{sec:RelScalCon} for notations).
After calculating the second variation of the anomaly and trivial term densities listed in equation \eqref{eq:GeneralExpreForAnomalyRel4d} with respect to the background metric, the results can be substituted into the RHS of the anomalous Ward identity \eqref{eq:WardWeylRel}.

The coefficients $\beta_a$, $\beta_b$ and $\beta_c$ can then be extracted by comparing the two sides of the identity. The values of the coefficients obtained using this procedure are the same as the ones given in \eqref{eq:CoeefsRel4d} and therefore agree with those found in the literature.

\section{The Free $z=2$ Lifshitz Scalar Field and its Scale Anomalies}
\label{Sec:z2LishitzFreeScalarResults}

In this section we study the Lifshitz anomaly of a Lifshitz $z=2$ scalar field in $2+1$ dimensions, using the method of split dimensional regularization and renomarlization described in subsection \ref{SplitDimensionalRegAndTheLifshitzCase}, applied to the two and three point correlation functions of the stress-energy tensor. Up to second order in the background fields, we find one anomaly in the two derivatives sector, and no anomalies in the four derivatives sector. This is in agreement with the results previously found in \cite{Baggio:2011ha} using a heat kernel calculation. The value of the anomaly coefficient also agrees with the result in \cite{Baggio:2011ha}. As expected, only the coefficients of the trivial terms which appear in the two and three point correlation functions 
depend on the regularization parameter $\alpha$, defined in subsection \ref{SplitDimensionalRegAndTheLifshitzCase} -- they are all regularization dependent, and can be removed by adding the appropriate counterterms to the effective action. We also show that trivial terms that correspond to a curved background structure that violates the Frobenius condition (that is, terms that vanish when the Frobenius condition is assumed) appear in the case we study with non-vanishing coefficients.
Therefore, the curved spacetime description of these terms, and their cancellation via a counterterm, requires giving up the foliation structure of the background manifold (see \cite{Arav:2016xjc} for further discussion).

\subsection{The $z=2$ Free Lifshitz Scalar Field in General Spacetime Dimensions}

The Lifshitz anomalies of a free Lifshitz scalar field with a dynamical critical exponent $z=2$ in $2+1$ dimensions have been considered in several previous works \cite{Baggio:2011ha,Griffin:2012qx}. 
The flat space action of the theory is given by:
\begin{equation}
\label{eq:LifshitzFreeAction3d}
S=\int{d^2xdt\left(\frac{1}{2}(\partial_t\phi)^2-\frac{1}{2}\kappa(\nabla^2\phi)^2\right) } .
\end{equation}

In order to apply the method of split dimensional regularization and renormalization described in subsection \ref{SplitDimensionalRegAndTheLifshitzCase}, one must first couple the theory described by the action \eqref{eq:LifshitzFreeAction3d} to a curved background manifold with a general number of space dimensions $d_s$ and time dimensions $d_t$.
This coupling must be done in a way that preserves the curved spacetime symmetries detailed in subsection \ref{subsubsec:LifshitzSymWardCuvedSpacetime}, as well as local rotations of the time directions, and smoothly reduces back to \eqref{eq:LifshitzFreeAction3d} in the limit of flat spacetime and $d_s\to 2, d_t \to 1$.

This can be accomplished in several ways (see appendix \ref{app:LifCurvedSpacetimeMultTimDims} for details). Here we choose to use the following curved spacetime action:
\begin{equation}
\label{eq:GeneralActionLifs}
\begin{aligned}
S & = \int d^{d_t+d_s}x \sqrt{|g|} \left\{ \frac{1}{2}\left[ \mathcal{L}_{n^{(i)}}\phi+\xi_1\, K_S^{(i)}\phi \right]^2 \right. \\
& \qquad \qquad \qquad \qquad \left. -\frac{\kappa}{2}\left[\bar{\nabla}^2\phi+\xi_2 \,a^\mu\bar{\nabla}_\mu\phi+\xi_3 \, a^2\phi+\xi_4 \,\bar{\nabla}_\mu a^\mu\phi \right]^2\right\},
\end{aligned}
\end{equation}
where $d_t$ and $d_s$ are the numbers of time and space dimensions respectively and $\{ n^{(i)}_\mu \} \, (i=1,\ldots,d_t)$ is an orthonormal set of 1-forms that represents the local $d_t$ time directions on the manifold. The coefficients $ \xi_1, \xi_2, \xi_3, \xi_4  $ are given by:
\begin{equation}
\begin{aligned}
&\xi_1 \equiv \frac{1}{d_s}\left(\frac{1}{2}d_{\text{lif}}-2\right), \qquad \quad \quad \qquad \qquad \xi_2\equiv \frac{d_t-1}{d_t}, \\
&  \xi_3 \equiv \frac{1}{4d_t^2}\left(\frac{1}{2}d_{\text{lif}}-2\right)\left(d_t-\frac{1}{2}d_s\right), \qquad \xi_4 \equiv \frac{1}{2d_t}\left(\frac{1}{2}d_{\text{lif}}-2\right),
\end{aligned}
\end{equation}
where $d_{\text{lif}}\equiv 2d_t+d_s$. 
The various background expressions and notations used in \eqref{eq:GeneralActionLifs} (the derivatives $\mathcal{L}_{n^{(i)}}$, ${{{\bar \nabla }_\mu }}$ and the expressions $a^\mu$, $K_S^{(i)}$) are defined in appendix \ref{app:LifCurvedSpacetimeMultTimDims} as a generalization of the definitions used in \cite{Arav:2014goa,Arav:2016xjc} to the case of multiple time directions. This action is indeed invariant under TPDs and anisotropic Weyl transformations, as well as under local time rotations of the form $ n_\mu^{(i)} \rightarrow \Lambda^{ij}(x) n_\mu^{(j)} $, where $ \Lambda^{ij}(x) $ is any orthogonal matrix in $d_t$ dimensions that depends on the spacetime coordinates $x$.

The action \eqref{eq:GeneralActionLifs} reduces in flat space to the following action:
\begin{equation}
\label{eq:FlatSpaceAction}
S=\int{d^{d_s+d_t}x \left(-\frac{1}{2}\hat{\partial}_\mu\phi\hat{\partial}^\mu\phi-\frac{1}{2}\kappa\bar{\nabla}^2\phi\bar{\nabla}^2\phi\right)},
\end{equation}
where, as mentioned in subsection \ref{SplitDimensionalRegAndTheLifshitzCase},
we use a flat spacetime metric of the form $\delta_{\mu\nu} = \operatorname{diag}(\hat\delta_{\hat\mu\hat\nu},\bar\delta_{\bar\mu\bar\nu})$, with $ \hat\delta_{\hat\mu\hat\nu} = \operatorname{diag}(-1,\ldots,-1)$ over the time dimensions and $ \bar\delta_{\bar\mu\bar\nu} = \operatorname{diag}(1,\ldots,1)$ over the space dimensions (see appendix \ref{app:conventions} for our notations and conventions).

The action \eqref{eq:FlatSpaceAction} is invariant under time and space rotations, as well as the following Lifshitz scaling transformation:
\begin{equation}
\label{eq:LifshitzScalingTrans}
{\hat{x}_{ \mu }} \to {\lambda ^{ - 2}}\,{\hat{x}_{ \mu }},\qquad {\bar{x}_{ \mu }} \to {\lambda ^{ - 1}}\,{\bar{x}_{ \mu }},\qquad \phi  \to {\lambda ^{\left( {2{d_t} + {d_s} - 4} \right)/2}} \,\phi ,
\end{equation} 
where $\lambda$ is a parameter of the scaling transformation, and $\kappa$ is dimensionless under the scaling transformation.
The flat space stress-energy tensor, as derived from the action \eqref{eq:GeneralActionLifs} by taking the variation of the action with respect to the vielbeins (according to the definition \eqref{CurvedLifshitzWard:StressTensorDef}), is given by:
\begin{align*}
{T^\mu }_\nu  &= {{\hat \partial }^\mu }\phi {\partial _\nu }\phi  + 2\kappa {{\bar \partial }^\mu }{\partial _\nu }\phi {{\bar \nabla }^2}\phi  - \frac{{\delta ^\mu _\nu }}{2}\left( {{{\hat \partial }_\sigma }\phi {{\hat \partial }^\sigma }\phi  + \kappa {{\bar \nabla }^2}\phi {{\bar \nabla }^2}\phi } \right) \\
&- \kappa \left( {{\partial ^\mu }{{\bar \partial }_\nu }\phi {{\bar \nabla }^2}\phi  + {\partial ^\mu }\phi {{\bar \nabla }^2}{{\bar \partial }_\nu }\phi  + {{\bar \partial }^\mu }{\partial _\nu }\phi {{\bar \nabla }^2}\phi  + {\partial _\nu }\phi {{\bar \nabla }^2}{{\bar \partial }^\mu }\phi } \right) \\
 &+ \frac{\kappa }{{4{d_t}}}\left( {\left( {2{d_t} + {d_s}} \right) - 4} \right)\left( {{{\hat \partial }_\nu }\phi {{\bar \partial }^\mu }{{\bar \nabla }^2}\phi  + \phi {{\bar \partial }^\mu }{{\hat \partial }_\nu }{{\bar \nabla }^2}\phi } \right)  \\
 &+ \frac{\kappa }{{4{d_t}}}\left( {\left( {{d_s} - 2{d_t}} \right)\left( {{{\hat \partial }_\nu }{{\bar \partial }^\mu }\phi {{\bar \nabla }^2}\phi  + {{\bar \partial }^\mu }\phi {{\bar \nabla }^2}{{\hat \partial }_\nu }\phi } \right)} \right) \\
 &+ \kappa {{\bar \delta }^\mu }_\nu \left( {{\bar\nabla^2}\phi {{\bar \nabla }^2}\phi  + {\bar\partial _\sigma }\phi {\bar\partial ^\sigma }{{\bar \nabla }^2}\phi } \right) \numthis
\label{eq:LifStressTensor} \\
& - \frac{\kappa }{{4{d_t}}}\hat \delta _\nu ^\mu \left( {\left( {2{d_s} - 4} \right){{\bar \partial }_\sigma }\phi {{\bar \partial }^\sigma }{{\bar \nabla }^2}\phi  + \left( {2{d_t} + {d_s} - 4} \right)\phi {{\bar \nabla }^4}\phi } \right)\\
& - \frac{\kappa }{{4{d_t}}}\hat \delta _\nu ^\mu \left( {{d_s} - 2{d_t}} \right){{\bar \nabla }^2}\phi {{\bar \nabla }^2}\phi  + \kappa \left( {{{\bar \partial }_\nu }{{\hat \partial }^\mu }\phi {{\bar \nabla }^2}\phi  + {{\hat \partial }^\mu }\phi {{\bar \nabla }^2}{{\bar \partial }_\nu }\phi } \right)\\
& + \frac{1}{{2{d_s}}}\left( {4 - \left( {2{d_t} + {d_s}} \right)} \right)\left( {{{\bar \partial }_\nu }\phi {{\hat \partial }^\mu }\phi  + \phi {{\hat \partial }^\mu }{{\bar \partial }_\nu }\phi } \right)\\
& - \frac{1}{{2{d_s}}}\bar \delta _\nu ^\mu \left( {4 - \left( {2{d_t} + {d_s}} \right)} \right)\left( {{{\hat \partial }_\sigma }\phi {{\hat \partial }^\sigma }\phi  + \phi {{\hat \partial }_\sigma }{{\hat \partial }^\sigma }\phi } \right).
\end{align*}
Note that this expression is symmetric in its spatial components (${T_{\bar \mu \bar \nu }} = {T_{\bar \nu \bar \mu }}$), and its temporal components (${T_{\hat \mu \hat \nu }} = {T_{\hat \nu \hat \mu }}$), indicating both time and space rotations invariance,  but not in its combined space/time components ($T_{\hat\mu\bar\nu} \neq T_{\bar\nu\hat\mu})$, since there is no Lorentz inveariance. It is also regular in the limit $d_s\to 2, d_t\to 1$. These two properties are crucial for the assumptions underlying the procedure outlined in subsection \ref{SplitDimensionalRegAndTheLifshitzCase} to be satisfied.\footnote{This stress-energy tensor is significantly different from the one found in \cite{Anselmi:2007ri}, which is not symmetric in its spatial components, and contains a coefficient that diverges 
in the limit $d_t\to 1$.} 

Using the equations of motion in flat space, given by:
\begin{equation}
\label{eq:EOMScalar}
\left( {{\hat{\partial _{\mu }}}^2 - \kappa {{\bar \nabla }^4}} \right)\phi  = 0,
\end{equation}
it is easily verified that the  stress-energy tensor \eqref{eq:LifStressTensor} satisfies the following Ward identities for any values of $d_t$ and $d_s$ (these are just the flat space versions of identities \eqref{CurvedLifshitzWard:TPDWardIdentTe}--\eqref{CurvedLifshitzWard:WeylWardIdent},  generalized to $d_t>1$):
\begin{align}
\label{eq:ImprovedConserv}
{\partial_\mu } T^\mu{}_\nu  &= 0, \\
\label{eq:ImprovedTreceless}
D^a_\mu T^\mu{}_a &\equiv \left( 2\hat\delta^a_\mu + \bar\delta^a_\mu \right) T^\mu{}_a = 0.
\end{align}

In the particular case of $d_s=2$, $d_t=1$, the stress-energy tensor \eqref{eq:LifStressTensor} reduces to the following expression:
\begin{equation}
\begin{aligned}
{T^\mu }_\nu  &= {{\hat \partial }^\mu }\phi {\partial _\nu }\phi  + 2\kappa {{\bar \partial }^\mu }{\partial _\nu }\phi {{\bar \nabla }^2}\phi  - \frac{{\delta ^\mu _\nu }}{2}\left( {{{\hat \partial }_\sigma }\phi {{\hat \partial }^\sigma }\phi  + \kappa {{\bar \nabla }^2}\phi {{\bar \nabla }^2}\phi } \right)\\
& - \kappa \left( {{{\bar \partial }^\mu }{{\bar \partial }_\nu }\phi {{\bar \nabla }^2}\phi  + {{\bar \partial }^\mu }\phi {{\bar \nabla }^2}{{\bar \partial }_\nu }\phi  + {{\bar \partial }^\mu }{\partial _\nu }\phi {{\bar \nabla }^2}\phi  + {\partial _\nu }\phi {{\bar \nabla }^2}{{\bar \partial }^\mu }\phi } \right)\\
& + \kappa {{\bar \delta }^\mu }_\nu \left( {{{\bar \nabla }^2}\phi {{\bar \nabla }^2}\phi  + {{\bar \partial }_\sigma }\phi {{\bar \partial }^\sigma }{{\bar \nabla }^2}\phi } \right).
\end{aligned}
\end{equation}

\subsection{The Lifshitz Anomaly of the Free Scalar}

In this subsection, we apply the procedure outlined in subsection \ref{SplitDimensionalRegAndTheLifshitzCase} to calculate the Lifshitz anomaly coefficients of the free Lifshitz $z=2$ scalar field in $2+1$ dimensions, using a split dimensional regularization scheme. The general steps of this calculation are as follows:
\begin{enumerate}
\item We consider the Feynman diagrams contributing to the flat space variational two and three point correlation functions of the stress-energy tensor, as given in \eqref{CorrLifshitzWard:VarToFeynmanCorr2Point}--\eqref{CorrLifshitzWard:VarToFeynmanCorr3Point}, in a general number of time and space dimensions.
\item For each of these diagrams, we extract the $\epslif$ pole residue of the diagram by expanding the integrand in powers of the external momenta as explained in subsection \ref{SplitDimensionalRegAndTheLifshitzCase}, utilizing the formulas \eqref{RegAndRenConf:SphericalSymmReplacement}--\eqref{RegAndRenConf:SymmMetricProdDef} and \eqref{RegAndRenLif:GeneralZIntegral}--\eqref{RegAndRenLif:GeneralZIntegralPole} (Note that we use the choice of $\epstil$ given in \eqref{RegAndRenLif:EpsTilChoiceDef}).
\item We substitute the pole residues into the LHS of Fourier transformed versions of the Ward identities \eqref{CorrLifshitzWard:WardIdentForVarCorrWeyl2Point}--\eqref{CorrLifshitzWard:WardIdentForVarCorrWeyl3Point}, and then use formula \eqref{RegAndRenLif:AnomalyFromRes} (taking the limit $(\epslif,\epstil)\to 0$) to obtain the anomalous contribution to these Ward identities. 
\item We calculate the first and second order variation of each of the possible independent anomaly and trivial term densities corresponding to this case\footnote{Note that we do not assume the Frobenius condition on the background 1-form $n_\mu$ here.}, as listed in \cite{Arav:2016xjc}, with respect to the vielbeins in the flat space limit. Note that we do not consider any $n$-point functions with $n>3$ in this work, and therefore only the coefficients of terms which are at most second order in the background fields can be calculated.
\item We substitute these expressions into the RHS of the Ward identities \eqref{CorrLifshitzWard:WardIdentForVarCorrWeyl2Point}--\eqref{CorrLifshitzWard:WardIdentForVarCorrWeyl3Point}, compare to the results of step 3, and extract the various coefficients.\footnote{When extracting these coefficients, one must be careful to take into account various dimensionally dependent identities that apply only in the physical dimensions of the theory, in this case $d_s=2, d_t=1$. For example, for $d_t=1$ the following identity applies: $ \hat P_\mu \hat P_\nu = \hat P_\rho \hat P ^\rho \hat \delta_{\mu\nu} $.}
\end{enumerate} 

For the purpose of calculating the pole residues of the correlation functions of the stress-energy tensor, it is convenient to drop terms
in \eqref{eq:LifStressTensor} that carry coefficients proportional to the parameter $\epslif$. 
These terms will only contribute expressions which are regular in the limit $(\epslif,\epstil)\to 0$ to the correlation functions.
Therefore, when calculating these pole residues, 
one can ignore such terms.
After dropping these terms, the stress-energy tensor takes the form:
\begin{equation}
\label{eq:StressTensorForPolesLif}
\begin{aligned}
{T^\mu }_\nu  &= {{\hat \partial }^\mu }\phi {{\hat \partial }_\nu }\phi  - \frac{{{{\hat \delta }^\mu }_\nu }}{2}\left( {{{\hat \partial }_\sigma }\phi {{\hat \partial }^\sigma }\phi  + \kappa \frac{{{d_s}}}{{2{d_t}}}{{\bar \nabla }^2}\phi {{\bar \nabla }^2}\phi  + \frac{\kappa }{{{d_t}}}\frac{{{d_s} - 2{d_t}}}{2}{{\bar \partial }_\sigma }\phi {{\bar \partial }^\sigma }{{\bar \nabla }^2}\phi } \right)\\
& + {{\hat \partial }^\mu }\phi {{\bar \partial }_\nu }\phi  + \kappa \frac{{2{d_t} + {d_s}}}{{4{d_t}}}{{\bar \partial }^\mu }{{\hat \partial }_\nu }\phi {{\bar \nabla }^2}\phi  - \kappa {{\hat \partial }_\nu }\phi {{\bar \nabla }^2}{{\bar \partial }^\mu }\phi  + \kappa \frac{{{d_s} - 2{d_t}}}{{4{d_t}}}{{\bar \partial }^\mu }\phi {{\bar \nabla }^2}{{\hat \partial }_\nu }\phi \\
& - \kappa {{\bar \partial }^\mu }\phi {{\bar \nabla }^2}{{\bar \partial }_\nu }\phi  - \kappa {{\bar \partial }_\nu }\phi {{\bar \nabla }^2}{{\bar \partial }^\mu }\phi  + {{\bar \delta }^\mu }_\nu \left( { - \frac{1}{2}{{\hat \partial }_\sigma }\phi {{\hat \partial }^\sigma }\phi  + \frac{\kappa }{2}{{\bar \nabla }^2}\phi {{\bar \nabla }^2}\phi  + \kappa {{\bar \partial }_\sigma }\phi {{\bar \partial }^\sigma }{{\bar \nabla }^2}\phi } \right)  \\
& + O(\epslif).
\end{aligned}
\end{equation} 
The expression for the Feynman diagram vertex that corresponds
to the stress-energy tensor \eqref{eq:StressTensorForPolesLif} is given in \eqref{eq:FeynmanVertexLifGen}. All other Feynman rules needed for the calculation of the relevant Feynman correlation functions can be found in the appendix \ref{FeynRulesRelDiagLif}. 

Note that although the correlation functions of the stress-energy tensor in this case seem like they might be IR divergent, similarly to the $d=2$ relativistic case as discussed in subsection \ref{SeriesExp} this is in fact not the case. The terms in the action \eqref{eq:GeneralActionLifs} of order $O(\epslif^0)$ contain only derivatives of the field $\phi$, and only the terms of order $O(\epslif)$ and higher contain $\phi$ with no derivatives. The stress-energy tensor \eqref{eq:StressTensorForPolesLif} and other operators defined as variations of the action with respect to the vielbeins (such as $ \frac{{{\delta ^2}S}}{{\delta \ve{b}{\alpha} \delta \ve{c}{\gamma}}} $) therefore have the same structure. These operators are therefore not Log correlated at $2+1$ dimensions, and their correlation functions are not expected to diverge in the IR. In terms of the corresponding Feynman diagrams regulated using split dimensional regularization, for any propagator in the 1-loop diagram with spacetime momentum $Q^\mu$, the $O(\epslif^0)$ part of each of its adjacent vertexes is of order $O(\hat Q)$ or $O(\bar Q)$. The terms in the integrand of the diagram that contribute to the IR divergence when $Q\to 0$ are of order $O(\epslif^2)$, and therefore vanish as $\epslif\to 0$ after integration is performed. This justifies the assumption made in subsection \ref{SplitDimensionalRegAndTheLifshitzCase}, that the correlation functions of the stress-energy tensor are free of IR divergences.

Since these calculations involve a very large number of terms, they were performed using a computer script that implements the previously described steps.
For the purpose of presenting the results of the calculations, we introduce the following notations for the Fourier transformed versions of the Ward identities of the flat space correlation functions:\footnote{In this section and the relevant appendixes, we denote the spacetime momenta using capital letters ($P$, $K$, $Q$).}
\begin{align}
\label{LifhitzScalarAnomaly:FeynmanTPDWardr2PointFT}
(2\pi)^{d_t+d_s}\delta\left(P+Q\right)\left[\mathcal{I}^{(2)}_D\right]^\rho_{ab}(Q,P)&\equiv\mathcal{FT}\left[(\mathcal{I}^{(2)}_D)^\rho_{ab}(x,y)\right],\\
\label{LifhitzScalarAnomaly:VarWeylWardr2PointFT}
(2\pi)^{d_t+d_s}\delta\left(P+Q\right)\left[\mathcal{W}^{(2)}_W)\right]^\rho_{b}(Q,P)&\equiv\mathcal{FT}\left[(\mathcal{W}^{(2)}_W)^\rho_b(x,y)\right],\\
\label{LifhitzScalarAnomaly:FeynmanTPDWardr3PointFT}
(2\pi)^{d_t+d_s}\delta\left(P+Q+K\right)\left[\mathcal{I}^{(3)}_{D}\right]^{\rho\alpha}_{abc}(Q,P,K) &\equiv \mathcal{FT}\left[(\mathcal{I}^{(3)}_{D})^{\rho\alpha}_{abc}(x,y,z)\right],\\
\label{LifhitzScalarAnomaly:VarWeylWardr3PointFT}
(2\pi)^{d_t+d_s}\delta\left(P+Q+K\right)\left[\mathcal{W}^{(3)}_W\right]^{\rho\alpha}_{bc}(Q,P,K)&\equiv\mathcal{FT}\left[(\mathcal{W}^{(3)}_W)^{\rho\alpha}_{bc}(x,y,z)\right],
\end{align}
where the expressions $(\mathcal{I}^{(2)}_D)^\rho_{ab}(x,y)$, $(\mathcal{W}^{(2)}_W)^\rho_b(x,y)$, $(\mathcal{I}^{(3)}_{D})^{\rho\alpha}_{abc}(x,y,z)$ and $(\mathcal{W}^{(3)}_W)^{\rho\alpha}_{bc}(x,y,z)$ are defined in equations \eqref{eq:ConWard2Point}, \eqref{CorrLifshitzWard:WardIdentForVarCorrWeyl2Point}, \eqref{eq:ConWard3Point} and \eqref{CorrLifshitzWard:WardIdentForVarCorrWeyl3Point} respectively. The conventions for the Fourier transforms are given in \eqref{eq:FourierLifTwoPoint} and \eqref{eq:FourierLifThreePoint}. 
The results we present here for the anomalous contributions to the Ward identities
$(\mathcal{W}^{(2)}_W)^\rho_b(x,y)$ and $(\mathcal{W}^{(3)}_W)^{\rho\alpha}_{bc}(x,y,z)$ are divided into two separate sectors according to the total number of derivatives: a two derivatives sector ($n_D=2$), and a four derivatives sector ($n_D=4$). This is in accordance with the definitions and the discussion in \cite{Arav:2014goa,Arav:2016xjc} (higher derivative sectors do not appear in the two and three point functions). 

At this point, we would like to reiterate that since
we study the correlation functions only up to the three point function level, we are able to study only the anomalies and trivial terms (as found in \cite{Arav:2016xjc}) that appear in this level, that is, only the ones which are at most second order in the background fields.

\subsubsection{Results for the Two Point Function}

The pole residues of the two point correlation function of the stress-energy tensor were extracted using the method described in subsection \ref{SplitDimensionalRegAndTheLifshitzCase}. Taking the weighted Lifshitz trace over these pole residues and following the previously mentioned steps of the calculation 
yields the following result for the anomalous contribution to the two point Ward identity:\footnote{The results for the anomalous Ward identities in this section are in $d_t=1$ time dimension. In this case, we use $\hat P$ and $\hat K$ to denote the time components of the momenta $P$ and $K$ respectively (see appendix \ref{LifshitzNotation}).}
\begin{equation}
\label{eq:TraceOverTwoPoint}
\begin{aligned}
& -\delta_{\alpha\alpha'}\delta^b_\beta\left[\mathcal{W}^{(2)}_W)\right]^{\alpha'}_{b}(Q,P)= -D^{\mu\nu}\left\langle {{T_{\mu \nu }}\left( { - P} \right){T_{\alpha \beta }}\left( P \right)} \right\rangle^{(\text{ren})}  = \\
& \qquad  - \frac{i\sqrt{\kappa} \alpha n_{\alpha} n_{\beta} \bar{P}_{\gamma} \bar{P}^{\gamma} \bar{P}_{\delta} \bar{P}^{\delta}}{24 \pi} + \frac{i (3 - 2 \alpha) n_{\alpha} \bar{P}_{\beta} \hat{P}}{96 \sqrt{\kappa} \pi} + \frac{i\sqrt{\kappa} \alpha n_{\beta} \bar{P}_{\alpha} \bar{P}_{\gamma} \bar{P}^{\gamma} \hat{P}}{24 \pi}  \\
& \qquad +\frac{i (-3 + 2 \alpha) \bar{\delta}_{\alpha \beta} \hat{P}^2}{96 \sqrt{\kappa} \pi}.
\end{aligned}
\end{equation}
Note that although the expression on the LHS of \eqref{eq:TraceOverTwoPoint} seems to depend on two different momenta $P$ and $Q$, due to the delta function $\delta(P+Q)$ in the definition \eqref{LifhitzScalarAnomaly:VarWeylWardr2PointFT} these momenta are not independent, and the RHS of \eqref{eq:TraceOverTwoPoint} involves only one independent momentum. 
The same comment holds for other similar identities in this section.
When comparing this result to the first order variation (in the flat space limit) of the cohomologically trivial terms\footnote{Only trivial terms are first order in the background fields.} (or coboundaries in cohomological terminology) in the two derivatives sector \eqref{eq:F1tDef} and in the four derivatives sector \eqref{eq:4DerTriv} with respect to the vielbeins, one finds the following results:

In the two derivatives sector we have:
\begin{equation}
\label{eq:TraceOverTwoPoint2Der}
\left.i\,{\delta_{\alpha\alpha'}\delta^b_\beta\left[\mathcal{W}^{(2)}_W)\right]^{\alpha'}_{b}(Q,P)}\right|_{n_D =2} = \frac{{2\alpha  - 3}}{{192{\sqrt{\kappa} }\pi }}[{\delta \mathcal{F}_1}]_{\alpha\beta}(P),
\end{equation}
where $[\delta X]_{\alpha \beta}(P)$ represents the Fourier transform of the first order variation of the expression $X$ with respect to the vielbeins in flat space, as defined in \eqref{eq:DefFirstVarFT}, and $\alpha$ is the regularization parameter defined in subsection \ref{SplitDimensionalRegAndTheLifshitzCase}.

In the four derivatives sector we get:
\begin{equation}
\label{eq:TraceOverTwoPoint4Der}
\begin{aligned}
&\left.i\,{\delta_{\alpha\alpha'}\delta^b_\beta\left[\mathcal{W}^{(2)}_W)\right]^{\alpha'}_{b}(Q,P)}\right|_{n_D =4} = \\ 
&\qquad -iC_0[{\delta \mathcal{F}_1}]_{\alpha\beta}(P)-\frac{\sqrt{\kappa}\alpha}{96\pi}[{\delta \mathcal{F}_1}]_{\alpha\beta}(P)+2iC_0[{\delta \mathcal{F}_3}]_{\alpha\beta}(P)\\
&\qquad +\frac{\sqrt{\kappa}\alpha}{48\pi}[{\delta \mathcal{F}_3}]_{\alpha\beta}(P)-iC_0[{\delta \mathcal{F}_6}]_{\alpha\beta}(P),
\end{aligned}
\end{equation} 
where $C_0$ is a free parameter, whose value cannot be extracted from the two point correlation function. 
This is due to the fact that, in the basis of first order trivial terms we are using here, the first order variations of the trivial term densities in the four derivatives sector are linearly dependent in the flat space limit (that is, there is a linear dependence between the first order variations of $\mathcal{F}_1$, $\mathcal{F}_3$ and $\mathcal{F}_6$ with respect to the vielbeins in flat space). When looking at higher point correlation functions, this dependence is removed and the coefficient $C_0$ can be extracted, as we indeed show in the next subsection studying the three point function. 

We have also confirmed the pole residue of the two point function satisfies the conservation Ward identity:
\begin{equation}
\left[\mathcal{I}^{(2)}_D\right]^\rho_{ab}(Q,P)=0 .
\end{equation}
This is expected from the argument made in subsection \ref{SplitDimensionalRegAndTheLifshitzCase} that the conservation Ward identity \eqref{eq:ConWard2Point} holds separately on the pole part and on the regular finite part of the correlation functions.
 
\subsubsection{Results for the Three Point Function and the Anomaly}

Using the previously mentioned calculation steps, we obtained the following result for the anomalous contribution to the three point Ward identity in the two derivative sector\footnote{We remind the reader that for the expressions in this subsection, the indexes $\alpha',e$ (and therefore $\alpha,\beta$) correspond to insertions at the spacetime point $y$, or the momentum $P$ after Fourier transform. Similarly, the indexes $\gamma',f$ (and therefore $\gamma,\delta$) correspond to insertions at  $z$, or the momentum $K$. See the definitions in \eqref{CorrLifshitzWard:WardIdentForVarCorrWeyl3Point}, \eqref{LifhitzScalarAnomaly:VarWeylWardr3PointFT} and \eqref{eq:FourierLifThreePoint}. }
\begin{align*}
&\left. -\delta_{\alpha\alpha'}\delta_{\gamma\gamma'}\delta^e_\beta\delta^f_\delta\left[\mathcal{W}^{(3)}_W\right]^{\alpha'\gamma'}_{ef}(Q,P,K)\right|_{n_D=2} =\\
& \qquad\frac{(-3 + 2 \alpha) \bar{K}_{\beta} \bar{K}_{\delta} n_{\alpha} n_{\gamma}}{96 \sqrt{\kappa} \pi} + \frac{(3 - 2 \alpha) \bar{K}_{\delta} \hat{K} n_{\alpha} n_{\beta} n_{\gamma}}{48 \sqrt{\kappa} \pi} + \frac{(-3 + 2 \alpha) \bar{K}_{\beta} \hat{K} n_{\gamma} \bar{\delta}_{\alpha \delta}}{96 \sqrt{\kappa} \pi} \\
& \qquad  + \frac{(3 - 2 \alpha) \hat{K}^2 \bar{\delta}_{\alpha \delta} \bar{\delta}_{\beta \gamma}}{96\sqrt{\kappa} \pi} + \frac{(3 - 2 \alpha) \bar{K}_{\beta} \hat{K} n_{\alpha} \bar{\delta}_{\gamma \delta}}{48 \sqrt{\kappa} \pi} + \frac{(3 - 2 \alpha) \hat{K} n_{\alpha} n_{\gamma} n_{\delta} \bar{P}_{\beta}}{96 \sqrt{\kappa} \pi}\\
& \qquad   + \frac{(-3 + 2 \alpha) \hat{K} n_{\gamma} \bar{\delta}_{\alpha \delta} \bar{P}_{\beta}}{96\sqrt{\kappa} \pi} + \frac{(3 - 4 \alpha) \hat{K} n_{\alpha} \bar{\delta}_{\gamma \delta} \bar{P}_{\beta}}{96 \sqrt{\kappa} \pi} + \frac{\hat{K} n_{\alpha} \bar{\delta}_{\beta \delta} \bar{P}_{\gamma}}{32 \sqrt{\kappa} \pi}\\
& \qquad  -  \frac{\bar{K}_{\beta} n_{\alpha} n_{\gamma} \bar{P}_{\delta}}{32 \sqrt{\kappa} \pi} + \frac{(3 - 2 \alpha) \hat{K} n_{\gamma} \bar{\delta}_{\alpha \beta} \bar{P}_{\delta}}{96 \sqrt{\kappa} \pi} + \frac{\alpha \hat{K} n_{\alpha} \bar{\delta}_{\beta \gamma} \bar{P}_{\delta}}{48\sqrt{\kappa} \pi} \\
& \qquad  -  \frac{\bar{K}^{\mu} n_{\alpha} n_{\gamma} \bar{\delta}_{\beta \delta} \bar{P}_{\mu}}{32 \sqrt{\kappa} \pi} + \frac{(3 - 2 \alpha) \bar{K}_{\delta} n_{\alpha} n_{\beta} n_{\gamma} \hat{P}}{96 \sqrt{\kappa} \pi} + \frac{(3 - 4 \alpha) \bar{K}_{\delta} n_{\gamma} \bar{\delta}_{\alpha \beta} \hat{P}}{96 \sqrt{\kappa} \pi}  \\
& \qquad  + \frac{\alpha \bar{K}_{\beta} n_{\gamma} \bar{\delta}_{\alpha \delta} \hat{P}}{48 \sqrt{\kappa} \pi} + \frac{(-3 + 2 \alpha) \bar{K}_{\delta} n_{\alpha} \bar{\delta}_{\beta \gamma} \hat{P}}{96 \sqrt{\kappa} \pi} + \frac{(3 - 4 \alpha) \hat{K} \bar{\delta}_{\alpha \delta} \bar{\delta}_{\beta \gamma} \hat{P}}{96 \sqrt{\kappa} \pi} + \frac{\bar{K}_{\alpha} n_{\gamma} \bar{\delta}_{\beta \delta} \hat{P}}{32 \sqrt{\kappa}\pi} \numthis \label{eq:TwoDerSectorResultsFeyn}\\
& \qquad  + \frac{(3 - 2 \alpha) \bar{K}_{\beta} n_{\alpha} \bar{\delta}_{\gamma \delta} \hat{P}}{96 \sqrt{\kappa} \pi} + \frac{(-3 + 2 \alpha) \hat{K} n_{\alpha} n_{\beta} \bar{\delta}_{\gamma \delta} \hat{P}}{96 \sqrt{\kappa} \pi} + \frac{(-3 + 4 \alpha) \hat{K} \bar{\delta}_{\alpha \beta} \bar{\delta}_{\gamma \delta} \hat{P}}{96 \sqrt{\kappa} \pi}\\
& \qquad  + \frac{(3 - 2 \alpha) n_{\alpha} n_{\gamma} n_{\delta} \bar{P}_{\beta} \hat{P}}{48 \sqrt{\kappa}\pi} + \frac{(-3 + 2 \alpha) n_{\gamma} \bar{\delta}_{\alpha \delta} \bar{P}_{\beta} \hat{P}}{96 \sqrt{\kappa} \pi} + \frac{(3 - 2 \alpha) n_{\gamma} \bar{\delta}_{\alpha \beta} \bar{P}_{\delta} \hat{P}}{48 \sqrt{\kappa} \pi} \\
& \qquad + \frac{(-3 + 2 \alpha) n_{\alpha} \bar{\delta}_{\beta \gamma} \bar{P}_{\delta} \hat{P}}{96 \sqrt{\kappa} \pi} + \frac{(-3 + 2 \alpha) n_{\gamma} n_{\delta} \bar{\delta}_{\alpha \beta} \hat{P}^2}{48 \sqrt{\kappa} \pi} + \frac{(3 - 2 \alpha) \bar{\delta}_{\alpha \delta} \bar{\delta}_{\beta \gamma} \hat{P}^2}{96 \sqrt{\kappa} \pi}\\
& \qquad + \frac{(-3 + 2 \alpha) \bar{K}_{\delta} \hat{K} n_{\alpha} \bar{\delta}_{\beta \gamma}}{96 \sqrt{\kappa} \pi}+ \frac{(-3 + 2 \alpha) \hat{K}^2 n_{\alpha} n_{\beta} \bar{\delta}_{\gamma \delta}}{48 \sqrt{\kappa} \pi} + \frac{(-3 + 4 \alpha) \bar{K}_{\delta} n_{\alpha} n_{\gamma} \bar{P}_{\beta}}{96 \sqrt{\kappa} \pi}  \\
& \qquad + \frac{(-3 + 2 \alpha) n_{\alpha} n_{\gamma} \bar{P}_{\beta} \bar{P}_{\delta}}{96 \sqrt{\kappa} \pi} + \frac{(-3 + 2 \alpha) \hat{K} n_{\gamma} n_{\delta} \bar{\delta}_{\alpha \beta} \hat{P}}{96 \sqrt{\kappa} \pi}-  \frac{\hat{K} \bar{\delta}_{\alpha \gamma} \bar{\delta}_{\beta \delta} \hat{P}}{32\sqrt{\kappa}\pi} .
\end{align*}
Writing this expression as a linear combination of the second order variations with respect to the vielbeins of the anomaly and trivial term densities in the two derivative sector (given in \eqref{eq:SecVarF1t}--\eqref{eq:SecVarA200}), we get:
\begin{equation}
\label{eq:Sol2DerSecCoeff}
\begin{split}
-\left. \left[\mathcal{W}^{(3)}_W\right]^{\alpha\gamma}_{ef}(Q,P,K)\right|_{n_D=2} &=\frac{1}{{32{\sqrt{\kappa}}\pi }}\left[\delta^2 \mathcal{A}^{\left( {2,0,0} \right)}\right]^{\alpha\gamma}_{ef}(P,K)\\
&+\frac{{2\alpha - 3}}{{96{\sqrt{\kappa}}\pi }}\left[\delta^2 \mathcal{F}_1\right]^{\alpha\gamma}_{ef}(P,K),
\end{split}
\end{equation}
where $[\delta^2 X]^{\alpha\gamma}_{ef}(P,K)$ represents the Fourier transform of the second order variation of the expression $X$ with respect to the vielbeins in flat space, as defined in \eqref{eq:DefSecVarFT}.
Note that, as expected, the coefficient of the anomaly $\mathcal{A}^{(2,0,0)}$ is independent of the regularization parameter $\alpha$ and agrees with the result found in \cite{Baggio:2011ha}, while the coefficient of the trivial term $\mathcal{F}_1$ depends on $\alpha$, and vanishes when $\alpha=\frac{3}{2}$. 

The full result for the anomalous contribution to the three point Ward identity \eqref{CorrLifshitzWard:WardIdentForVarCorrWeyl3Point} in the four derivative sector is given in \eqref{eq:4DerLeftWardRes} in the appendix. 
Comparing this result to the second order variations of the anomaly and trivial term densities in this sector we get the following linear combination: 
\begin{equation}
\label{eq:FourDerAnoFinalRes}
\begin{aligned}
&-\left. \left[\mathcal{W}^{(3)}_W\right]^{\alpha\gamma}_{ef}(Q,P,K)\right|_{n_D=4} = - \frac{\sqrt{\kappa}\alpha }{{160 \pi }}{\left[\delta^2 \mathcal{F}_1\right]^{\alpha\gamma}_{ef}(P,K)}+ \frac{\sqrt{\kappa}\alpha }{{240 \pi }}{\left[\delta^2 \mathcal{F}_2\right]^{\alpha\gamma}_{ef}(P,K)}  \\
 & \qquad + \frac{\sqrt{\kappa}\alpha }{{80 \pi }}{\left[\delta^2 \mathcal{F}_3\right]^{\alpha\gamma}_{ef}(P,K)} 
 + \frac{\alpha \sqrt{\kappa}}{{240 \pi }}{\left[\delta^2 \mathcal{F}_6\right]^{\alpha\gamma}_{ef}(P,K)}  \\
 & \qquad + \left( {\frac{\alpha\sqrt{\kappa} }{{30 \pi }} - C} \right){\left[\delta^2 \mathcal{F}_8\right]^{\alpha\gamma}_{ef}(P,K)}  + \left( {\frac{\alpha\sqrt{\kappa} }{{30 \pi }} - 2C} \right){\left[\delta^2 \mathcal{F}_9\right]^{\alpha\gamma}_{ef}(P,K)} \\
 &\qquad+ \frac{{\alpha\sqrt{\kappa} }}{{24 \pi }}{\left[\delta^2 \mathcal{F}_{11}\right]^{\alpha\gamma}_{ef}(P,K)}+ C{\left[\delta^2 \mathcal{F}_{12}\right]^{\alpha\gamma}_{ef}(P,K)},
\end{aligned}
\end{equation}
where the trivial terms (the $\mathcal{F}_i$-s) are defined in \eqref{eq:4DerTriv} and the anomalies are defined in \eqref{eq:4DerAnomaliesFullExpr} or \eqref{eq:4DerAnomaliesExpr}. 
$C$ is again a free parameter that cannot be extracted from the three point correlation function, similar to 
the appearance of the free parameter $C_0$ in the two point level \eqref{eq:TraceOverTwoPoint4Der}. Note that the three point function fully determines the coefficients of $\mathcal{F}_1$, $\mathcal{F}_3$ and $\mathcal{F}_6$, and therefore the value of the free parameter $C_0$ from equation \eqref{eq:TraceOverTwoPoint4Der}. Alternatively, the result in \eqref{eq:FourDerAnoFinalRes} can be written in terms of the second order variations of the scalars ($\phi$'s) defined in \eqref{eq:FPDInv4Der}:
\begin{align*}
&-\left. \left[\mathcal{W}^{(3)}_W\right]^{\alpha\gamma}_{ef}(Q,P,K)\right|_{n_D=4} = \\
& \qquad\frac{{\alpha\sqrt{\kappa} {{\left[\delta^2 \phi_{3}\right]^{\alpha\gamma}_{ef}(P,K)}}}}{{24 \pi }} + \frac{{\alpha \sqrt{\kappa} {{\left[\delta^2 \phi_{4}\right]^{\alpha\gamma}_{ef}(P,K)}}}}{{24 \pi }} + \frac{{\alpha\sqrt{\kappa} {{\left[\delta^2 \phi_{9}\right]^{\alpha\gamma}_{ef}(P,K)}}}}{{60 \pi }} \\
& \qquad + \frac{{\alpha\sqrt{\kappa} {{\left[\delta^2 \phi_{10}\right]^{\alpha\gamma}_{ef}(P,K)}}}}{{15 \pi }} + \frac{{\alpha\sqrt{\kappa} {{\left[\delta^2 \phi_{11}\right]^{\alpha\gamma}_{ef}(P,K)}}}}{{8 \pi }} +
\frac{{\alpha\sqrt{\kappa} {{\left[\delta^2 \phi_{12}\right]^{\alpha\gamma}_{ef}(P,K)}}}}{{24 \pi }} \numthis \\
&\qquad - \frac{{{\alpha\sqrt{\kappa}{\left[\delta^2 \phi_{28}\right]^{\alpha\gamma}_{ef}(P,K)}} }}{{12 \pi }} - \frac{{{\alpha\sqrt{\kappa}{\left[\delta^2 \phi_{29}\right]^{\alpha\gamma}_{ef}(P,K)}} }}{{12 \pi }} 
+ \frac{{7\alpha\sqrt{\kappa} {{\left[\delta^2 \phi_{32}\right]^{\alpha\gamma}_{ef}(P,K)}}}}{{60 \pi }} \\
&\qquad  - \frac{{\alpha\sqrt{\kappa} {{\left[\delta^2 \phi_{33}\right]^{\alpha\gamma}_{ef}(P,K)}}}}{{30 \pi }} - \frac{{7\alpha\sqrt{\kappa} {{\left[\delta^2 \phi_{38}\right]^{\alpha\gamma}_{ef}(P,K)}}}}{{30 \pi }} - \frac{{\alpha\sqrt{\kappa} {{\left[\delta^2 \phi_{39}\right]^{\alpha\gamma}_{ef}(P,K)}}}}{{15 \pi }}.
\end{align*}

It is apparent from these results that the coefficients of the 3 possible anomalies in the four derivative sector that contribute to the three point correlation function vanish. This is consistent with previous results 
(see \cite{Griffin:2012qx,Baggio:2011ha}), that considered only the case where the Frobenius condition is satisfied (and therefore only the anomaly ${A_4}^{\left( {0,4,0} \right)} = {\left( {\hat R + {{\bar \nabla }_\alpha }{a^\alpha }} \right)^2}$).
All non-vanishing terms are proportional to $\alpha$, and therefore represent only trivial terms that can be removed by adding local counterterms to the effective action. 
It is also important to note that these trivial terms contain contributions from $\phi_{28}, \phi_{29}, \phi_{32}, \phi_{33}, \phi_{38}$ and $\phi_{39}$. 
These are terms that vanish in the Frobenius case (for which $K_A=0$, see \cite{Arav:2016xjc} for details), and therefore their coefficients cannot be extracted from a curved spacetime coupling that assumes the existence of a foliation structure. We conclude that for $\alpha \neq 0$, violating the Frobenius condition is essential for describing the obtained trivial terms in curved spacetime, and for constructing the appropriate counterterms to cancel them.

Finally, we have again verified that the pole residues of the various two and three point functions satisfy the conservation Ward identity:
\begin{equation}
\left[\mathcal{I}^{(3)}_{D}\right]^{\rho\alpha}_{afe}(Q,P,K)=0,
\end{equation}
which is consistent with the argument made in subsection \ref{SplitDimensionalRegAndTheLifshitzCase} that the conservation Ward identity \eqref{eq:ConWard3Point} holds separately on the pole part and on the regular finite part of the correlation functions (so that TPD invariance is not anomalous).

\section{Summary and Outlook}
\label{SummaryandOutlook}

In this work we developed a general scheme for field theory calculations of Lifshitz scale anomalies.
We analyzed the general structure of correlation functions of the stress-energy tensor in Lifshitz field theories and
constructed  the corresponding anomalous Ward identities.
We presented a subtle ambiguity in the definition of the anomaly coefficients and clarified it.
Our framework for calculating the anomaly coefficients was based on 
a split dimensional regularization 
where space (momentum) and time (frequency) integrals are 
regulated separately, and pole residue calculations  that allowed to extract the anomaly
coefficients without  a full calculation of the correlation functions.

In order to implement the calculational scheme we had to analyze the coupling of a non-relativistic, Lifshitz invariant field theory in $d_t$ time dimensions and $d_s$ space dimensions to a curved spacetime manifold.
This generalized  the curved spacetime structure introduced in \cite{Arav:2014goa,Arav:2016xjc} to the case of multiple time directions.

We considered as a particular example the $z=2$ free scalar field theory in $2+1$ spacetime dimensions. We showed
 that the only non-zero anomaly coefficient (of those appearing in the three point functions) is in the two derivatives sector, which 
agrees with the heat kernel calculation in \cite{Baggio:2011ha}. In order to account for some of the trivial terms arising from this calculation, we found it necessary to give up the Frobenius
condition (and the corresponding foliation structure) of the curved spacetime description of the theory. This is because these terms are not in the span of possible anomalous contributions one obtains when assuming the Frobenius condition.
 
There are many directions for further studies that  follow from our analysis.
It would be interesting to use the general scheme developed here to calculate the anomaly coefficients of other Lifshitz field theories.
One can also generalize the discussion and consider non-relativistic field theories that exhibit non-relativistic
boost invariance. In such cases one has in addition to the B-type scale anomalies also A-type ones \cite{Arav:2016xjc}
(see also \cite{Jensen:2014hqa}).
In the relativistic CFT case, scale anomaly coefficients multiply universal terms in entanglement entropies.
It would be of interest to analyze the entanglement entropy structure in Lifshitz field theories and 
the role of the scale anomaly coefficients.
In the CFT case, A-type anomaly charges exhibit RG properties, i.e.\ a decrease from the UV to the IR. 
The non-relativistic versions of these are still lacking.
Finally, it would be interesting to ask whether there are experimental observables of the anomaly charges
in non-relativistic systems such as low energy condensed matter ones.
One potential path to consider is the hydrodymanics of such systems \cite{Hoyos:2013eza,Hoyos:2013qna} and 
the role of scale anomalies in such descriptions.

\section*{Acknowledgments}
We would like to thank Itamar Hason, Carlos Hoyos, Zohar Komargodski, Adam Schwimmer and Stefan Theisen
for valuable discussions and comments.
Many of the calculations in this paper were performed using xAct \cite{Garcia} and xTras \cite{Nutma:2013zea}, tensor computer algebra packages for \textit{Mathematica}.
This work is supported in part by the I-CORE program
of Planning and Budgeting Committee (grant number 1937/12), the US-Israel Binational Science Foundation, GIF and the ISF Center of Excellence. A.R.M gratefully acknowledges the support of the Adams Fellowship Program of the Israel Academy of Sciences and Humanities. I.A is thankful for the support of the Alexander Zaks fellowship program.

\appendix

\section{Notations and Conventions}
\label{app:conventions}
In this appendix we describe the notations and conventions used in this paper.

\subsection{Notations and Conventions in Curved Spacetime}
\label{app:ConventionsCurvedSpacetime}

Throughout this paper we use Greek letters ($\mu ,\nu ,\rho,\ldots$) to denote spacetime indexes, both in the relativistic and non-relativistic cases.
For the relativistic case, we use a spacetime metric $g_{\mu\nu}$ with signature $\operatorname{diag}(1,-1,-1,\ldots)$ and denote the flat space metric by $ \eta_{\mu\nu} $. For the non-relativistic case, we use a signature of $ \operatorname{diag}(-1,-1,\ldots,1,1,\ldots) $ (with negative signs for the time dimensions and positive for the space dimensions), and denote the flat spacetime metric by $ \delta_{\mu\nu} $. 

In both cases we use the standard torsionless Levi-Civita connection associated with the spacetime metric $g_{\mu\nu}$. That is, 
the covariant derivative of a vector $A^{\mu}$ is given by:
\begin{equation}
{\nabla _\nu }{A^\mu } \equiv {\partial _\nu }{A^\mu } + \Gamma _{\nu \rho }^\mu {A^\rho },
\end{equation}
where the Christoffel symbols are given by:
\begin{equation}
\Gamma _{\nu \rho }^\mu  = \frac{1}{2}{g^{\mu \kappa }}\left[ { - {\partial _\kappa }{g_{\nu \rho }} + {\partial _\nu }{g_{\kappa \rho }} + {\partial _\rho }{g_{\kappa \nu }}} \right].
\end{equation}
We use the following convention for the Riemann tensor:
\begin{equation}
{R^\lambda }_{\mu \kappa \nu } = {\partial _\kappa }\Gamma _{\mu \nu }^\lambda -{\partial _\nu }\Gamma _{\mu \kappa }^\lambda +\Gamma _{\kappa \eta }^\lambda \Gamma _{\mu \nu }^\eta - \Gamma _{\nu \eta }^\lambda \Gamma _{\mu \kappa }^\eta   ,
\end{equation}
while the Ricci tensor and scalar are given by:
\begin{equation}
{R_{\mu \nu }} = {R^\gamma }_{\mu \gamma \nu }, \qquad R = {g^{\mu \nu }}{R_{\mu \nu }}.
\end{equation}
When using the vielbein formalism, the vielbeins $ \ve{a}{\mu} $ are defined in the relativistic case such that:
\begin{equation}
g_{\mu\nu} = \eta_{ab} \ve{a}{\mu} \ve{b}{\nu},
\end{equation}
and in the non-relativistic case:
\begin{equation}
g_{\mu\nu} = \delta_{ab} \ve{a}{\mu} \ve{b}{\nu}.
\end{equation}
We use the following notations for the determinants of the metric and the vielbeins:
\begin{equation}
g \equiv \det \left( g_{\mu\nu} \right), \qquad e \equiv \det \left( \ve{a}{\mu} \right).
\end{equation}
We use the following formulas for variations of the metric:
\begin{equation}
\delta {g_{\mu \nu }} =  - {g_{\mu \alpha }}{g_{\nu \beta }}\delta {g^{\alpha \beta }}, \qquad \delta {g^{\mu \nu }} =  - {g^{\mu \alpha }}{g^{\nu \beta }}\delta {g_{\alpha \beta }},
\end{equation}
\begin{equation}
\frac{{\delta {g^{\alpha \beta }}\left( x \right)}}{{\delta {g_{\mu \nu }}\left( {{x_1}} \right)}} =  - \frac{1}{2}\left( {{g^{\alpha \mu }}{g^{\beta \nu }} + {g^{\alpha \nu }}{g^{\beta \mu }}} \right)\delta \left( {x - {x_1}} \right),
\end{equation}
\begin{equation}
\frac{{\delta {g^{\sigma \rho }}}}{{\delta {e^c}_\alpha }} =  - {g^{\sigma \lambda }}{g^{\tau \rho }}\frac{{\delta {g_{\lambda \tau }}}}{{\delta {e^c}_\alpha }},
\end{equation}
\begin{equation}
\frac{{\delta {g_{\rho \nu }}}}{{\delta {e^c}_\alpha }} = {\eta _{ac}}\left( {{\delta ^\alpha }_\rho {e^a}_\nu  + {\delta ^\alpha }_\nu {e^a}_\rho } \right),
\end{equation}
\begin{equation}
\delta \sqrt { |g|}  =  - \frac{1}{2}\sqrt {  |g|} {g_{a\beta }}\delta {g^{\alpha \beta }}, \qquad\delta \sqrt { | g|}  = \frac{1}{2}\sqrt { | g|} {g^{\alpha \beta }}\delta {g_{\alpha \beta }}.
\end{equation}
Finally, for reference we give here the expressions for the Weyl tensor squared and the Euler density of the background manifold in the ($3+1$)-dimensional case:
\begin{align}
\label{eq:FWeylDef}
W^2 &= {R^{\alpha \beta \gamma \delta }}{R_{\alpha \beta \gamma \delta }} - 2{R^{\alpha \beta }}{R_{\alpha \beta }} + \frac{1}{3}{R^2}, \\
\label{eq:GEulerDef}
E_4 &= {R^{\alpha \beta \gamma \delta }}{R_{\alpha \beta \gamma \delta }} - 4{R^{\alpha \beta }}{R_{\alpha \beta }} + {R^2}.
\end{align}

\subsection{Notations and Conventions for the Relativistic Scalar Case}
\label{sec:RelScalCon}
As mentioned in appendix \ref{app:ConventionsCurvedSpacetime}, we use Greek letters ($\mu,\nu,\rho,\ldots$) to denote spacetime indexes. In the relativistic case, we use a flat spacetime metric of the form: $\eta_{\mu\nu} = \operatorname{diag}(1,-1,-1,\ldots)$.
We use the following conventions for the Fourier transforms of two and three point correlation functions in the relativistic case: 
\begin{align}
\label{eq:Fourier2PoingConv}
\mathcal{FT} \left[ I_{(2)}(x_1,x_2) \right] &\equiv  \int {{d^d}{x_1}{d^d}{x_2}}\, I_{(2)}(x_1,x_2) \,{e^{ - i\left( { - k \cdot {x_1} - q \cdot {x_2}} \right)}}  ,\\
\label{eq:Fourier3PoingConv}
\mathcal{FT} \left[ I_{(3)}(x_1,x_2,x_3) \right] &\equiv \int {{d^d}{x_1}{d^d}{x_2}{d^d}{x_3}}\, I_{(3)}(x_1,x_2,x_3) \,{e^{ - i\left( { - k \cdot {x_1} - q \cdot {x_2} - p \cdot {x_3}} \right)}} ,
\end{align}
where the lower-case letters $p,k,q$ denote spacetime momenta.
We also use the following notations for the Fourier transformed two and three point correlation functions of the stress-energy tensor and the variations of the action in flat space:
\begin{align}
\label{eq:Fourier2PoingTTConv}
&{\left( {2\pi } \right)^d}{\delta}\left( { - k - q} \right)\left\langle {{T^{\mu \nu }}\left( { - q} \right){T^{\rho \sigma }}\left( q \right)} \right\rangle \equiv \mathcal{FT}\left[ \left\langle {{T^{\mu \nu }}\left( {{x_1}} \right){T^{\rho \sigma }}\left( {{x_2}} \right)} \right\rangle \right],\\
\begin{split}
&{\left( {2\pi } \right)^d}{\delta}\left( { - k - p - q} \right)\left\langle {{T^{\mu \nu }}\left( { - p - q} \right){T^{\rho \sigma }}\left( q \right){T^{\alpha \beta }}\left( p \right)} \right\rangle \\
&\qquad\qquad\qquad\qquad\qquad \equiv \mathcal{FT}\left[ \left\langle {{T^{\mu \nu }}\left( {{x_1}} \right){T^{\rho \sigma }}\left( {{x_2}} \right){T^{\alpha \beta }}\left( {{x_3}} \right)} \right\rangle \right] ,
\end{split}\\
\begin{split}\label{eq:TwoPointLikeRelativistic}
&(2\pi)^d \delta(-k-p-q)
\left\langle {\frac{{{\delta ^2}S}}{{\delta {g_{\mu \nu }}\delta {g_{\rho \sigma }}}}(k,q)\frac{{\delta S}}{{\delta {g_{\alpha \beta }}}}}(p) \right\rangle \\
&\qquad\qquad\qquad\qquad\qquad \equiv \mathcal{FT}\left[ {\left\langle {\frac{{{\delta ^2}S}}{{\delta {g_{\mu \nu }}\left( {{x_1}} \right)\delta {g_{\rho \sigma }}\left( {{x_2}} \right)}}\frac{{\delta S}}{{\delta {g_{\alpha \beta }}\left( {{x_3}} \right)}}} \right\rangle } \right].
\end{split}
\end{align}
 
Finally, we use the following notation for the Fourier transformed second variation of the expression $X$ with respect to the background metric, evaluated in flat spacetime: 
\begin{equation}
(2\pi)^d \delta(-k-p-q)
{\left[ {\delta^2 X\left( {q,p} \right)} \right]^{\rho \sigma \alpha \beta }} \equiv \mathcal{FT}\left[ {{{\left. {\frac{{{\delta ^2}X (x_1)}}{{\delta {g_{\rho \sigma }}\left( {{x_2}} \right)\delta {g_{\alpha \beta }}\left( {{x_3}} \right)}}} \right|}_{{g_{\mu \nu }} = {\eta _{\mu \nu }}}}} \right] .
\end{equation}
 
\subsection{Notations and Conventions for the Lifshitz $z=2$ Scalar Case}
\label{LifshitzNotation} 
As explained in subsection \ref{SplitDimensionalRegAndTheLifshitzCase}, in the non-relativistic case we define the theory on a manifold $ \mathcal{M} = \mathcal{M}_t \times \mathcal{M}_s $, where $ \mathcal{M}_t $ is a $d_t$-dimensional time manifold and $ \mathcal{M}_s  $ is a $d_s$-dimensional space manifold, such that it is invariant both under rotations in the time manifold and in the space manifold separately. 

Spacetime coordinates are denoted by $ x^\mu = (x^{\hat\mu},x^{\bar\mu}) $ where $ \hat\mu = 1,\ldots,d_t $ are time indexes, $ \bar\mu = 1,\ldots,d_s $ are space indexes and $\mu=1,\ldots,d_t+d_s$ are spacetime indexes. 
We define a flat metric $ \hat \delta_{\hat\mu\hat\nu} = \operatorname{diag}(-1,\ldots,-1)$ on $ \mathcal{M}_t$, and $ \bar \delta_{\bar\mu\bar\nu} = \operatorname{diag}(1,\ldots,1)$ on $\mathcal{M}_s$. We also define the time projector on $\mathcal{M}$ as $\hat\delta_{\mu\nu} = \operatorname{diag}(\hat\delta_{\hat\mu\hat\nu},0) $ and similarly the space projector as $ \bar\delta_{\mu\nu} = \operatorname{diag}(0,\bar\delta_{\bar\mu\bar\nu})$, so that $ \delta^\mu_\nu = \hat\delta^\mu_\nu + \bar\delta^\mu_\nu $. 
Given a vector $v^\mu$ on $\mathcal{M}$, we denote its time projection by $\hat v^\mu \equiv \hat\delta^\mu_\nu v^\nu$, and its space projection by $\bar v^\mu \equiv \bar\delta^\mu_\nu v^\nu$.
In the case of $d_t=1$ time dimension, we use $\hat v$ (with no index) to denote the time component of the vector $v^\mu$, i.e.\ $\hat v_\mu = -\hat v n_\mu$ where $\hat v \equiv v^\mu n_\mu $ and $n_\mu = (1,0,0,\ldots)$.

We use capital letters $P,K,Q$ to denote spacetime momenta. 
The notations $\bar{P}$ and $\hat{P}$ then refer to the spatial and temporal projections of the momentum $P$, respectively.
We use the following conventions for the Fourier transforms of two and three point correlation functions in the non-relativistic case: 
\begin{align}
\label{eq:FourierLifTwoPoint}
&\mathcal{FT}\left[ I_{(2)}\left(x = \left( {\hat x,\bar x} \right), y = \left( {\hat y,\bar y} \right) \right) \right] \equiv
\int {d^{{d_t} + {d_s}}}x\,{d^{{d_t} + {d_s}}}y\, I_{(2)}\left(x , y \right) e^{i(Q\cdot x+P\cdot y)} , \\
\label{eq:FourierLifThreePoint}
&\mathcal{FT}\left[ I_{(3)}(x,y,z) \right] \equiv
\int {{d^{{d_t} + {d_s}}}x\,{d^{{d_t} + {d_s}}}y\,{d^{{d_t} + {d_s}}}z\,} I_{(3)}(x,y,z) e^{i(Q\cdot x+P\cdot y+K\cdot z)} .
\end{align}

We also use the following notations for the Fourier transformed two and three point correlation functions of the stress-energy tensor and the variations of the action with respect to the vielbeins in flat space:
\begin{align}
&{\left( {2\pi } \right)^{{d_t} + {d_s}}}\delta \left( { - P - Q} \right)\left\langle {{T^{\mu \nu }}\left( P \right){T^{\rho \sigma }}\left( { - P} \right)} \right\rangle 
\equiv \mathcal{FT}\left[ \left\langle {{T^{\mu \nu }}\left( {x} \right){T^{\alpha \beta }}\left( {y} \right)} \right\rangle \right],\\
\begin{split}
&{\left( {2\pi } \right)^{{d_t} + {d_s}}}\delta \left( {  - P - K - Q } \right)\left\langle {{T^{\mu \nu }}\left( { - P - K} \right){T^{\alpha \beta }}\left( P \right){T^{\gamma \delta }}\left( K \right)} \right\rangle \\
&\qquad\qquad\qquad\qquad\qquad\qquad
\equiv \mathcal{FT}\left[ \left\langle {{T^{\mu \nu }}\left( x \right){T^{\alpha \beta }}\left( y \right){T^{\gamma \delta }}\left( z \right)} \right\rangle \right],
\end{split}\\
\begin{split}
\label{eq:LifshitzTwoPointLikeDiagExpr}
&(2\pi)^{d_t+d_s}
\delta(-P-K-Q)\left\langle\frac{{{\delta ^2}S}}{{\delta \ve{b}{\alpha} \delta \ve{c}{\gamma}}}(P,K) \frac{\delta S}{\delta \ve{a}{\mu}}(Q) \right\rangle \\
&\qquad\qquad\qquad\qquad\qquad\qquad
\equiv \mathcal{FT}\left[ {\left\langle {\frac{{{\delta ^2}S}}{{\delta {\ve{b}{\alpha}}\left( {{y}} \right)\delta {\ve{c}{\gamma}}\left( {{z}} \right)}}\frac{{\delta S}}{{\delta {\ve{a}{\mu}}\left( {{x}} \right)}}} \right\rangle } \right].
\end{split}
\end{align}

Finally, our notations for the various Ward identities of the flat space correlation functions \eqref{eq:ConWard2Point}--\eqref{CorrLifshitzWard:WardIdentForVarCorrWeyl3Point} are as follows: We use $\mathcal{I}$ to denote the Ward identities \eqref{eq:ConWard2Point}--\eqref{CorrLifshitzWard:WardIdentForFeynmanCorrWeyl3Point} derived using a change of the variables in the path integral, whereas $\mathcal{W}$ denotes Ward identities \eqref{CorrLifshitzWard:WardIdentForVarCorrWeyl2Point}--\eqref{CorrLifshitzWard:WardIdentForVarCorrWeyl3Point} derived from variations of the curved spacetime Ward identities. The subscript refers to the relevant symmetry: $D$ corresponds to TPD symmetry, whereas $W$ corresponds to anisotropic Weyl symmetry. The superscript $(n)$ refers to the number of points in the correlation function. The notation for the Fourier transforms of these Ward identities is given in \eqref{LifhitzScalarAnomaly:FeynmanTPDWardr2PointFT}--\eqref{LifhitzScalarAnomaly:VarWeylWardr3PointFT}.

\section{Non-Relativistic Curved Spacetime with Multiple Time Directions}
\label{app:LifCurvedSpacetimeMultTimDims}

In this appendix we discuss the coupling of a non-relativistic, Lifshitz invariant field theory in $d_t$ time dimensions and $d_s$ space dimensions to a curved spacetime manifold. We first generalize the curved spacetime structure introduced in \cite{Arav:2014goa,Arav:2016xjc} to the case of multiple time directions. We then discuss the local symmetries of the theory over curved spacetime. Finally we construct the curved spacetime action that corresponds to the free $z=2$ Lifshitz scalar.

\subsection{Curved Spacetime Definitions}
\label{app:LifCurvedSpacetimeMultTimDimsDefs}

Consider a non-relativistic field theory defined over a spacetime manifold with $d_t$ time dimensions and $d_s$ space dimensions, such that it is invariant both under time rotations and space rotations. In order to define the theory over a curved spacetime manifold, we generalize the structure introduced in \cite{Arav:2014goa,Arav:2016xjc}. We require the background manifold to be equipped with a metric $g_{\mu\nu}$, or alternatively vielbeins $\ve{a}{\mu}$, as well as a distribution of dimension $d_s$, corresponding to the space directions at each point of the manifold. 

This distribution can be represented by the cotangent subbundle of 1-forms that annihilate space tangent vectors. Suppose this subbundle is spanned by a basis of $d_t$ linearly independent 1-forms $t_\mu^{(i)}\, (i=1,\ldots, d_t)$ that correspond to the $d_t$ time directions, so that a vector $V^\alpha$ is space tangent if and only if $t_\mu^{(i)} V^\mu = 0$ for all $1 \leq i \leq d_t$. Then physical quantities, such as the curved spacetime action of the theory, will depend on the subbundle spanned by $\{t_\mu^{(i)}\}$, but not on the choice of basis. We therefore expect them to be invariant under transformations of the form: $ t_\mu^{(i)} \rightarrow L^{ij}(x) t_\mu^{(j)} $, where $ L^{ij}$ is an invertible matrix that depends on the spacetime coordinate $x$.  Alternatively, we can choose an orthonormal basis of 1-forms $ n_\mu^{(i)}$, satisfying:
\begin{equation}\label{AppNonRelCurvedMultTime:OrthonormalCond}
g^{\mu\nu} n_\mu^{(i)} n_\nu^{(j)} = -\delta_{ij},
\end{equation}
and require invariance under local rotations of the time directions -- transformations of the form $ n_\mu^{(i)} \rightarrow \Lambda^{ij}(x) n_\mu^{(j)} $, where $ \Lambda^{ij} $ is an orthogonal matrix that depends on the coordinate $x$.

Using the set of 1-forms $n_\mu^{(i)}$ we can make several definitions on the background manifold.
A tensor $\bar T_{\alpha\beta\ldots}$ is space tangent if it satisfies:
\begin{equation}
n^\alpha_{(i)} \bar T_{\alpha\beta\gamma} = n^\beta_{(i)} \bar T_{\alpha\beta\gamma} = \ldots = 0,
\end{equation}
for any $1\leq i \leq d_t$. Any tensor can be rendered space tangent by projecting it on the space directions using the space projector (or spatial metric), defined as:\footnote{A summing convention is assumed for repeated time indexes ($i,j,\ldots$).}
\begin{equation}
P_{\mu\nu} = g_{\mu\nu} + n_\mu^{(i)} n_\nu^{(i)}.
\end{equation}
The covariant derivative of the 1-form $n_\mu^{(i)}$ can be decomposed as follows:
\begin{equation}
\nabla_\alpha n_\beta^{(i)} = (K_S)_{\alpha\beta}^{(i)} + (K_A)_{\alpha\beta}^{(i)} - a_\beta^{(ij)} n_\alpha^{(j)} -  b_\alpha^{(ij)} n_\beta^{(j)} + c^{(ijk)} n_\alpha^{(j)} n_\beta^{(k)},
\end{equation}
where $ (K_S)_{\alpha\beta}^{(i)} $, $ (K_A)_{\alpha\beta}^{(i)} $, $ a_\alpha^{(ij)} $ and $ b_\alpha^{(ij)}$ are space tangent tensors. $ (K_S)_{\alpha\beta}^{(i)} $ is symmetric, and given by:
\begin{equation}
(K_S)_{\alpha\beta}^{(i)} = P^{\alpha'}_\alpha P^{\beta'}_\beta \nabla_{(\alpha'} n^{(i)}_{\beta')} =  \frac{1}{2} \Lieb{n^{(i)}} P_{\alpha\beta},
\end{equation}
where the space projected Lie derivative $\Lieb{n^{(i)}}$ of a space tangent tensor $ \bar T_{\alpha\beta\ldots} $ is defined as follows:
\begin{equation}
\Lieb{n^{(i)}} \bar T_{\alpha\beta\ldots} \equiv P^{\alpha'}_{\alpha} P^{\beta'}_{\beta}\ldots \Lie{n^{(i)}} \bar T_{\alpha'\beta'\ldots}.
\end{equation}
We denote its trace by $K_S^{(i)} \equiv (K_S)^{(i)}{}^{\mu}_{\mu} $. $(K_A)^{(i)}_{\alpha\beta}$ is antisymmetric and given by:
\begin{equation}
(K_A)^{(i)}_{\alpha\beta} = P^{\alpha'}_\alpha P^{\beta'}_\beta \nabla_{[\alpha'} n^{(i)}_{\beta']}.
\end{equation}
$a^{(ij)}_\beta$ is given by:
\begin{equation}
a^{(ij)}_\beta = P^{\beta'}_\beta n^\alpha_{(j)} \nabla_\alpha n_{\beta'}^{(i)},
\end{equation}
and we define the generalized acceleration vector $a_\beta$ as trace over the time indexes: $ a_\beta \equiv a^{(ii)}_\beta $. $ b_\alpha^{(ij)} $ is given by:
\begin{equation}
b_\alpha^{(ij)} = P^{\alpha'}_\alpha n^\beta_{(j)} \nabla_{\alpha'} n^{(i)}_\beta ,
\end{equation}
and from \eqref{AppNonRelCurvedMultTime:OrthonormalCond} it is easy to show that $ b_\alpha^{(ij)} $ is antisymmetric in its time indexes, i.e.\  $ b_\alpha^{(ji)} = -b_\alpha^{(ij)} $. One can also obtain the following expression for the space projected Lie derivative of the 1-form $n_\mu^{(j)}$ in the direction of $n^\mu_{(i)}$:
\begin{equation}
\Lieb{n^{(i)}} n^{(j)}_\mu = a^{(ji)}_\mu - b^{(ji)}_\mu ,
\end{equation}
so that the generalized acceleration vector is also given by:
\begin{equation}
a_\mu \equiv a^{(ii)}_\mu = \Lieb{n^{(i)}} n^{(i)}_\mu .
\end{equation}
Finally, $ c^{(ijk)} $ is a scalar given by:
\begin{equation}
c^{(ijk)} = n^\alpha_{(j)} n^\beta_{(k)} \nabla_\alpha n^{(i)}_\beta,
\end{equation}
that satisfies $ c^{(kji)} = -c^{(ijk)} $ (again from \eqref{AppNonRelCurvedMultTime:OrthonormalCond}).

Next we define the space tangent covariant derivative of a space tangent tensor $ \bar T_{\alpha\beta\ldots} $ as follows:
\begin{equation}
\wn_\mu \bar T_{\alpha\beta\ldots} \equiv P^{\mu'}_\mu P^{\alpha'}_\alpha P^{\beta'}_\beta \ldots \nabla_{\mu'} \bar T_{\alpha'\beta'\ldots} .
\end{equation}
Note that the spatial metric $P_{\alpha\beta}$ is covariantly constant under this derivative:
\begin{equation}
\wn_\mu P_{\alpha\beta} = 0.
\end{equation}
Operating with the commutation of two space tangent derivatives on a space tangent tensor $\bar T_{\alpha\beta\ldots} $, one obtains the following expression:
\begin{equation}
\left[\wn_\mu, \wn_\nu\right] \bar T_{\alpha\beta\ldots} = \widetilde R_{\alpha\rho\mu\nu} \bar T^\rho {}_{\beta\ldots} + \widetilde R_{\beta\rho\mu\nu} \bar T_\alpha {}^\rho{}_{\ldots} + \ldots + 2 (K_A)_{\mu\nu}^{(i)} \Lieb{n^{(i)}} \bar T_{\alpha\beta\ldots} ,
\end{equation}
where $ \widetilde R_{\alpha\rho\mu\nu} $ is a space tangent tensor defined by:
\begin{equation}
\widetilde R_{\alpha\rho\mu\nu} \equiv P^{\alpha'}_\alpha P^{\rho'}_\rho P^{\mu'}_\mu P^{\nu'}_\nu R_{\alpha'\rho'\mu'\nu'} - 2 (K_A)_{\mu\nu}^{(i)} K_{\alpha\rho}^{(i)} - K_{\mu\alpha}^{(i)} K_{\nu\rho}^{(i)} + K_{\nu\alpha}^{(i)} K_{\mu\rho}^{(i)},
\end{equation}
$ K_{\alpha\beta}^{(i)} \equiv (K_S)_{\alpha\beta}^{(i)} + (K_A)_{\alpha\beta}^{(i)} $ is the total space tangent component of $ \nabla_\alpha n_\beta^{(i)} $ and $ R_{\alpha\rho\mu\nu} $ is the standard Riemann curvature associated with the covariant derivative $ \nabla_\mu $ (see appendix \ref{app:conventions} for our conventions). Similarly to the one time direction case (see \cite{Arav:2016xjc}), the tensor $ \widetilde R_{\alpha\rho\mu\nu} $ does not have all of the standard symmetries of the Riemann tensor. It is therefore useful to define a modified Riemann tensor:
\begin{equation}
\begin{split}
\widehat R_{\alpha\rho\mu\nu} \equiv
\widetilde R_{\alpha\rho\mu\nu}
+2(K_A)_{\mu\nu}^{(i)}(K_S)_{\alpha\rho}^{(i)}
&+(K_A)_{\mu\alpha}^{(i)}(K_S)_{\nu\rho}^{(i)}
+(K_S)_{\mu\alpha}^{(i)}(K_A)_{\nu\rho}^{(i)}\\
&-(K_A)_{\nu\alpha}^{(i)}(K_S)_{\mu\rho}^{(i)}
-(K_S)_{\nu\alpha}^{(i)}(K_A)_{\mu\rho}^{(i)},
\end{split}
\end{equation}
which satisfies the usual Riemann tensor symmetries except for the second Bianchi identity.
We then define the equivalents of the Ricci tensor and scalar for this modified Riemann tensor $\widehat R_{\alpha\rho\mu\nu}$ as follows:
\begin{equation}
\begin{split}
&\widehat R_{\rho\nu} \equiv \widehat R^{\mu}{}_{\rho\mu\nu}  = P^{\alpha\mu} \widehat R_{\alpha\rho\mu\nu} ,\\
&\widehat R \equiv \widehat R^{\nu}_{\nu} = P^{\rho\nu} \widehat R_{\rho\nu}.
\end{split}
\end{equation}
Note that from the above definitions, one gets the following identity for the divergence of a space tangent vector $\bar V^\mu$:
\begin{equation}
\label{AppNonRelCurvedMultTime:TotDerFormula}
\nabla_\mu \bar V^\mu = (\wn_\mu + a_\mu ) \bar V^\mu .
\end{equation}

\subsection{Symmetries Over Curved Spacetime}
\label{app:LifCurvedSpacetimeMultTimDimsSym}

Next we turn to discuss the symmetries of the curved spacetime field theory. Like in the one time direction case, the symmetries of the flat space Lifshitz theory translate to local symmetries over curved spacetime:

First, as mentioned in subsection \ref{subsubsec:LifshitzSymWardCuvedSpacetime} for the one time direction case, we require time-direction preserving diffeomorphism (TPD) invariance that corresponds to space rotation symmetry in flat space. In this case, these are diffeomorphisms with a parameter $\xi$ that satisfies: $ \mathcal{L}_\xi t_\alpha^{(i)} = M^{ij}(x) t_\alpha^{(j)} $ where $M^{ij}$ is some spacetime dependent invertible $d_t \times d_t$ matrix. Similarly to the one time direction case (see \cite{Arav:2014goa,Arav:2016xjc}), we can extend these to the full diffeomorphism group by having the 1-forms $t_\mu^{(i)}$ transform appropriately:
\begin{equation}
\delta^D_\xi g_{\mu\nu} = \nabla_\mu \xi_\nu + \nabla_\nu \xi_\mu,
\quad
\delta^D_\xi t_\alpha^{(i)} = \Lie{\xi} t_\alpha^{(i)} = \xi^\beta \nabla_\beta t_\alpha^{(i)} + \nabla_\alpha \xi^\beta t_\beta^{(i)}  .
\end{equation}

Second, as previously mentioned, we require invariance under local time rotations of the form $ n_\mu^{(i)} \rightarrow \Lambda^{ij}(x) n_\mu^{(j)} $, where $ \Lambda^{ij} $ is a spacetime dependent orthogonal matrix. In infinitesimal form, these transformations are given by:
\begin{equation}
\delta^T_\omega n^{(i)}_\alpha = \omega_{ij} n^{(j)}_\alpha, 
\qquad \delta^T_\omega g_{\mu\nu} = \delta^T_\omega P_{\mu\nu} = 0,
\end{equation}
where $\omega^{ij}$ is a transformation parameter that satisfies $\omega^{ji}=-\omega^{ij}$. From this transformation and the definition of the derivatives $\Lieb{n^{(i)}}$, $\wn_\mu$ we obtain the following for a space tangent tensor $\bar T_{\alpha\beta\ldots} $:
\begin{equation}
\begin{split}
\delta^T_\omega \Lieb{n^{(i)}} \bar T_{\alpha\beta\ldots} &= \Lieb{n^{(i)}} \delta^T_\omega \bar T_{\alpha\beta\ldots} + \omega^{ij} \Lieb{n^{(j)}} \bar T_{\alpha\beta\ldots}, \\
\delta^T_\omega \wn_\mu \bar T_{\alpha\beta\ldots} &= \wn_\mu \delta^T_\omega \bar T_{\alpha\beta\ldots}.
\end{split}
\end{equation}
Using these formulas and the various definitions from appendix \ref{app:LifCurvedSpacetimeMultTimDimsDefs}, the following time rotation transformation rules can be derived:
\begin{align*}
\delta^T_\omega (K_S)_{\alpha\beta}^{(i)} &= \omega^{ij} (K_S)_{\alpha\beta}^{(j)},\\
\delta^T_\omega K_S^{(i)} &= \omega^{ij} K_S^{(j)},\\
\delta^T_\omega (K_A)_{\alpha\beta}^{(i)} &= \omega^{ij} (K_A)_{\alpha\beta}^{(j)},\numthis\\
\delta^T_\omega a_\mu^{(ij)} &= \omega^{ik} a_\mu^{(kj)} + \omega^{jk} a_\mu^{(ik)},\\
\delta^T_\omega a_\mu &= 0, \\
\delta^T_\omega \widehat R_{\alpha\rho\mu\nu} &= \delta^T_\omega \widehat R_{\rho\nu} = \delta^T_\omega \widehat R = 0.
\end{align*}

Finally we require anisotropic Weyl invariance, which is the local version of Lifshitz scale invariance in flat space \eqref{eq:LifshitzScalingTrans}. In the case of multiple time directions, the infinitesimal anisotropic Weyl transformation is given by:
\begin{equation}
\dels P_{\mu\nu} = 2\sigma P_{\mu\nu}, \qquad
\dels n_{(i)}^\mu = -z\sigma n_{(i)}^\mu, \qquad
\dels n_\mu^{(i)} = z\sigma n_\mu^{(i)}.
\end{equation}
From this transformation and the definition of the derivatives $\Lieb{n^{(i)}}$, $\wn_\mu$ we get the following formulas for a space tangent tensor $\bar T_{\alpha\beta\ldots} $:
\begin{equation}
\begin{split}
&
\dels \Lieb{n^{(i)}} \bar T_{\alpha\beta\gamma\ldots} =
-z\sigma \Lieb{n^{(i)}} \bar T_{\alpha\beta\gamma\ldots} +
 \Lieb{n^{(i)}} \dels \bar T_{\alpha\beta\gamma\ldots},
\\
&
\dels (\wn_\mu \bar T_{\alpha\beta\gamma\ldots}) = \wn_\mu  (\dels \bar T_{\alpha\beta\gamma\ldots})
- I[\bar T] \wn_\mu \sigma \bar T_{\alpha\beta\ldots}
\\
&~~~~~~~~~~~~~~~~~~~~~~~~
- ( \wn_\alpha \sigma ) \bar T_{\mu\beta\gamma \ldots} +\wn_\rho \sigma P_{\mu\alpha} \bar T^\rho{}_{\beta\ldots} - \ldots,
\end{split}
\end{equation}
where $I[\bar T]$ is the rank of the tensor $\bar T_{\alpha\beta\ldots} $. Using these formulas and the various definitions from appendix \ref{app:LifCurvedSpacetimeMultTimDimsDefs}, the following anisotropic Weyl transformation rules can be derived:
\begin{equation}
\begin{split}\label{eq:Weyl_trans_laws}
&
\dels (K_S)_{\mu\nu}^{(i)} = (2-z) \sigma (K_S)_{\mu\nu}^{(i)} +P_{\mu\nu} \Lieb{n^{(i)}} \sigma,
\\
&
\dels K_S^{(i)} = -z \sigma K_S^{(i)} + d_s \Lieb{n^{(i)}} \sigma, 
\\
&
\dels a_\mu = z d_t \wn_\mu\sigma,
\\
&
\dels (K_A)_{\mu\nu}^{(i)} = z \sigma (K_A)_{\mu\nu}^{(i)},
\\
&
\dels \widehat R_{\alpha\rho\mu\nu} = 2\sigma \widehat R_{\alpha\rho\mu\nu}
+ P_{\alpha\nu} \wn_{(\rho} \wn_{\mu)} \sigma
- P_{\alpha\mu} \wn_{(\rho} \wn_{\nu)} \sigma
\\
&
~~~~~~~~~~~~~~~~~~~~~~~~~~~+
P_{\rho\mu} \wn_{(\alpha} \wn_{\nu)} \sigma
-
P_{\rho\nu} \wn_{(\alpha} \wn_{\mu)} \sigma,
\\
&
 \dels \widehat{R}_{\alpha\mu} = (2-d_s) \wn_{(\alpha} \wn_{\mu)} \sigma -P_{\alpha\mu} \bar \nabla^2 \sigma,
\\
&
\dels \widehat R = -2 \sigma \widehat R -2(d_s-1) \bar \nabla^2 \sigma.
\end{split}
\end{equation}

\subsection{Action of the Free $z=2$ Scalar}

Consider the free Lifshitz scalar in $d_s+d_t$ dimensions with a dynamical critical exponent of $z=2$. Its flat space action is given in \eqref{eq:FlatSpaceAction}. In order to couple it to a curved spacetime manifold, we have to define its curved spacetime action $ S(g_{\mu\nu},t_\alpha^{(i)},\phi) $ such that it is invariant under TPD, local time rotation and anisotropic Weyl transformations, and it reduces to the action \eqref{eq:FlatSpaceAction} in the flat space limit. We also require it to be regular in the physical limit $ d_t \to 1 $, for the sake of using the split dimensional regularization procedure as described in subsection \ref{SplitDimensionalRegAndTheLifshitzCase}.

Suppose that $\phi$ has a scaling dimension $s$ under anisotropic Weyl transformations, that is:
\begin{equation}
\dels \phi = s\, \sigma \phi.
\end{equation}
For the temporal part of the action, using the transformations given in appendix \ref{app:LifCurvedSpacetimeMultTimDimsSym} we can find a linear combination of $ \mathcal{L}_{n^{(i)}} \phi $ and $ K_S^{(i)} \phi $ which is covariant under both local time rotations and anisotropic Weyl transformations (it transforms with no contribution from derivatives of the parameters $\omega^{ij}$ and $\sigma$):
\begin{equation}
\left( \mathcal{L}_{n^{(i)}} - \frac{s}{d_s} K_S^{(i)} \right) \phi.
\end{equation}
For the spatial part of the action, we can find a linear combination of $\wn^2 \phi$, $a^\mu \wn_\mu \phi$, $a^2 \phi$ and $\wn_\mu a^\mu \,\phi$ which is anisotropic-Weyl-covariant, given by:
\begin{equation}
\left[ \wn^2 + \frac{2-d_s-2s}{z d_t} a^\mu \wn_\mu - \frac{s(2-d_s-s)}{z^2 d_t^2} a^2 - \frac{s}{z d_t} \wn_\mu a^\mu \right] \phi,
\end{equation}
or alternatively, we can use $\wn^2 \phi$, $a^\mu \wn_\mu \phi$, $a^2 \phi$ and $ \widehat R \phi $  to obtain the following anisotropic-Weyl-covariant linear combination:\footnote{Of course, any linear combination of these two options could also be used.}
\begin{equation}
\left[ \wn^2 + \frac{2-d_s-2s}{zd_t} a^\mu \wn_\mu - \frac{s(2-d_s-2s)}{2z^2d_t^2} a^2 - \frac{s}{2(1-d_s)}\widehat R \right] \phi.
\end{equation}

Finally, for the action to be anisotropic-Weyl-invariant with $z=2$, the dimension of the scalar field is required to satisfy:
\begin{equation}
2(s-2)+2d_t+d_s = 0 \Rightarrow s=2-d_t-\frac{1}{2}d_s .
\end{equation}

Combining these expressions, we arrive at two possible options for the curved spacetime action of the free $z=2$ scalar (corresponding to the two options for the spatial part). The first option is given by:
\begin{equation}
\begin{aligned}
S & = \int d^{d_t+d_s}x \sqrt{|g|} \left\{ \frac{1}{2}\left[ \mathcal{L}_{n^{(i)}}\phi+\xi_1\, K_S^{(i)}\phi \right]^2 \right. \\
& \qquad \qquad \qquad \qquad \left. -\frac{\kappa}{2}\left[\bar{\nabla}^2\phi+\xi_2 \,a^\mu\bar{\nabla}_\mu\phi+\xi_3 \, a^2\phi+\xi_4 \,\bar{\nabla}_\mu a^\mu\phi \right]^2\right\},
\end{aligned}
\end{equation}
where: 
\begin{equation}
\begin{alignedat}{2}
\xi_1 &\equiv \frac{1}{d_s}\left(\frac{1}{2}d_{\text{lif}}-2\right), 
&\qquad \xi_2 &\equiv \frac{d_t-1}{d_t}, \\
\xi_3 &\equiv \frac{1}{4d_t^2}\left(\frac{1}{2}d_{\text{lif}}-2\right)\left(d_t-\frac{1}{2}d_s\right), &\qquad \xi_4 &\equiv \frac{1}{2d_t}\left(\frac{1}{2}d_{\text{lif}}-2\right),
\end{alignedat}
\end{equation}
and $d_\text{lif} \equiv 2 d_t + d_s $. The second option is given by:
\begin{equation}
\begin{aligned}
S & = \int d^{d_t+d_s}x \sqrt{|g|} \left\{ \frac{1}{2}\left[ \mathcal{L}_{n^{(i)}}\phi+\xi_1' \, K_S^{(i)}\phi \right]^2 \right. \\
& \qquad \qquad \qquad \qquad \left. -\frac{\kappa}{2}\left[\bar{\nabla}^2\phi+\xi_2' \,a^\mu\bar{\nabla}_\mu\phi+\xi_3' \, a^2\phi+\xi_4' \, \widehat R \phi \right]^2\right\},
\end{aligned}
\end{equation}
where:
\begin{equation}
\begin{alignedat}{2}
\xi_1' &\equiv \frac{1}{d_s}\left(\frac{1}{2}d_{\text{lif}}-2\right),
&\qquad 
\xi_2' &\equiv \frac{d_t-1}{d_t}, \\
\xi_3' &\equiv \frac{d_t-1}{4d_t^2}\left(\frac{1}{2}d_{\text{lif}}-2\right), &\qquad \xi_4' &\equiv - \frac{1}{2(d_s-1)} \left(\frac{1}{2}d_{\text{lif}}-2\right).
\end{alignedat}
\end{equation}
These actions are indeed invariant under TPDs, local time rotations and anisotropic Weyl transformations. They are also regular in the $d_t \to 1$ limit, as required.

In this work we have chosen to use the first option. Note, however, that for the purpose of the Lifshitz anomalies calculation in $2+1$ dimensions (as done in section \ref{Sec:z2LishitzFreeScalarResults}), the results would be the same for both options, since the difference between them is proportional to $\epslif$ and therefore does not contribute to the $\epslif$ pole residues of the flat space correlation functions (see subsection \ref{SplitDimensionalRegAndTheLifshitzCase} for details).

\section{Relativistic Scalar Field -- Feynman Rules, Vertexes and Integrals}
\label{app:DetailsCalcScalar}

In this appendix we give some details for the calculations of the two and three point correlation functions which are required for computing the conformal anomaly coefficients of the relativistic scalar field in two and four spacetime dimensions, as explained in section~\ref{sec:ConformalAnom}.  

\subsection{Feynman Rules and Diagrams}
\label{FeynRulesRelDiagRel}
The propagator of the relativistic scalar field is given by:
\begin{equation}
\left\langle \phi\phi \right\rangle (q)= \frac{i}{q^2-m^2+i\epsilon},
\end{equation}
where, as explained in subsection \ref{ReviewOfDimensionalRegConformalCase}, $m$ is an IR mass regulator later taken to be zero.
The (Fourier transformed) two and three point Feynman correlation functions of the stress-energy tensor are given by the expressions:
\begin{equation}
\label{eq:RelativisticTwoPointGeneralExp}
\left\langle {{T^{\mu \nu }}\left( p \right){T^{\rho \sigma }}\left( { - p} \right)} \right\rangle  = \frac{(i)^2}{2}\int {\frac{{{d^d}q}}{{{{\left( {2\pi } \right)}^d}}}\frac{{{V_T}^{\mu \nu }\left( {q,p - q} \right)}}{{\left[ {{q^2} - {m^2}+i\epsilon} \right]}}\frac{{{V_T}^{\rho \sigma }\left( {q,p - q} \right)}}{{[ {{{\left( {p - q} \right)}^2} - {m^2}+i\epsilon} ]}}} ,
\end{equation}
and
\begin{equation}
\label{eq:RelativisticThreePointGeneralExp}
\begin{aligned}
& \left\langle {{T^{\mu \nu }}\left( {k =  - p - q} \right){T^{\rho \sigma }}\left( q \right){T^{\alpha \beta }}\left( p \right)} \right\rangle_F  = \\ 
 & \qquad (i^3)\int {\frac{{{d^d}l}}{{{{\left( {2\pi } \right)}^d}}}\frac{{{V_T}^{\mu \nu }\left( { - l,l - p - q} \right)}}{{\left[ {{l^2} - {m^2}+i\epsilon} \right]}}\frac{{{V_T}^{\rho \sigma }\left( {l,q - l} \right)}}{{\left[ {{{\left( {q - l} \right)}^2} - {m^2}+i\epsilon} \right]}}} \\
& \qquad \qquad \qquad  \times \frac{{{V_T}^{\alpha \beta }\left( {l - q, - l + p + q} \right)}}{{\left[ {{{\left( {p + q - l} \right)}^2} - {m^2}+i\epsilon} \right]}}.
\end{aligned} 
\end{equation}
These expressions correspond to Feynman diagrams of the form given in figure \ref{fig:diagramLif2p} for the two point function of the stress-energy tensor, and figure \ref{fig:diagramLif3p} for the three point function of the stress-energy tensor. The vertexes $ V_T^{\mu\nu} $ are given by: 
\begin{equation}
\label{eq:VertexRelStress}
\begin{aligned}
{V_T}^{\mu \nu }\left( {q,p} \right) & = -\left( {1 + \frac{{d - 2}}{{2\left( {1 - d} \right)}}} \right){A_T}^{\mu \nu }\left( {q,p} \right) - \left( {\frac{{d - 2}}{{2\left( {1 - d} \right)}}} \right){C_T}^{\mu \nu }\left( {q,p} \right) \\
& + \left( {\frac{1}{2} + \frac{{d - 2}}{{2\left( {1 - d} \right)}}} \right){B_T}^{\mu \nu }\left( {q,p} \right) + \left( {\frac{{d - 2}}{{2\left( {1 - d} \right)}}} \right){D_T}^{\mu \nu }\left( {q,p} \right),
\end{aligned}
\end{equation}
where $A_T$, $B_T$, $C_T$ and $D_T$ are defined by:
\begin{align}
\label{eq:VertexRelStressOpA}
{A_T}^{\mu \nu }\left( {q,p} \right) &= {\left( i \right)^2}\left( {{q_\mu }{p_\nu } + {q_\nu }{p_\mu }} \right), \\
\label{eq:VertexRelStressOpB}
{B_T}^{\mu \nu }\left( {q,p} \right) &= 2{\left( i \right)^2}{\eta _{\mu \nu }}\left( {q \cdot p} \right),\\
\label{eq:VertexRelStressOpC}
{C_T}^{\mu \nu }\left( {q,p} \right) &= {\left( i \right)^2}\left( {{q_\mu }{q_\nu } + {p_\mu }{p_\nu }}\right),\\
\label{eq:VertexRelStressOpD}
{D_T}^{\mu \nu }\left( {q,p} \right) &= {\left( i \right)^2}{\eta _{\mu \nu }}\left( {{q^2} + {p^2}} \right).
\end{align}
Note that terms which carry coefficients of order $O(\varepsilon)$ do not contribute to the pole residues of the correlation functions, and can therefore be ignored for the purpose of our calculations.
For example, the terms $C_T$ and $D_T$ in equation \eqref{eq:VertexRelStress} can be neglected when calculating the pole residues around two spacetime dimensions. However, these terms cannot be neglected in four dimensions since their coefficients are no longer proportional to $\varepsilon$. 

The expression for the correlation function \eqref{eq:TwoPointLikeRelativistic} (which corresponds to a Feynman diagram of the form given in figure \ref{fig:diagram_lifLongVer}) is the following:
\begin{equation}
\label{eq:RelLongVertex}
\left\langle {\frac{{{\delta ^2}S}}{{\delta {g_{\mu \nu }}\delta {g_{\rho \sigma }}}}(k,q)\frac{{\delta S}}{{\delta {g_{\alpha \beta }}}}}(p) \right\rangle  = \frac{(i)^2}{2}\int {\frac{{{d^d}l}}{{{{\left( {2\pi } \right)}^d}}}\frac{{{V^{\mu \nu \rho \sigma }}\left( {p+l,-l} \right)}}{{\left[ {{l^2} - {m^2}+i\epsilon} \right]}}\frac{{{V_T}^{\alpha \beta }\left( {-l,p + l} \right)}}{{\left[ {{{\left( {p + l} \right)}^2} - {m^2}+i\epsilon} \right]}}},
\end{equation}
where the vertex $ V^{\mu \nu \rho \sigma } $ is defined by:
\begin{equation}
{V^{\mu \nu \rho \sigma }}\left( {p,q} \right) \equiv {{V_0}^{\mu \nu \rho \sigma }}\left( {p,q} \right)+{{V_1}^{\mu \nu \rho \sigma }}\left( {p,q} \right),
\end{equation}
where:
\begin{equation}
\begin{aligned}
&{{V_0}^{\mu \nu \rho \sigma }}\left( {p,q} \right) \equiv \\
& \qquad - \tfrac{1}{8} \eta_{\mu \sigma} \eta_{\nu \rho} p^{\alpha} q_{\alpha} -  \tfrac{1}{8} \eta_{\mu \rho} \eta_{\nu \sigma} p^{\alpha} q_{\alpha} + \tfrac{1}{8} \eta_{\mu \nu} \eta_{\rho \sigma} p^{\alpha} q_{\alpha} -  \tfrac{1}{8} \eta_{\rho \sigma} p_{\nu} q_{\mu} + \tfrac{1}{8} \eta_{\nu \sigma} p_{\rho} q_{\mu}\\
&\qquad  + \tfrac{1}{8} \eta_{\nu \rho} p_{\sigma} q_{\mu} -  \tfrac{1}{8} \eta_{\rho \sigma} p_{\mu} q_{\nu} + \tfrac{1}{8} \eta_{\mu \sigma} p_{\rho} q_{\nu} + \tfrac{1}{8} \eta_{\mu \rho} p_{\sigma} q_{\nu} + \tfrac{1}{8} \eta_{\nu \sigma} p_{\mu} q_{\rho} + \tfrac{1}{8} \eta_{\mu \sigma} p_{\nu} q_{\rho}\\
&\qquad  -  \tfrac{1}{8} \eta_{\mu \nu} p_{\sigma} q_{\rho} + \tfrac{1}{8} \eta_{\nu \rho} p_{\mu} q_{\sigma} + \tfrac{1}{8} \eta_{\mu \rho} p_{\nu} q_{\sigma} -  \tfrac{1}{8} \eta_{\mu \nu} p_{\rho} q_{\sigma},
\end{aligned}
\end{equation}
\begin{align*}
&{{V_1}^{\mu \nu \rho \sigma }}\left( {p,q} \right) \equiv \\
& \qquad - \frac{\eta_{\mu \sigma} \eta_{\nu \rho} p_{\alpha} p^{\alpha}}{32 (-1 + d)} + \frac{d \eta_{\mu \sigma} \eta_{\nu \rho} p_{\alpha} p^{\alpha}}{64 (-1 + d)} -  \frac{\eta_{\mu \rho} \eta_{\nu \sigma} p_{\alpha} p^{\alpha}}{32 (-1 + d)} + \frac{d \eta_{\mu \rho} \eta_{\nu \sigma} p_{\alpha} p^{\alpha}}{64 (-1 + d)}\\
&\qquad  -  \frac{\eta_{\mu \nu} \eta_{\rho \sigma} p_{\alpha} p^{\alpha}}{16 (-1 + d)} + \frac{d \eta_{\mu \nu} \eta_{\rho \sigma} p_{\alpha} p^{\alpha}}{32 (-1 + d)} -  \frac{3 \eta_{\rho \sigma} p_{\mu} p_{\nu}}{16 (-1 + d)} + \frac{3 d \eta_{\rho \sigma} p_{\mu} p_{\nu}}{32 (-1 + d)} + \frac{\eta_{\nu \sigma} p_{\mu} p_{\rho}}{32 (-1 + d)}\\
&\qquad  -  \frac{d \eta_{\nu \sigma} p_{\mu} p_{\rho}}{64 (-1 + d)} + \frac{\eta_{\mu \sigma} p_{\nu} p_{\rho}}{32 (-1 + d)} -  \frac{d \eta_{\mu \sigma} p_{\nu} p_{\rho}}{64 (-1 + d)} + \frac{\eta_{\nu \rho} p_{\mu} p_{\sigma}}{32 (-1 + d)} -  \frac{d \eta_{\nu \rho} p_{\mu} p_{\sigma}}{64 (-1 + d)}\\
&\qquad  + \frac{\eta_{\mu \rho} p_{\nu} p_{\sigma}}{32 (-1 + d)} -  \frac{d \eta_{\mu \rho} p_{\nu} p_{\sigma}}{64 (-1 + d)} -  \frac{3 \eta_{\mu \nu} p_{\rho} p_{\sigma}}{16 (-1 + d)} + \frac{3 d \eta_{\mu \nu} p_{\rho} p_{\sigma}}{32 (-1 + d)} -  \frac{\eta_{\mu \sigma} \eta_{\nu \rho} p^{\alpha} q_{\alpha}}{8 (-1 + d)}\\
&\qquad  + \frac{d \eta_{\mu \sigma} \eta_{\nu \rho} p^{\alpha} q_{\alpha}}{16 (-1 + d)} -  \frac{\eta_{\mu \rho} \eta_{\nu \sigma} p^{\alpha} q_{\alpha}}{8 (-1 + d)} + \frac{d \eta_{\mu \rho} \eta_{\nu \sigma} p^{\alpha} q_{\alpha}}{16 (-1 + d)} -  \frac{\eta_{\mu \sigma} \eta_{\nu \rho} q_{\alpha} q^{\alpha}}{32 (-1 + d)}\\
&\qquad  + \frac{d \eta_{\mu \sigma} \eta_{\nu \rho} q_{\alpha} q^{\alpha}}{64 (-1 + d)} -  \frac{\eta_{\mu \rho} \eta_{\nu \sigma} q_{\alpha} q^{\alpha}}{32 (-1 + d)} + \frac{d \eta_{\mu \rho} \eta_{\nu \sigma} q_{\alpha} q^{\alpha}}{64 (-1 + d)} -  \frac{\eta_{\mu \nu} \eta_{\rho \sigma} q_{\alpha} q^{\alpha}}{16 (-1 + d)}\\
&\qquad  + \frac{d \eta_{\mu \nu} \eta_{\rho \sigma} q_{\alpha} q^{\alpha}}{32 (-1 + d)} -  \frac{\eta_{\rho \sigma} p_{\nu} q_{\mu}}{8 (-1 + d)} + \frac{d \eta_{\rho \sigma} p_{\nu} q_{\mu}}{16 (-1 + d)} + \frac{\eta_{\nu \sigma} p_{\rho} q_{\mu}}{16 (-1 + d)} -  \frac{d \eta_{\nu \sigma} p_{\rho} q_{\mu}}{32 (-1 + d)} \numthis \\
&\qquad  + \frac{\eta_{\nu \rho} p_{\sigma} q_{\mu}}{16 (-1 + d)} -  \frac{d \eta_{\nu \rho} p_{\sigma} q_{\mu}}{32 (-1 + d)} -  \frac{\eta_{\rho \sigma} p_{\mu} q_{\nu}}{8 (-1 + d)} + \frac{d \eta_{\rho \sigma} p_{\mu} q_{\nu}}{16 (-1 + d)} + \frac{\eta_{\mu \sigma} p_{\rho} q_{\nu}}{16 (-1 + d)}\\
&\qquad  -  \frac{d \eta_{\mu \sigma} p_{\rho} q_{\nu}}{32 (-1 + d)} + \frac{\eta_{\mu \rho} p_{\sigma} q_{\nu}}{16 (-1 + d)} -  \frac{d \eta_{\mu \rho} p_{\sigma} q_{\nu}}{32 (-1 + d)} -  \frac{3 \eta_{\rho \sigma} q_{\mu} q_{\nu}}{16 (-1 + d)} + \frac{3 d \eta_{\rho \sigma} q_{\mu} q_{\nu}}{32 (-1 + d)}\\
&\qquad  + \frac{\eta_{\nu \sigma} p_{\mu} q_{\rho}}{16 (-1 + d)} -  \frac{d \eta_{\nu \sigma} p_{\mu} q_{\rho}}{32 (-1 + d)} + \frac{\eta_{\mu \sigma} p_{\nu} q_{\rho}}{16 (-1 + d)} -  \frac{d \eta_{\mu \sigma} p_{\nu} q_{\rho}}{32 (-1 + d)} -  \frac{\eta_{\mu \nu} p_{\sigma} q_{\rho}}{8 (-1 + d)}\\
&\qquad  + \frac{d \eta_{\mu \nu} p_{\sigma} q_{\rho}}{16 (-1 + d)} + \frac{\eta_{\nu \sigma} q_{\mu} q_{\rho}}{32 (-1 + d)} -  \frac{d \eta_{\nu \sigma} q_{\mu} q_{\rho}}{64 (-1 + d)} + \frac{\eta_{\mu \sigma} q_{\nu} q_{\rho}}{32 (-1 + d)} -  \frac{d \eta_{\mu \sigma} q_{\nu} q_{\rho}}{64 (-1 + d)}\\
&\qquad  + \frac{\eta_{\nu \rho} p_{\mu} q_{\sigma}}{16 (-1 + d)} -  \frac{d \eta_{\nu \rho} p_{\mu} q_{\sigma}}{32 (-1 + d)} + \frac{\eta_{\mu \rho} p_{\nu} q_{\sigma}}{16 (-1 + d)} -  \frac{d \eta_{\mu \rho} p_{\nu} q_{\sigma}}{32 (-1 + d)} -  \frac{\eta_{\mu \nu} p_{\rho} q_{\sigma}}{8 (-1 + d)}\\
&\qquad  + \frac{d \eta_{\mu \nu} p_{\rho} q_{\sigma}}{16 (-1 + d)} + \frac{\eta_{\nu \rho} q_{\mu} q_{\sigma}}{32 (-1 + d)} -  \frac{d \eta_{\nu \rho} q_{\mu} q_{\sigma}}{64 (-1 + d)} + \frac{\eta_{\mu \rho} q_{\nu} q_{\sigma}}{32 (-1 + d)} -  \frac{d \eta_{\mu \rho} q_{\nu} q_{\sigma}}{64 (-1 + d)}\\
&\qquad  -  \frac{3 \eta_{\mu \nu} q_{\rho} q_{\sigma}}{16 (-1 + d)} + \frac{3 d \eta_{\mu \nu} q_{\rho} q_{\sigma}}{32 (-1 + d)}.
\end{align*}

\subsection{Massless Integrals}
\label{MasslessIntegrals}
The full evaluation of the two point correlation function of the stress-energy tensor as given in equation \eqref{eq:TwoPoint2DFullRes} requires the use of the following dimensionally-regulated integrals (taken from \cite{Capper:1973pv}, and transformed into the Lorentzian signature conventions):
\begin{align}
\begin{split}
{I_1} &\equiv \int {\frac{{{d^d}q}}{{{{\left( {2\pi } \right)}^d}}}} \frac{1}{{{q^2}{{\left( {q - p} \right)}^2}}} \\
&= \frac{(-1)^{d/2-2} \, i}{{{{\left( {4\pi } \right)}^{d/2}}}}\frac{{\Gamma \left( {d/2 - 1} \right)\Gamma \left( {d/2 - 1} \right)\Gamma \left( {2 - d/2} \right)}}{{\Gamma \left( {d - 2} \right)}}{\left( {{p^2}} \right)^{d/2 - 2}},
\end{split}\\
&\quad \int {\frac{{{d^d}q}}{{{{\left( {2\pi } \right)}^d}}}} \frac{{{q_\mu }}}{{{q^2}{{\left( {q - p} \right)}^2}}} = {I_2}{p_\mu },\\
&\quad \int {\frac{{{d^d}q}}{{{{\left( {2\pi } \right)}^d}}}} \frac{{{q_\mu }{q_\nu }}}{{{q^2}{{\left( {q - p} \right)}^2}}} = {I_3}{\eta _{\mu \nu }} + {I_4}{p_\mu }{p_\nu },\\
&\quad \int {\frac{{{d^d}q}}{{{{\left( {2\pi } \right)}^d}}}} \frac{{{q_\mu }{q_\nu }{q_\gamma }}}{{{q^2}{{\left( {q - p} \right)}^2}}} = {I_5}{p_\mu }{p_\nu }{p_\gamma } + {I_6}{E_{\mu \nu \gamma }}, \\
&\quad \int {\frac{{{d^d}q}}{{{{\left( {2\pi } \right)}^d}}}} \frac{{{q_\mu }{q_\nu }{q_\gamma }{q_\sigma }}}{{{q^2}{{\left( {q - p} \right)}^2}}} = {I_7}{p_\mu }{p_\nu }{p_\gamma }{p_\sigma } + {I_8}{G_{\mu \nu \gamma \sigma }} + {I_9}{H_{\mu \nu \gamma \sigma }},
\end{align}
where:
\begin{align}
{E_{\mu \nu \gamma }} &\equiv {\eta _{\mu \nu }}{p_\gamma } + {\eta _{\mu \gamma }}{p_\nu } + {\eta _{\nu \gamma }}{p_\mu },\\
\begin{split}
{G_{\mu \nu \gamma \sigma }} &\equiv {\eta _{\mu \nu }}{p_\gamma }{p_\sigma } + {\eta _{\mu \gamma }}{p_\nu }{p_\sigma } + {\eta _{\mu \sigma }}{p_\gamma }{p_\nu } + {\eta _{\nu \gamma }}{p_\mu }{p_\sigma } \\
 &+ {\eta _{\nu \sigma }}{p_\mu }{p_\gamma } + {\eta _{\gamma \sigma }}{p_\mu }{p_\nu },
\end{split}\\
{H_{\mu \nu \gamma \sigma }} &= {\eta _{\mu \nu }}{\eta _{\gamma \sigma }} + {\eta _{\mu \sigma }}{\eta _{\nu \gamma }} + {\eta _{\mu \gamma }}{\eta _{\nu \sigma }},
\end{align}
and:
\begin{align}
{I_2} &= \frac{1}{2}{I_1}, \\
{I_3} &=  - \frac{{{p^2}}}{{4\left( {d - 1} \right)}}{I_1}, \\
{I_4} &= \frac{d}{{4\left( {d - 1} \right)}}{I_1}, \\
{I_5} &= \frac{{d + 2}}{{8\left( {d - 1} \right)}}{I_1}, \\
{I_6} &=  - \frac{{{p^2}}}{{8\left( {d - 1} \right)}}{I_1}, \\
{I_7} &= \frac{{\left( {d + 2} \right)\left( {d + 4} \right)}}{{16\left( {{d^2} - 1} \right)}}{I_1}, \\
{I_8} &=  - \frac{{\left( {d + 2} \right)}}{{16\left( {{d^2} - 1} \right)}}{p^2}{I_1}, \\
{I_9} &= \frac{1}{{16\left( {{d^2} - 1} \right)}}{\left( {{p^2}} \right)^2}{I_1}.
\end{align}

\section{Lifshitz $z=2$ Scalar Field -- Feynman Rules, Vertexes and Integrals}
\label{FeynRulesRelDiagLif}

In this appendix we give some details for the calculation of the two and three point correlation functions which are required for computing the Lifshitz anomaly coefficients 
of a $z=2$ free scalar field in $2+1$ dimensions, as explained in section \ref{Sec:z2LishitzFreeScalarResults}. These include the expressions for the Feynman diagrams, the Feynman rules for the propagator and all the vertexes needed.

\begin{figure}[h!]
\centering
\includegraphics[width=50mm]{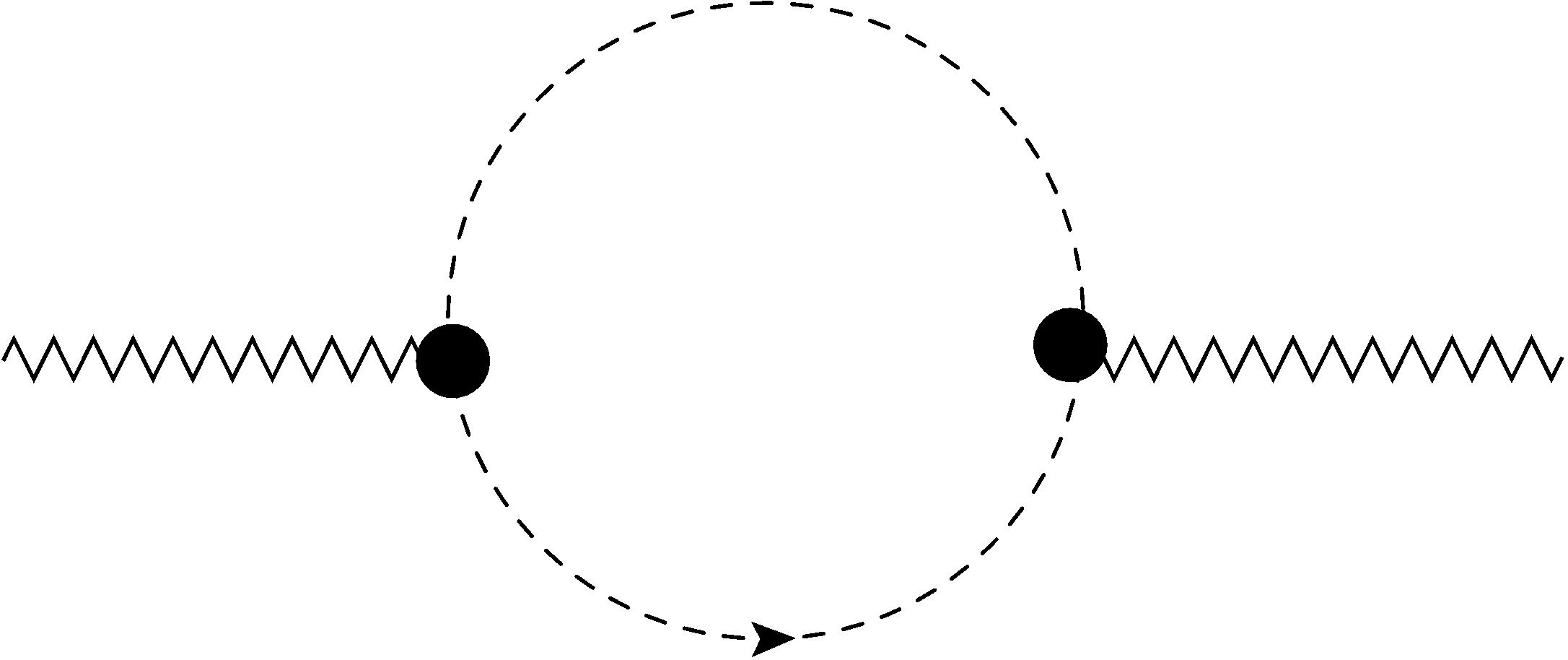}
\caption{The Feynman diagram corresponding to the two point correlation function of the stress-energy tensor. 
Zigzag lines represent external momenta associated with the stress-energy tensor insertions. Dashed lines represent the propagators of the scalar $\phi$ ``running'' in the loop. The expression corresponding to the vertexes is given in equation \eqref{eq:FeynmanVertexLifGen}. }
\label{fig:diagramLif2p}
\end{figure}

The propagator of the Lifshitz $z=2$ scalar is given by:
\begin{equation}
\left\langle {\phi \phi } \right\rangle \left( Q \right) = - \frac{i}{i\epsilon+m^2 + \kappa (\bar{Q}_{\alpha} \bar{Q}^{\alpha})^2 + (\hat{Q}_{\alpha} \hat{Q}^{\alpha})}.
\end{equation}
We denote the external momentum of the two point Feynman diagram by ${P_\mu } = \left( {{\hat{P}_{\mu }},{\bar{P}_\mu }} \right)$, and the ``running'' loop momentum by ${Q_\mu } = \left( {{\hat{Q}_{\mu }},{\bar{Q}_\mu }} \right)$. The expression for the two point function Feynman diagram, as illustrated in figure \ref{fig:diagramLif2p}, is given by:
\begin{equation}
\label{TwoPointFunctLifshitz}
\begin{aligned}
&\left\langle {{T^{\mu \nu }}\left( P \right){T^{\alpha \beta }}\left( { - P} \right)} \right\rangle  = \\
&\qquad -\frac{1}{2}\int {\frac{{{d^{{d_t}}}\hat{Q} }}{{{{\left( {2\pi } \right)}^{{d_t}}}}}} \int {\frac{{{d^{{d_s}}}\bar{Q}}}{{{{\left( {2\pi } \right)}^{{d_s}}}}}} \frac{{{V^{\mu \nu }}\left( {Q,P - Q} \right)}}{[i\epsilon+ m^2 + \kappa (\bar{Q}_{\alpha} \bar{Q}^{\alpha})^2 + (\hat{Q}_{\alpha} \hat{Q}^{\alpha})]}\\
& \qquad \qquad \qquad \qquad \cdot \frac{{{V^{\alpha \beta }}\left( {Q,P - Q} \right)}}{[i\epsilon+m^2 + \kappa ((\bar{Q}-\bar{P})_{\alpha} (\bar{Q}-\bar{P})^{\alpha})^2 + ((\hat{Q}-\hat{P})_\alpha (\hat{Q}-\hat{P})^{\alpha})]},
\end{aligned}
\end{equation}
where (as in the relativistic calculation) $m$ is an IR regulator later taken to be zero. 
\begin{figure}[h!]
\centering
\includegraphics[width=45mm]{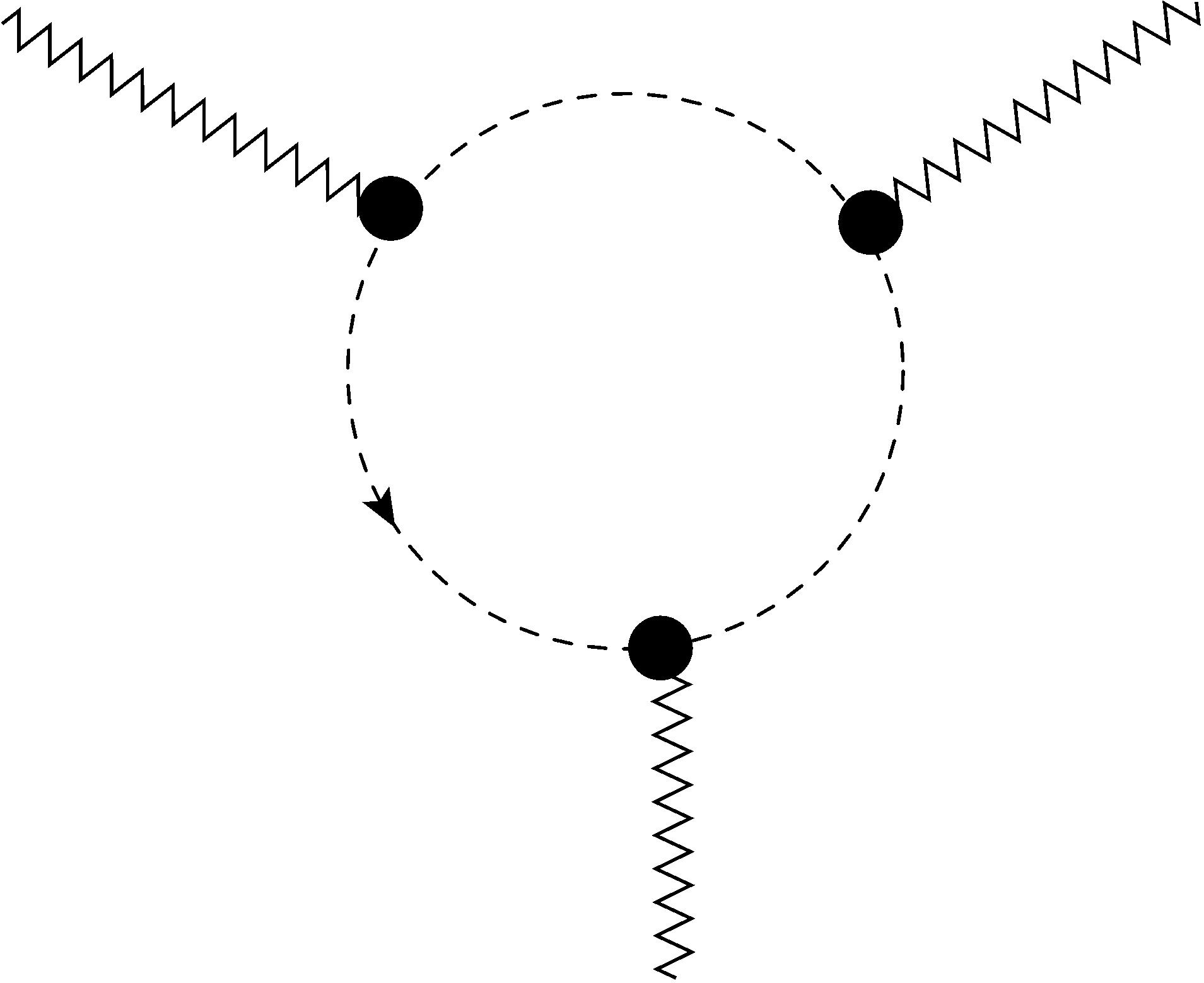}
\caption{The Feynman diagram corresponding to the three point correlation function of the stress-energy tensor. Zigzag lines represent external momenta associated with the stress-energy tensor insertions. Dashed lines represent the propagators of the scalar $\phi$ ``running'' in the loop. The expression corresponding to the vertexes is given in equation \eqref{eq:FeynmanVertexLifGen}. }
\label{fig:diagramLif3p}
\end{figure}
\begin{figure}[h!]
\centering
\includegraphics[width=50mm]{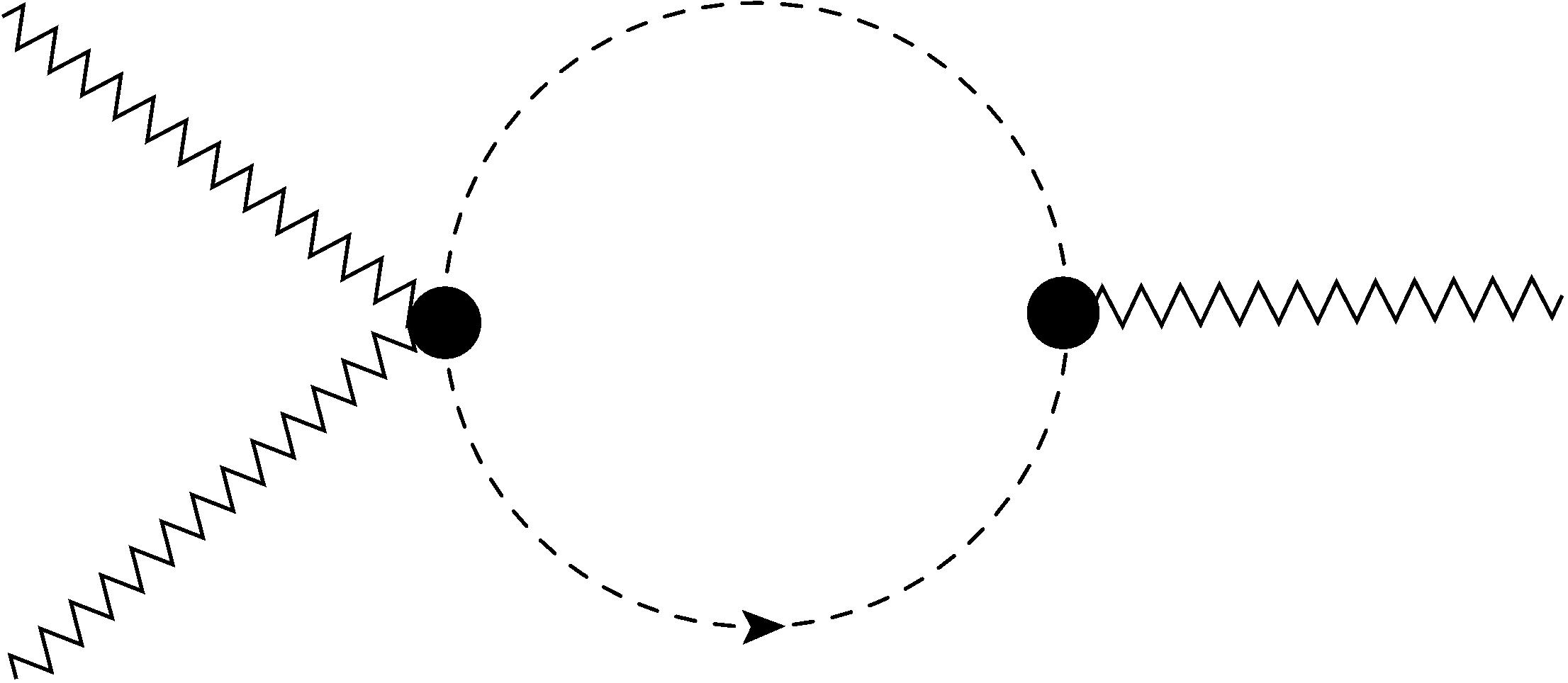}
\caption{The Feynman diagram corresponding to the correlation function \eqref{eq:ExprForDiagLong}.  Zigzag lines represent the external momenta. Dashed lines represent the propagators of the scalar $\phi$ ``running'' in the loop. The expression for the right vertex is given in equation \eqref{eq:FeynmanVertexLifGen}. The expression for the left vertex is constructed from several terms which are detailed in this appendix.}
\label{fig:diagram_lifLongVer}
\end{figure}

It is straightforward to calculate the expression corresponding to the vertex $V_{\mu\nu}$ from the stress-energy tensor given in \eqref{eq:StressTensorForPolesLif}. The result is:  
\begin{equation}
\begin{split}
\label{eq:FeynmanVertexLifGen}
&V^\alpha{}_\beta(P,K)= 
- \kappa \bar{K}_{\beta} \bar{K}_{\gamma} \bar{K}^{\gamma} \bar{P}^{\alpha} + \frac{(d_s - 2 d_t) \kappa \bar{K}_{\gamma} \bar{K}^{\gamma} \hat{K}_{\beta} \bar{P}^{\alpha}}{4 d_t} -  \kappa \bar{K}^{\alpha} \bar{K}_{\gamma} \bar{K}^{\gamma} \bar{P}_{\beta} -  \hat{K}^{\alpha} \bar{P}_{\beta}\\
&\qquad  + \frac{(d_s + 2 d_t) \kappa \bar{K}^{\alpha} \hat{K}_{\beta} \bar{P}_{\gamma} \bar{P}^{\gamma}}{4 d_t} -  \kappa \bar{K}_{\beta} \bar{P}^{\alpha} \bar{P}_{\gamma} \bar{P}^{\gamma} -  \kappa \hat{K}_{\beta} \bar{P}^{\alpha} \bar{P}_{\gamma} \bar{P}^{\gamma} -  \kappa \bar{K}^{\alpha} \bar{P}_{\beta} \bar{P}_{\gamma} \bar{P}^{\gamma}\\
&\qquad  + \kappa \bar{K}_{\gamma} \bar{K}^{\gamma} \bar{K}^{\delta} \bar{\delta}^{\alpha}{}_{\beta} \bar{P}_{\delta} -  \frac{(d_s - 2 d_t) \kappa \bar{K}_{\gamma} \bar{K}^{\gamma} \bar{K}^{\delta} \hat{\delta}^{\alpha}{}_{\beta} \bar{P}_{\delta}}{4 d_t} + \kappa \bar{K}_{\gamma} \bar{K}^{\gamma} \bar{\delta}^{\alpha}{}_{\beta} \bar{P}_{\delta} \bar{P}^{\delta} \\
&\qquad  -  \frac{d_s \kappa \bar{K}_{\gamma} \bar{K}^{\gamma} \hat{\delta}^{\alpha}{}_{\beta} \bar{P}_{\delta} \bar{P}^{\delta}}{2 d_t} + \kappa \bar{K}^{\gamma} \bar{\delta}^{\alpha}{}_{\beta} \bar{P}_{\gamma} \bar{P}_{\delta} \bar{P}^{\delta}-  \frac{(d_s - 2 d_t) \kappa \bar{K}^{\gamma} \hat{\delta}^{\alpha}{}_{\beta} \bar{P}_{\gamma} \bar{P}_{\delta} \bar{P}^{\delta}}{4 d_t}\\
&\qquad   -  \bar{K}_{\beta} \hat{P}^{\alpha} -  \hat{K}_{\beta} \hat{P}^{\alpha} -  \kappa \bar{K}^{\alpha} \bar{K}_{\gamma} \bar{K}^{\gamma} \hat{P}_{\beta} + \hat{K}^{\gamma} \hat{\delta}^{\alpha}{}_{\beta} \hat{P}_{\gamma}+ \hat{K}^{\gamma} \bar{\delta}^{\alpha}{}_{\beta} \hat{P}_{\gamma}\\
&\qquad  -  \hat{K}^{\alpha} \hat{P}_{\beta} + \frac{(d_s + 2 d_t) \kappa \bar{K}_{\gamma} \bar{K}^{\gamma} \bar{P}^{\alpha} \hat{P}_{\beta}}{4 d_t} + \frac{(d_s - 2 d_t) \kappa \bar{K}^{\alpha} \bar{P}_{\gamma} \bar{P}^{\gamma} \hat{P}_{\beta}}{4 d_t} .
\end{split}
\end{equation}
The expression \eqref{eq:FeynmanVertexLifGen} corresponds to the vertexes in figures \ref{fig:diagramLif2p} and \ref{fig:diagramLif3p}, and to the right vertex in figure \ref{fig:diagram_lifLongVer}.

The expression for the Feynman diagram corresponding to the three point function of the stress-energy tensor, as illustrated in figure \ref{fig:diagramLif3p}, is given by:
\begin{equation}
\label{eq:LifThreePointExprDiag}
\begin{aligned}
& \left\langle {{T^{\mu \nu }}\left( {L =  - P - K} \right){T^{\alpha \beta }}\left( P \right){T^{\rho \sigma }}\left( K \right)} \right\rangle  = \\
& \qquad i\int {\frac{{{d^{{d_t}}}\hat{Q} }}{{{{\left( {2\pi } \right)}^{{d_t}}}}}} \int {\frac{{{d^{{d_s}}}\bar{Q}}}{{{{\left( {2\pi } \right)}^{{d_s}}}}}} \, \frac{{{V^{\alpha \beta }}\left( {Q,P - Q} \right)}}{{[i\epsilon+{m^2} + \kappa {{({{\bar Q}_\alpha }{{\bar Q}^\alpha })}^2} + ({{\hat Q}_\alpha }{{\hat Q}^\alpha })]}}\\
& \qquad \qquad \qquad \qquad \, \cdot \frac{{{V^{\rho \sigma }}\left( { - P + Q,K + P - Q} \right)}}{{[{i\epsilon+m^2} + \kappa {{({{(\bar Q - \bar P)}_\alpha }{{(\bar Q - \bar P)}^\alpha })}^2} + ((\hat Q - \hat P)_\alpha {{(\hat Q - \hat P)}^\alpha })]}}\\
& \cdot \frac{{{V^{\mu \nu }}\left( { - K - P + Q, - Q} \right)}}{{[i\epsilon+{m^2} + \kappa {{({{(\bar Q - \bar P - \bar K)}_\alpha }{{(\bar Q - \bar P - \bar K)}^\alpha })}^2} + ((\hat Q - \hat P - \hat K)_\alpha {{(\hat Q - \hat P - \hat K)}^\alpha })]}}.
\end{aligned}
\end{equation}
The expression for the Feynman diagram corresponding to the correlation function \eqref{eq:LifshitzTwoPointLikeDiagExpr}, as illustrated in figure \ref{fig:diagram_lifLongVer}, is given by:
\begin{equation}
\label{eq:ExprForDiagLong}
\begin{aligned}
&\left\langle\frac{{{\delta ^2}S}}{{\delta \ve{b}{\alpha} \delta \ve{c}{\gamma}}}(P,K) \frac{\delta S}{\delta \ve{a}{\mu}}(-P-K) \right\rangle = \\
&\qquad -\frac{1}{2}\int {\frac{{{d^{{d_t}}}\hat{Q} }}{{{{\left( {2\pi } \right)}^{{d_t}}}}}} \int {\frac{{{d^{{d_s}}}\bar{Q}}}{{{{\left( {2\pi } \right)}^{{d_s}}}}}} \frac{{{{\left[ {{\delta ^2}S} \right]^{\alpha\gamma}_{bc}}}\left( {P,K,P+K-Q,Q} \right)}}{[i\epsilon+ m^2 + \kappa (\bar{Q}_{\alpha} \bar{Q}^{\alpha})^2 + (\hat{Q}_{\alpha} \hat{Q}^{\alpha})]}\\
&  \cdot \frac{{{V^{\mu}{}_a}\left( {P+K-Q, Q} \right)}}{[i\epsilon+m^2 + \kappa ((\bar{Q}-\bar{P}-\bar{K})_{\alpha} (\bar{Q}-\bar{P}-\bar{K})^{\alpha})^2 + ((\hat{Q}-\hat{P}-\hat{K})_\alpha (\hat{Q}-\hat{P}-\hat{K})^{\alpha})]},
\end{aligned}
\end{equation}
where the expression ${\left[ {{\delta ^2}S} \right]_{\alpha \beta \gamma \delta }}\left( {P,K,Q,L} \right)\equiv {\delta_{\alpha\alpha'}\delta_{\gamma\gamma'}\delta^b_\beta\delta^c_\delta}{{{\left[ {{\delta ^2}S} \right]^{\alpha'\gamma'}_{bc}}}\left( {P,K,P+K-Q,Q} \right)}$ corresponds to the insertion of the second variation of the action $ \frac{{{\delta ^2}S}}{{\delta \ve{b}{\alpha} \delta \ve{c}{\gamma}}} $ in flat space. $P$ and $K$ in this expression refer to the external momenta, whereas the two momenta $Q$ and $L$ correspond to the two scalar propagator lines attached to the vertex, as illustrated in figure \ref{fig:diagram_lifLongVer}.

The expression for the vertex ${\left[ {{\delta ^2}S} \right]_{\alpha \beta \gamma \delta }}\left( {P,K,Q,L} \right)$ is quite long, and the remainder of this appendix is dedicated to the details of its calculation. 
For convenience, we split it into several parts as follows.
Ignoring the terms of order $O(\epslif)$, the action \eqref{eq:GeneralActionLifs} takes the form:
\begin{equation}
\label{eq:ActionLifShortVer}
S = \int {{d^{{d_t+d_s}}}x}\left( e \right)\left( {\frac{1}{2}{{\left( {{{ \mathcal{L}}_n}^{\left( i \right)}\phi } \right)}^2} - \frac{\kappa }{2}{{\left( {{{\bar \nabla }^2}\phi  + \left( {\frac{{{d_t} - 1}}{{{d_t}}}} \right){a^\mu }{{\bar \nabla }_\mu }\phi } \right)}^2}} \right).
\end{equation}
Let us define:
\begin{equation}
\label{eq:S1Def}
{S_I} \equiv \int {{d^{{d_t+d_s}}}x}\left( e \right)\left( {\frac{1}{2}{{\left( {{{ \mathcal{L}}_n}^{\left( i \right)}\phi } \right)}^2}} \right),
\end{equation}
and 
\begin{equation}
\label{eq:S2Def}
{S_{II}} \equiv  - \frac{\kappa }{2}\int {{d^{{d_t+d_s}}}x} \left( e \right){\left( {{{\bar \nabla }^2}\phi  + \left( {\frac{{{d_t} - 1}}{{{d_t}}}} \right){a^\mu }{{\bar \nabla }_\mu }\phi } \right)^2},
\end{equation}
so that:
\begin{equation}
S = {S_I} + {S_{II}}.
\end{equation}

The contribution of the first part of the action \eqref{eq:S1Def} to the vertex ${\left[ {{\delta ^2}S} \right]_{\alpha \beta \gamma \delta }}\left( {P,K,Q,L} \right)$ can be easily derived:
\begin{equation}
\label{eq:LongVertexS1}
\begin{aligned}
&{\left[ {{\delta ^2}{S_I}} \right]_{\alpha \beta \gamma \delta }}\left( {P,K,Q,L} \right)= \\
&\qquad \tfrac{1}{2} \bigl(- \delta_{\gamma \delta} \hat{L}_{\alpha} Q_{\beta} + \delta_{\alpha \delta} \hat{L}_{\gamma} Q_{\beta} + L_{\delta} \hat{\delta}_{\alpha \gamma} Q_{\beta}\\
&\qquad  + \delta_{\gamma \beta} \hat{L}_{\alpha} Q_{\delta} -  \delta_{\alpha \beta} \hat{L}_{\gamma} Q_{\delta} + L_{\beta} \hat{\delta}_{\alpha \gamma} Q_{\delta} -  \delta_{\gamma \delta} L_{\beta} \hat{Q}_{\alpha}\\
&\qquad  + \delta_{\gamma \beta} L_{\delta} \hat{Q}_{\alpha} + \delta_{\alpha \delta} L_{\beta} \hat{Q}_{\gamma} -  \delta_{\alpha \beta} L_{\delta} \hat{Q}_{\gamma} -  \tfrac{1}{2} (\delta_{\alpha \delta} \delta_{\gamma \beta}\\
&\qquad  -  \delta_{\alpha \beta} \delta_{\gamma \delta}) \hat{L}_{\mu} \hat{Q}^{\mu} -  \tfrac{1}{2} (\delta_{\alpha \delta} \delta_{\gamma \beta}\\
&\qquad  -  \delta_{\alpha \beta} \delta_{\gamma \delta}) \hat{L}^{\nu} \hat{Q}_{\nu}\bigr).
\end{aligned}
\end{equation}

For the contribution of the second part of the action \eqref{eq:S2Def}, it is convenient to use the following identity (which follows from \eqref{AppNonRelCurvedMultTime:TotDerFormula}):
\begin{equation}
\label{eq:LaplacianTrick}
{{\bar \nabla }^2}\phi  = \nabla_\mu \wn^\mu \phi - {a^\nu }{{\bar \nabla }_\nu }\phi .
\end{equation}
$S_{II}$ can then be written as follows:
\begin{equation}
\label{eq:S1OurSigns}
{S_{II}} = 
 - \frac{\kappa }{2}\int {{d^{{d_t+d_s}}}x} \left( e \right) I_3^2 \equiv 
 - \frac{\kappa }{2}\int {{d^{{d_t+d_s}}}x} \left( e \right)\left( {{I_1} - \frac{1}{{{d_t}}}{I_2}} \right)^2,
\end{equation}
where $I_1$, $I_2$ and $I_3$ are defined by:
\begin{equation}
\label{eq:I1DefLifAction}
{I_1} \equiv \nabla_\mu \wn^\mu \phi = \frac{1}{e}{\partial _\mu }\left( {e{P^{ab}}{e^\mu }_a{e^\nu }_b{\partial _\nu }\phi } \right),
\end{equation}
\begin{equation}
\label{eq:I2DefLifAction}
{I_2} \equiv {a^\nu }{{\bar \nabla }_\nu }\phi ,
\end{equation}
and
\begin{equation}
I_3 \equiv I_1 - \frac{1}{d_t} I_2 .
\end{equation}
The second variation of the expression $S_{II}$ with respect to the vielbeins, evaluated in flat space, can be decomposed as follows:
\begin{equation}
\label{eq:DecompOfSII}
\begin{aligned}
&\left.\frac{\delta^2S_{II}}{\delta\ve{b}{\alpha}(y)\delta\ve{c}{\gamma}(z)}\right|_\text{flat space}=\\
& \qquad -\frac{\kappa}{2}\int d^{d_t+d_s}x \left[ \frac{\delta^2 e(x)}{\delta\ve{b}{\alpha}(y)\delta\ve{c}{\gamma}(z)}\left(\bar\partial_{\mu} \bar\partial^{\mu}\phi(x)\right)^2 \right. \\
&\qquad \qquad \qquad \qquad \quad \left. + 2\frac{\delta e(x)}{\delta\ve{b}{\alpha}(y)}\left(\bar\partial_{\mu} \bar\partial^{\mu}\phi(x)\right)\frac{\delta I_3(x)}{\delta\ve{c}{\gamma}(z)}  \right.\\
& \qquad \qquad \qquad \qquad \quad \left. + 2\frac{\delta e(x)}{\delta\ve{c}{\gamma}(z)}\left(\bar\partial_{\mu} \bar\partial^{\mu}\phi(x)\right)\frac{\delta I_3(x)}{\delta\ve{b}{\alpha}(y)}  +2\frac{\delta I_3(x)}{\delta\ve{b}{\alpha}(y)} \frac{\delta I_3(x)}{\delta\ve{c}{\gamma}(z)} \right.\\
& \qquad \qquad \qquad \qquad \quad\left. +2\left(\bar\partial_{\mu} \bar\partial^{\mu}\phi(x)\right)\frac{\delta^2 I_3(x)}{\delta\ve{b}{\alpha}(y)\delta\ve{c}{\gamma}(z)}   \right],
\end{aligned}
\end{equation}
where all variations with respect to the vielbeins in this expression are evaluated at flat space.
Each instance of the first order variations of $I_3$ that appear in the second, third and fourth terms in \eqref{eq:DecompOfSII} contributes an expression of the following form to the vertex:
\begin{equation}
\label{eq:FirstVarI3}
{\left[ {\delta \left( {{I_3}} \right)} \right]_{\alpha \beta }}\left( {P,Q} \right) 
\equiv - \bar{P}_{\alpha} Q_{\beta} -  P_{\beta} \bar{Q}_{\alpha} + \frac{\hat{P}_{\beta} \bar{Q}_{\alpha}}{d_t} + 2 Q_{\beta} \bar{Q}_{\alpha} + \delta_{\alpha \beta} \bar{P}^{\gamma} \bar{Q}_{\gamma} -  \frac{\hat{\delta}_{\alpha \beta} \bar{P}^{\gamma} \bar{Q}_{\gamma}}{d_t},
\end{equation}
where the momentum $P$ here refers to one of the external momenta associated with the vertex, and $Q$ to the momentum of one of the scalar propagator lines attached to it. The contribution of the expression $ \int d^{d_t+d_s}x \, \bar\partial_{\mu} \bar\partial^{\mu}\phi(x) \frac{\delta^2 I_1(x)}{\delta\ve{b}{\alpha}(y)\delta\ve{c}{\gamma}(z)}  $ (contained in the last term of \eqref{eq:DecompOfSII}) to the vertex is given by:
\begin{align*}
&{\left[ {{\delta ^2}{\tilde{I}_1}} \right]_{\alpha \beta \gamma \delta }}\left( {P,K,Q,L} \right) \equiv\\
& \qquad \tfrac{1}{2} \biggl(- \Bigl(-2 (- \delta_{\gamma \delta} L_{\beta} \bar{L}_{\alpha} + \delta_{\gamma \beta} L_{\delta} \bar{L}_{\alpha}) - 2 (\delta_{\alpha \delta} L_{\beta} \bar{L}_{\gamma} -  \delta_{\alpha \beta} L_{\delta} \bar{L}_{\gamma}) -  (- \delta_{\alpha \delta} \delta_{\gamma \beta}\\
&\qquad  + \delta_{\alpha \beta} \delta_{\gamma \delta}) \bar{L}_{\rho} \bar{L}^{\rho} -  (\delta_{\alpha \delta} \delta_{\gamma \beta} + \delta_{\alpha \beta} \delta_{\gamma \delta}) \bar{L}_{\rho} \bar{L}^{\rho} + \delta_{\gamma \delta} (-2 L_{\beta} \bar{L}_{\alpha} + \delta_{\alpha \beta} \bar{L}_{\rho} \bar{L}^{\rho})\\
&\qquad  + \delta_{\alpha \beta} (-2 L_{\delta} \bar{L}_{\gamma} + \delta_{\gamma \delta} \bar{L}_{\rho} \bar{L}^{\rho}) - 2 L_{\beta} L_{\delta} \bar{\delta}_{\alpha \gamma} -  \delta_{\gamma \delta} P_{\rho} (- \delta_{\beta}{}^{\rho} \bar{L}_{\alpha} + \delta_{\alpha \beta} \bar{L}^{\rho}\\
&\qquad  -  L_{\beta} \bar{\delta}^{\rho}{}_{\alpha}) -  \delta_{\alpha \beta} K_{\rho} (- \delta_{\delta}{}^{\rho} \bar{L}_{\gamma} + \delta_{\gamma \delta} \bar{L}^{\rho} -  L_{\delta} \bar{\delta}^{\rho}{}_{\gamma}) + (K_{\rho} + P_{\rho}) \bigl((\delta_{\delta}{}^{\rho} \delta_{\gamma \beta}\\
&\qquad  -  \delta_{\beta}{}^{\rho} \delta_{\gamma \delta}) \bar{L}_{\alpha} + (- \delta_{\delta}{}^{\rho} \delta_{\alpha \beta} + \delta_{\beta}{}^{\rho} \delta_{\alpha \delta}) \bar{L}_{\gamma} + (- \delta_{\alpha \delta} \delta_{\gamma \beta} + \delta_{\alpha \beta} \delta_{\gamma \delta}) \bar{L}^{\rho}\\
&\qquad  + (\delta_{\delta}{}^{\rho} L_{\beta} + \delta_{\beta}{}^{\rho} L_{\delta}) \bar{\delta}_{\gamma \alpha} + (- \delta_{\gamma \delta} L_{\beta} + \delta_{\gamma \beta} L_{\delta}) \bar{\delta}^{\rho}{}_{\alpha} + (\delta_{\alpha \delta} L_{\beta}\\
&\qquad  -  \delta_{\alpha \beta} L_{\delta}) \bar{\delta}^{\rho}{}_{\gamma}\bigr)\Bigr) \bar{Q}_{\mu} \bar{Q}^{\mu} -  \bar{L}_{\mu} \bar{L}^{\mu} \Bigl(-2 \bar{\delta}_{\alpha \gamma} Q_{\beta} Q_{\delta} - 2 (- \delta_{\gamma \delta} Q_{\beta} \bar{Q}_{\alpha} + \delta_{\gamma \beta} Q_{\delta} \bar{Q}_{\alpha})\numthis\label{eq:SecVarI1}\\
&\qquad  - 2 (\delta_{\alpha \delta} Q_{\beta} \bar{Q}_{\gamma} -  \delta_{\alpha \beta} Q_{\delta} \bar{Q}_{\gamma}) -  (- \delta_{\alpha \delta} \delta_{\gamma \beta} + \delta_{\alpha \beta} \delta_{\gamma \delta}) \bar{Q}_{\nu} \bar{Q}^{\nu} -  (\delta_{\alpha \delta} \delta_{\gamma \beta}\\
&\qquad  + \delta_{\alpha \beta} \delta_{\gamma \delta}) \bar{Q}_{\nu} \bar{Q}^{\nu} -  \delta_{\gamma \delta} P_{\nu} (- \bar{\delta}^{\nu}{}_{\alpha} Q_{\beta} -  \delta_{\beta}{}^{\nu} \bar{Q}_{\alpha} + \delta_{\alpha \beta} \bar{Q}^{\nu}) -  \delta_{\alpha \beta} K_{\nu} (- \bar{\delta}^{\nu}{}_{\gamma} Q_{\delta}\\
&\qquad  -  \delta_{\delta}{}^{\nu} \bar{Q}_{\gamma} + \delta_{\gamma \delta} \bar{Q}^{\nu}) + (K_{\nu} + P_{\nu}) \bigl(\bar{\delta}_{\gamma \alpha} (\delta_{\delta}{}^{\nu} Q_{\beta} + \delta_{\beta}{}^{\nu} Q_{\delta}) + \bar{\delta}^{\nu}{}_{\gamma} (\delta_{\alpha \delta} Q_{\beta} -  \delta_{\alpha \beta} Q_{\delta})\\
&\qquad  + \bar{\delta}^{\nu}{}_{\alpha} (- \delta_{\gamma \delta} Q_{\beta} + \delta_{\gamma \beta} Q_{\delta}) + (\delta_{\delta}{}^{\nu} \delta_{\gamma \beta} -  \delta_{\beta}{}^{\nu} \delta_{\gamma \delta}) \bar{Q}_{\alpha} + (- \delta_{\delta}{}^{\nu} \delta_{\alpha \beta}\\
&\qquad  + \delta_{\beta}{}^{\nu} \delta_{\alpha \delta}) \bar{Q}_{\gamma} + (- \delta_{\alpha \delta} \delta_{\gamma \beta} + \delta_{\alpha \beta} \delta_{\gamma \delta}) \bar{Q}^{\nu}\bigr) + \delta_{\gamma \delta} (-2 Q_{\beta} \bar{Q}_{\alpha} + \delta_{\alpha \beta} \bar{Q}_{\nu} \bar{Q}^{\nu})\\
&\qquad  + \delta_{\alpha \beta} (-2 Q_{\delta} \bar{Q}_{\gamma} + \delta_{\gamma \delta} \bar{Q}_{\nu} \bar{Q}^{\nu})\Bigr)\biggr).
\end{align*}
Finally, the contribution of the expression $ \int d^{d_t+d_s}x \, \bar\partial_{\mu} \bar\partial^{\mu}\phi(x) \frac{\delta^2 I_2(x)}{\delta\ve{b}{\alpha}(y)\delta\ve{c}{\gamma}(z)}  $ (also contained in the last term of \eqref{eq:DecompOfSII}) to the vertex is given by:
\begin{align*}
&{\left[ {{\delta ^2}{\tilde{I}_2}} \right]_{\alpha \beta \gamma \delta }}\left( {P,K,Q,L} \right) \equiv\\
& \qquad \tfrac{1}{2} \biggl(- \Bigl(- (\delta_{\rho \alpha} \bar{L}_{\beta} -  L_{\beta} \bar{\delta}_{\rho \alpha}) (\delta_{\gamma}{}^{\rho} \hat{K}_{\delta} -  K^{\rho} \hat{\delta}_{\gamma \delta}) -  (\delta_{\rho \gamma} \bar{L}_{\delta}\\
&\qquad  -  L_{\delta} \bar{\delta}_{\rho \gamma}) (- P^{\rho} \hat{\delta}_{\alpha \beta} + \delta_{\alpha}{}^{\rho} \hat{P}_{\beta}) + i \bar{L}_{\rho} \bigl(i (- \delta_{\gamma}{}^{\rho} K_{\beta} \hat{\delta}_{\alpha \delta} + \delta_{\gamma \beta} K^{\rho} \hat{\delta}_{\alpha \delta}\\
&\qquad  -  \delta_{\alpha}{}^{\rho} P_{\delta} \hat{\delta}_{\gamma \beta} + \delta_{\alpha \delta} P^{\rho} \hat{\delta}_{\gamma \beta}) - i (\delta_{\beta}{}^{\rho} \delta_{\gamma \alpha} \hat{K}_{\delta} + \delta_{\alpha}{}^{\rho} \delta_{\gamma \beta} \hat{K}_{\delta}\\
&\qquad  -  \delta_{\beta}{}^{\rho} K_{\alpha} \hat{\delta}_{\gamma \delta} -  \delta_{\alpha}{}^{\rho} K_{\beta} \hat{\delta}_{\gamma \delta}) - i (- \delta_{\delta}{}^{\rho} P_{\gamma} \hat{\delta}_{\alpha \beta} -  \delta_{\gamma}{}^{\rho} P_{\delta} \hat{\delta}_{\alpha \beta}\\
&\qquad  + \delta_{\gamma}{}^{\rho} \delta_{\alpha \delta} \hat{P}_{\beta} + \delta_{\delta}{}^{\rho} \delta_{\gamma \alpha} \hat{P}_{\beta})\bigr)\Bigr) \bar{Q}_{\mu} \bar{Q}^{\mu} -  \bar{L}_{\mu} \bar{L}^{\mu} \Bigl(- (\delta_{\gamma}{}^{\nu} \hat{K}_{\delta}\numthis \label{eq:SecVarI2}\\
&\qquad  -  K^{\nu} \hat{\delta}_{\gamma \delta}) (- \bar{\delta}_{\nu \alpha} Q_{\beta} + \delta_{\nu \alpha} \bar{Q}_{\beta}) -  (- P^{\nu} \hat{\delta}_{\alpha \beta} + \delta_{\alpha}{}^{\nu} \hat{P}_{\beta}) (- \bar{\delta}_{\nu \gamma} Q_{\delta} + \delta_{\nu \gamma} \bar{Q}_{\delta})\\
&\qquad  + i \bigl(i (- \delta_{\gamma}{}^{\nu} K_{\beta} \hat{\delta}_{\alpha \delta} + \delta_{\gamma \beta} K^{\nu} \hat{\delta}_{\alpha \delta} -  \delta_{\alpha}{}^{\nu} P_{\delta} \hat{\delta}_{\gamma \beta} + \delta_{\alpha \delta} P^{\nu} \hat{\delta}_{\gamma \beta})\\
&\qquad  - i (\delta_{\beta}{}^{\nu} \delta_{\gamma \alpha} \hat{K}_{\delta} + \delta_{\alpha}{}^{\nu} \delta_{\gamma \beta} \hat{K}_{\delta} -  \delta_{\beta}{}^{\nu} K_{\alpha} \hat{\delta}_{\gamma \delta} -  \delta_{\alpha}{}^{\nu} K_{\beta} \hat{\delta}_{\gamma \delta})\\
&\qquad  - i (- \delta_{\delta}{}^{\nu} P_{\gamma} \hat{\delta}_{\alpha \beta} -  \delta_{\gamma}{}^{\nu} P_{\delta} \hat{\delta}_{\alpha \beta} + \delta_{\gamma}{}^{\nu} \delta_{\alpha \delta} \hat{P}_{\beta}\\
&\qquad  + \delta_{\delta}{}^{\nu} \delta_{\gamma \alpha} \hat{P}_{\beta})\bigr) \bar{Q}_{\nu}\Bigr)\biggr).
\end{align*}

Using the expressions \eqref{eq:FirstVarI3}--\eqref{eq:SecVarI2} and the standard relations:
\begin{align}
{\left. {\frac{{\delta e\left( x \right)}}{{\delta {e^a}_\mu \left( {y} \right)}}} \right|_{\text{flat space}}} &= {\delta _a}^\mu \delta \left( {x - y} \right),\\
{\left. {\frac{{{\delta ^2}e\left( x \right)}}{{\delta {e^a}_\mu \left( {{y}} \right)\delta {e^b}_\rho \left( {{z}} \right)}}} \right|_{\text{flat space}}} &= \left( {{\delta ^\mu }_a{\delta ^\rho }_b - {\delta ^\mu }_b{\delta ^\rho }_a} \right)\delta \left( {x - y} \right)\delta \left( {x - z} \right),
\end{align}
the total contribution of $ S_{II} $ to the vertex ${\left[ {{\delta ^2}S} \right]_{\alpha \beta \gamma \delta }}\left( {P,K,Q,L} \right)$ can be assembled as follows:
\begin{equation}
\begin{aligned}
&{\left[ {{\delta ^2}{S_{II}}} \right]_{\alpha \beta \gamma \delta }}\left( {P,K,Q,L} \right)= \\
& \qquad \tfrac{1}{2} \kappa \delta_{\alpha \delta} \delta_{\beta \gamma} \bar{L}_{\mu} \bar{L}^{\mu} \bar{Q}_{\nu} \bar{Q}^{\nu} -  \tfrac{1}{2} \kappa \delta_{\alpha \beta} \delta_{\gamma \delta} \bar{L}_{\mu} \bar{L}^{\mu} \bar{Q}_{\nu} \bar{Q}^{\nu}\\
& \qquad +\frac{\kappa}{2}\left( \bar{Q}_{\nu}\bar{Q}^{\nu}{\left[ {\delta \left( {{I_3}} \right)} \right]_{\gamma \delta }}\left( {K,L} \right)+\bar{L}_\nu\bar{L}^\nu{\left[ {\delta \left( {{I_3}} \right)} \right]_{\gamma \delta }}\left( {K,L} \right)\right)\delta_{\alpha\beta}\\
& \qquad +\frac{\kappa}{2}\left( \bar{Q}_{\nu}\bar{Q}^{\nu}{\left[ {\delta \left( {{I_3}} \right)} \right]_{\alpha \beta }}\left( {P,L} \right)+\bar{L}_\nu\bar{L}^\nu{\left[ {\delta \left( {{I_3}} \right)} \right]_{\alpha \beta }}\left( {P,L} \right)\right)\delta_{\gamma\delta}\\
& \qquad -\frac{\kappa}{2}\left( {\left[ {\delta \left( {{I_3}} \right)} \right]_{\alpha \beta }}\left( {P,Q} \right){\left[ {\delta \left( {{I_3}} \right)} \right]_{\gamma \delta }}\left( {K,L} \right) +{\left[ {\delta \left( {{I_3}} \right)} \right]_{\alpha \beta }}\left( {P,L} \right){\left[ {\delta \left( {{I_3}} \right)} \right]_{\gamma \delta }}\left( {K,Q} \right)\right)\\
&\qquad -\kappa{\left[ {{\delta ^2}{\tilde{I}_1}} \right]_{\alpha \beta \gamma \delta }}\left( {P,K,Q,L} \right) +\frac{\kappa}{d_t}{\left[ {{\delta ^2}{\tilde{I}_2}} \right]_{\alpha \beta \gamma \delta }}\left( {P,K,Q,L} \right).
\end{aligned}
\end{equation}

\section{Variations of the Anomaly and Trivial Term Densities}
\label{SecVarPhis}

In \cite{Arav:2016xjc}, the possible forms of Lifshitz scale anomalies (and the cohomologically trivial terms) were derived for the case of $2+1$ dimensions with a dynamical critical exponent of $z=2$, using a cohomological formulation of the Wess-Zumino consistency condition in curved spacetime. In this appendix we detail the first and second order variations of these anomaly and trivial term densities\footnote{We use the results of the more general non-Frobenius case (See \cite{Arav:2016xjc} for details and discussion).} with respect to the background vielbeins, evaluated in flat spacetime. These variations are used in conjunction with the anomalous Ward identities for the flat space correlation functions given in \eqref{CorrLifshitzWard:WardIdentForVarCorrWeyl2Point}--\eqref{CorrLifshitzWard:WardIdentForVarCorrWeyl3Point} to extract the anomaly and trivial term coefficients in section \ref{Sec:z2LishitzFreeScalarResults}. The appendix is divided into two parts, for the two and four derivatives sectors respectively. 

Throughout this appendix we use the following notation for the Fourier transformed variation of some scalar $X$ with respect to the vielbeins:
\begin{equation}
\label{eq:DefFirstVarFT}
(2\pi)^{d_t+d_s} \delta(-P-Q) 
{\left[ {\delta X} \right]_{\alpha \beta }}\left( P \right) \equiv {\delta ^b_\beta} {\delta _{\alpha \alpha '}}\mathcal{FT}\left[ {\frac{{\delta X(x)}}{{\delta \ve{b}{\alpha'}\left( {{y}} \right)}}} \right],
\end{equation}
\begin{equation}
\label{eq:DefSecVarFT}
(2\pi)^{d_t+d_s} \delta(-P-K-Q)
{\left[ {{\delta ^2}X} \right]_{\alpha \beta \gamma \delta }}\left( {P,K} \right) \equiv {\delta _{\alpha \alpha '}}{\delta _{\gamma \gamma '}}{\delta ^b_\beta} {\delta ^c_\delta}\, \mathcal{FT}\left[ {\frac{{{\delta ^2}X(x)}}{{\delta \ve{b}{\alpha'}\left( {{y}} \right)\delta \ve{c}{\gamma'}\left( {{z}} \right)}}} \right],
\end{equation}
where the variations are evaluated in flat space, and the Fourier transforms use the conventions \eqref{eq:FourierLifTwoPoint}--\eqref{eq:FourierLifThreePoint}. Note that the expressions in this appendix are all in $d_t=1$ time dimension, and therefore the time components of the momenta are given by:
\begin{equation}
\hat P_\mu = -\hat P n_\mu, \qquad \hat K_\mu = -\hat K n_\mu,
\end{equation}
where $ \hat P \equiv P_\mu n^\mu $ and $ \hat K \equiv K_\mu n^\mu $, as explained in appendix \ref{LifshitzNotation}.

\subsection{The Two Derivatives Sector} \label{TwoDerSecFTVars}

Taken from \cite{Arav:2016xjc}, this sector contains the following scalar terms: 
\begin{equation}
\label{eq:FPDTwoDerSec}
{\phi _1} = \tr\left( {{K_S^2}} \right), \quad {\phi _2} = {K_S^2}, \quad {\phi _3} = \mathcal{L}_nK_S.
\end{equation}
Out of these terms, the only one which is first order in the background fields is $\phi_3$, whose first order variation with respect to the vielbeins (evaluated in flat space and Fourier transformed) is given by:
\begin{equation}
\label{eq:FirstVarPhi3t}
\begin{aligned}
{\left[ \delta \phi_3 \right]_{\alpha \beta}} = n_{\alpha} \bar{P}_{\beta} \hat{P} -  \bar{\delta}_{\alpha \beta} \hat{P}^2.
\end{aligned}
\end{equation}

The second order variations of these terms with respect to the vielbeins (evaluated in flat space and Fourier transformed) are as follows:
\begin{align}
\begin{split}
\label{eq:SecVarPhi1t}
&{\left[ {{\delta ^2}\phi _1} \right]_{\alpha \beta \gamma \delta }}\left( {P,K} \right) = \hat{K} n_{\alpha} \bar{\delta}_{\beta \delta} \bar{P}_{\gamma} -  \bar{K}_{\beta} n_{\alpha} n_{\gamma} \bar{P}_{\delta} + \hat{K} n_{\alpha} \bar{\delta}_{\beta \gamma} \bar{P}_{\delta} -  \bar{K}^{\mu} n_{\alpha} n_{\gamma} \bar{\delta}_{\beta \delta} \bar{P}_{\mu}\\
& \qquad  + \bar{K}_{\beta} n_{\gamma} \bar{\delta}_{\alpha \delta} \hat{P} -  \hat{K} \bar{\delta}_{\alpha \delta} \bar{\delta}_{\beta \gamma} \hat{P} + \bar{K}_{\alpha} n_{\gamma} \bar{\delta}_{\beta \delta} \hat{P} -  \hat{K} \bar{\delta}_{\alpha \gamma} \bar{\delta}_{\beta \delta} \hat{P},
\end{split}\\
\begin{split}
\label{eq:SecVarPhi2t}
&{\left[ {{\delta ^2}\phi _2} \right]_{\alpha \beta \gamma \delta }}\left( {P,K} \right)=-2 \bar{K}_{\delta} n_{\alpha} n_{\gamma} \bar{P}_{\beta} + 2 \hat{K} n_{\alpha} \bar{\delta}_{\gamma \delta} \bar{P}_{\beta} + 2 \bar{K}_{\delta} n_{\gamma} \bar{\delta}_{\alpha \beta} \hat{P} - 2 \hat{K} \bar{\delta}_{\alpha \beta} \bar{\delta}_{\gamma \delta} \hat{P},\\
\end{split}\\
\begin{split}
\label{eq:SecVarPhi3t}
&{\left[ {{\delta ^2}\phi _3} \right]_{\alpha \beta \gamma \delta }}\left( {P,K} \right) =- \bar{K}_{\beta} \bar{K}_{\delta} n_{\alpha} n_{\gamma} + 2 \bar{K}_{\delta} \hat{K} n_{\alpha} n_{\beta} n_{\gamma} -  \bar{K}_{\beta} \hat{K} n_{\gamma} \bar{\delta}_{\alpha \delta}\\
&\qquad  + \hat{K}^2 \bar{\delta}_{\alpha \delta} \bar{\delta}_{\beta \gamma} + 2 \bar{K}_{\beta} \hat{K} n_{\alpha} \bar{\delta}_{\gamma \delta} - 2 \hat{K}^2 n_{\alpha} n_{\beta} \bar{\delta}_{\gamma \delta} + \hat{K} n_{\alpha} n_{\gamma} n_{\delta} \bar{P}_{\beta}\\
&\qquad  -  \hat{K} n_{\gamma} \bar{\delta}_{\alpha \delta} \bar{P}_{\beta} + \hat{K} n_{\gamma} \bar{\delta}_{\alpha \beta} \bar{P}_{\delta} -  \hat{K} n_{\alpha} \bar{\delta}_{\beta \gamma} \bar{P}_{\delta} -  n_{\alpha} n_{\gamma} \bar{P}_{\beta} \bar{P}_{\delta}\\
&\qquad  + \bar{K}_{\delta} n_{\alpha} n_{\beta} n_{\gamma} \hat{P} -  \hat{K} n_{\gamma} n_{\delta} \bar{\delta}_{\alpha \beta} \hat{P} -  \bar{K}_{\beta} n_{\gamma} \bar{\delta}_{\alpha \delta} \hat{P} -  \bar{K}_{\delta} n_{\alpha} \bar{\delta}_{\beta \gamma} \hat{P}\\
&\qquad  + 2 \hat{K} \bar{\delta}_{\alpha \delta} \bar{\delta}_{\beta \gamma} \hat{P} + \bar{K}_{\beta} n_{\alpha} \bar{\delta}_{\gamma \delta} \hat{P} -  \hat{K} n_{\alpha} n_{\beta} \bar{\delta}_{\gamma \delta} \hat{P} + 2 n_{\alpha} n_{\gamma} n_{\delta} \bar{P}_{\beta} \hat{P}\\
&\qquad  -  n_{\gamma} \bar{\delta}_{\alpha \delta} \bar{P}_{\beta} \hat{P} + 2 n_{\gamma} \bar{\delta}_{\alpha \beta} \bar{P}_{\delta} \hat{P} -  n_{\alpha} \bar{\delta}_{\beta \gamma} \bar{P}_{\delta} \hat{P} - 2 n_{\gamma} n_{\delta} \bar{\delta}_{\alpha \beta} \hat{P}^2\\
&\qquad  + \bar{\delta}_{\alpha \delta} \bar{\delta}_{\beta \gamma} \hat{P}^2 -  \bar{K}_{\delta} \hat{K} n_{\alpha} \bar{\delta}_{\beta \gamma}.
\end{split}
\end{align}
As found in \cite{Arav:2016xjc}, this sector contains a trivial term (coboundary) with the following density: 
\begin{equation}
\label{eq:F1tDef}
\mathcal{F}_1=-2\phi_2-2\phi_3,
\end{equation}
as well as an anomaly term with the following density: 
\begin{equation}
\label{eq:AnomalyTwoDerCurv}
{\mathcal{A}^{\left( {2,0,0} \right)}} = {\phi_1} - \frac{1}{2}{\phi_2}.
\end{equation}
The second variation of the trivial term $\mathcal{F}_1$ with respect to the vielbeins (evaluated in flat space and Fourier transformed) is therefore given by the following expression:
\begin{equation}
\label{eq:SecVarF1t}
\begin{aligned}
&{\left[ {{\delta ^2}\mathcal{F} _1} \right]_{\alpha \beta \gamma \delta }}\left( {P,K} \right) = 2 \bar{K}_{\beta} \bar{K}_{\delta} n_{\alpha} n_{\gamma} - 4 \bar{K}_{\delta} \hat{K} n_{\alpha} n_{\beta} n_{\gamma} + 2 \bar{K}_{\beta} \hat{K} n_{\gamma} \bar{\delta}_{\alpha \delta} \\
&\qquad  - 2 \hat{K}^2 \bar{\delta}_{\alpha \delta} \bar{\delta}_{\beta \gamma} - 4 \bar{K}_{\beta} \hat{K} n_{\alpha} \bar{\delta}_{\gamma \delta} + 4 \hat{K}^2 n_{\alpha} n_{\beta} \bar{\delta}_{\gamma \delta} + 4 \bar{K}_{\delta} n_{\alpha} n_{\gamma} \bar{P}_{\beta}\\
&\qquad  - 2 \hat{K} n_{\alpha} n_{\gamma} n_{\delta} \bar{P}_{\beta} + 2 \hat{K} n_{\gamma} \bar{\delta}_{\alpha \delta} \bar{P}_{\beta} - 4 \hat{K} n_{\alpha} \bar{\delta}_{\gamma \delta} \bar{P}_{\beta} - 2 \hat{K} n_{\gamma} \bar{\delta}_{\alpha \beta} \bar{P}_{\delta}\\
&\qquad  + 2 \hat{K} n_{\alpha} \bar{\delta}_{\beta \gamma} \bar{P}_{\delta} + 2 n_{\alpha} n_{\gamma} \bar{P}_{\beta} \bar{P}_{\delta} - 2 \bar{K}_{\delta} n_{\alpha} n_{\beta} n_{\gamma} \hat{P} - 4 \bar{K}_{\delta} n_{\gamma} \bar{\delta}_{\alpha \beta} \hat{P}\\
&\qquad  + 2 \hat{K} n_{\gamma} n_{\delta} \bar{\delta}_{\alpha \beta} \hat{P} + 2 \bar{K}_{\beta} n_{\gamma} \bar{\delta}_{\alpha \delta} \hat{P} + 2 \bar{K}_{\delta} n_{\alpha} \bar{\delta}_{\beta \gamma} \hat{P} - 4 \hat{K} \bar{\delta}_{\alpha \delta} \bar{\delta}_{\beta \gamma} \hat{P}\\
&\qquad  - 2 \bar{K}_{\beta} n_{\alpha} \bar{\delta}_{\gamma \delta} \hat{P} + 2 \hat{K} n_{\alpha} n_{\beta} \bar{\delta}_{\gamma \delta} \hat{P} + 4 \hat{K} \bar{\delta}_{\alpha \beta} \bar{\delta}_{\gamma \delta} \hat{P} - 4 n_{\alpha} n_{\gamma} n_{\delta} \bar{P}_{\beta} \hat{P}\\
&\qquad  + 2 n_{\gamma} \bar{\delta}_{\alpha \delta} \bar{P}_{\beta} \hat{P} - 4 n_{\gamma} \bar{\delta}_{\alpha \beta} \bar{P}_{\delta} \hat{P} + 2 n_{\alpha} \bar{\delta}_{\beta \gamma} \bar{P}_{\delta} \hat{P} + 4 n_{\gamma} n_{\delta} \bar{\delta}_{\alpha \beta} \hat{P}^2\\
&\qquad  - 2 \bar{\delta}_{\alpha \delta} \bar{\delta}_{\beta \gamma} \hat{P}^2 + 2 \bar{K}_{\delta} \hat{K} n_{\alpha} \bar{\delta}_{\beta \gamma},
\end{aligned}
\end{equation}
whereas the second variation of the anomaly density $\mathcal{A}^{\left( {2,0,0} \right)}$ is given by:
\begin{equation}
\label{eq:SecVarA200}
\begin{aligned}
&{\left[ {{\delta ^2}{\mathcal{A}^{\left( {2,0,0} \right)}}} \right]_{\alpha \beta \gamma \delta }}\left( {P,K} \right) =\bar{K}_{\delta} n_{\alpha} n_{\gamma} \bar{P}_{\beta} -  \hat{K} n_{\alpha} \bar{\delta}_{\gamma \delta} \bar{P}_{\beta} + \hat{K} n_{\alpha} \bar{\delta}_{\beta \delta} \bar{P}_{\gamma} \\
&\qquad  -  \bar{K}^{\mu} n_{\alpha} n_{\gamma} \bar{\delta}_{\beta \delta} \bar{P}_{\mu} -  \bar{K}_{\delta} n_{\gamma} \bar{\delta}_{\alpha \beta} \hat{P} + \bar{K}_{\beta} n_{\gamma} \bar{\delta}_{\alpha \delta} \hat{P} -  \hat{K} \bar{\delta}_{\alpha \delta} \bar{\delta}_{\beta \gamma} \hat{P} + \bar{K}_{\alpha} n_{\gamma} \bar{\delta}_{\beta \delta} \hat{P}\\
&\qquad  -  \hat{K} \bar{\delta}_{\alpha \gamma} \bar{\delta}_{\beta \delta} \hat{P} + \hat{K} \bar{\delta}_{\alpha \beta} \bar{\delta}_{\gamma \delta} \hat{P}-  \bar{K}_{\beta} n_{\alpha} n_{\gamma} \bar{P}_{\delta} + \hat{K} n_{\alpha} \bar{\delta}_{\beta \gamma} \bar{P}_{\delta}.
\end{aligned}
\end{equation}

\subsection{The Four Derivatives Sector}
\label{FourDerSecFTVars}
 
The independent scalar terms in the four derivatives sector are given in \cite{Arav:2016xjc}. Out of those, the ones which are at most second order in the background fields are the following:\footnote{Note that we changed the notations of $\widetilde{\nabla}$ and $\tilde\epsilon^{\mu\nu}$ from \cite{Arav:2016xjc} to $\bar\nabla$ and $\bar\epsilon^{\mu\nu}$ respectively, to better fit the notations of this paper.}
\begin{equation}
\label{eq:FPDInv4Der}
\begin{array}{l}
{\phi _1} = {{\hat R}^2}, \qquad {\phi _3} = {a^\alpha }{{\bar \nabla }_\alpha }\hat R, \qquad {\phi _4} = \hat R{{\bar \nabla }_\alpha }{a^\alpha },\\
{\phi _5} = {{\bar \nabla }^2}\hat R, \qquad {\phi _9} = {\left( {{{\bar \nabla }_\alpha }{a^\alpha }} \right)^2},\qquad {\phi _{10}} = {{\bar \nabla }_{(\alpha }}{a_{\beta )}}{{\bar \nabla }^{(\alpha }}{a^{\beta )}}, \qquad{\phi _{11}} = {a^\alpha }{{\bar \nabla }_\alpha }{{\bar \nabla }_\beta }{a^\beta },\\
{\phi _{12}} = {{\bar \nabla }^2}{{\bar \nabla }_\alpha }{a^\alpha }, \qquad {\phi _{27}} = {{\bar \epsilon }^{\mu \nu }}{{\bar \nabla }_\mu }{K_A}{{\bar \nabla }_\nu }{K_S}, \qquad {\phi _{28}} = {K_A}{{\bar \nabla }_\alpha }{{\bar \nabla }_\beta }{{\tilde K}_S}^{\alpha \beta },\\
{\phi _{29}} = {{\bar \nabla }_\alpha }{K_A}{{\bar \nabla }_\beta }{{\tilde K}_S}^{\alpha \beta }, \qquad {\phi _{30}} = {{\bar \nabla }_\alpha }{{\bar \nabla }_\beta }{K_A}{{\tilde K}_S}^{\alpha \beta }, \qquad {\phi _{32}} = {{\bar \epsilon }^{\mu \nu }}{\mathcal{L}_n}{{\bar \nabla }_\mu }{K_A}{a^\nu },\\
{\phi _{33}} = {{\bar \epsilon }^{\mu \nu }}{{\bar \nabla }_\mu }{K_A}{\mathcal{L}_n}{a^\nu }, \qquad {\phi _{38}} = {K_A}{\mathcal{L}_n}^2{K_A}, \qquad {\phi _{39}} = {\left( {{\mathcal{L}_n}{K_A}} \right)^2}.
\end{array}
\end{equation}
Note that terms which are of higher order in the background fields will not show up in our analysis, since we consider here only the two and three point correlation functions, as explained in section \ref{Sec:z2LishitzFreeScalarResults}.

Out of the scalar terms in \eqref{eq:FPDInv4Der}, the only ones which are first order in the background fields are $\phi_5$ and $\phi_{12}$. Their first order variations with respect to the vielbeins (evaluated in flat space and Fourier transformed) are given by:
\begin{align}
\label{eq:FirstVarPhi5}
&{\left[ {\delta {\phi _5}} \right]_{\alpha \beta }}\left( P \right)=2 \bar{P}_{\alpha} \bar{P}_{\beta} \bar{P}_{\gamma} \bar{P}^{\gamma} - 2 \bar{\delta}_{\alpha \beta} \bar{P}_{\gamma} \bar{P}^{\gamma} \bar{P}_{\delta} \bar{P}^{\delta},\\
\label{eq:FirstVarPhi12}
&{\left[ {\delta {\phi _{12}}} \right]_{\alpha \beta }}\left( P \right)=- n_{\alpha} n_{\beta} \bar{P}_{\gamma} \bar{P}^{\gamma} \bar{P}_{\delta} \bar{P}^{\delta} + n_{\beta} \bar{P}_{\alpha} \bar{P}_{\gamma} \bar{P}^{\gamma} \hat{P}.
\end{align}
The second order variation of the terms in \eqref{eq:FPDInv4Der} with respect to the vielbeins (evaluated in flat space and Fourier transformed) are as follows:


As found in \cite{Arav:2016xjc}, this sector contains 12 independent trivial terms (coboundaries). The 10 of them which are at most second order in the background fields have the following densities:\footnote{The expressions in equations \eqref{eq:4DerTriv}--\eqref{eq:4DerAnomaliesExpr} are taken from equations (D.12)--(D.13) in \cite{Arav:2016xjc}. The ellipsis ($\ldots$) in these expressions stands for terms which are more than second order in the background fields, i.e.\ terms whose second order variations with respect to the vielbeins vanish in flat space. These terms will therefore not contribute to the two and three point anomalous Ward identities.}
\begin{align*}
{\mathcal{F}_1} &=  - 8{\Phi_3} - 4{\phi_4} - 4{\phi_5}+\ldots\, , \\
{\mathcal{F}_2} &=  - 4{\phi_3} - 4{\phi_4} - 4{\phi_{10}} - 4{\phi_{11}} + 8{\phi_{31}} - 8{\phi_{32}} - 8{\phi_{39}}+\ldots \, ,\\
{\mathcal{F}_3} &= 2{\phi_4} - 2{\phi_5} + 2{\phi_9} + 4{\phi_{10}} + 8{\phi_{11}} + 2{\phi_{12}} - 8{\phi_{31}} + 8{\phi_{32}} + 8{\phi_{39}}+\ldots \, ,\\
{\mathcal{F}_5} &=  - 2{\phi_2} + 4{\phi_9} - 4{\phi_{10}} + 8{\phi_{31}} - 8{\phi_{32}} - 8{\phi_{39}} +\ldots \, ,\\
{\mathcal{F}_6} &= 2{\phi_3} + 2{\phi_4} - 2{\phi_9} + 8{\phi_{10}} + 10{\phi_{11}} + 4{\phi_{12}} - 16{\phi_{31}} \\
&\quad + 12{\phi_{32}} - 8{\phi_{33}} - 24{\phi_{38}} +\ldots \, , \numthis
\label{eq:4DerTriv}\\
{\mathcal{F}_8} &=  - 2{\phi_{27}} - 2{\phi_{32}} - 2{\phi_{33}} - 4{\phi_{38}} - 4{\phi_{39}} +\ldots \, ,\\
{\mathcal{F}_9} &= 2{\phi_{27}} + 2{\phi_{32}} + 2{\phi_{33}} +\ldots \, ,\\
{\mathcal{F}_{10}} &= 2{\phi_{28}} + 4{\phi_{29}} + 2{\phi_{30}} +\ldots \, ,\\
{\mathcal{F}_{11}} &=  - 2{\phi_{28}} - 2{\phi_{29}} +\ldots \, ,\\
{\mathcal{F}_{12}} &= 2{\phi_{27}} + 2{\phi_{32}} + 2{\phi_{33}} - 4{\phi_{38}} - 4{\phi_{39}} +\ldots \, .
\end{align*}
This sector also contains 4 independent anomaly terms. The 3 of them which are at most second order in the background fields have the following densities:
\begin{equation}
\label{eq:4DerAnomaliesExpr}
\begin{array}{l}
{\mathcal{A}_2}^{\left( {0,4,0} \right)} =\phi_{38}+\ldots = {K_A}{\mathcal{L}_n}^2{K_A}+\ldots \, ,\\
{\mathcal{A}_3}^{\left( {0,4,0} \right)} = \phi_{30}+\ldots={{\tilde K}_S}^{\alpha \beta }{{\bar \nabla }_\alpha }{{\bar \nabla }_\beta }{K_A}+\ldots \, ,\\
{\mathcal{A}_4}^{\left( {0,4,0} \right)} = \phi_1+2\phi_4+\phi_9={\left( {\hat R + {{\bar \nabla }_\alpha }{a^\alpha }} \right)^2}.
\end{array}
\end{equation}
The second variations of these anomaly and trivial term densities can now be obtained from linear combinations of the expressions \eqref{eq:SecVarPhi1}--\eqref{eq:SecVarPhi39}.

Finally, we give here the complete result for the anomalous contribution to the three point Ward identity \eqref{CorrLifshitzWard:WardIdentForVarCorrWeyl3Point} in the four derivative sector:
%

\end{document}